\documentclass[a4paper,11pt]{article}
\pdfoutput=1 % if your are submitting a pdflatex (i.e. if you have
\usepackage{jheppub} % for details on the use of the package, please
\usepackage[T1]{fontenc} % if needed

\usepackage{epstopdf}
\usepackage{graphicx}
\usepackage{epsfig}
\usepackage{dcolumn}  
\usepackage{bm}    
\usepackage{caption}
\usepackage{subcaption}
\usepackage{amssymb} 
\usepackage{tikz}
\usetikzlibrary{arrows}
\usepackage{xcolor}
\usepackage{epstopdf}
\usepackage{graphicx}
\usepackage{epsfig}
\usepackage{amsmath,bm}
\usepackage{amsfonts}  
\usepackage{amsmath}  
\usepackage{slashed}  
\usepackage{enumitem}
\usepackage[mathscr]{euscript}
\usepackage{tabu}
\usepackage{epsfig}
\makeatletter
\g@addto@macro\bfseries{\boldmath}
\makeatother

\def\Re{R\'{e}nyi }
\def\l1{{{1-loop}}}

\def\n1{\Bigg|_{n=1}}

\def\n{{(n)}}

\usepackage[T1]{fontenc} % if needed
\usepackage{tikz}
\usepackage{amsmath,amssymb}
\usepackage{relsize}
\usepackage{latexsym}
\usepackage{leftidx}
\usepackage{xcolor}
\usepackage{diagbox}
\usepackage[T1]{fontenc}
\usepackage{array}
\usepackage{makecell}
\usepackage{csquotes}
\usepackage{tikz}
\usepackage{enumitem}
\usepackage{setspace}
\usepackage{multirow}
\usepackage{amsmath,amssymb}
\usepackage{relsize}
\usepackage{latexsym}
\usepackage{leftidx}
\usepackage{xcolor}
\usepackage{csquotes}
\usepackage{tikz}
\usepackage{enumitem}
\usetikzlibrary{decorations.markings}
\usetikzlibrary{decorations.pathmorphing}
\usetikzlibrary{decorations.markings}
\usetikzlibrary{decorations.pathmorphing}
\usepackage{pifont}
\usepackage{hyperref}

\usepackage{bookmark}

 \title{\textbf{\textsf{Entanglement entropy of local  gravitational quenches
}}}
  \author{Justin R. David, Jyotirmoy Mukherjee}
\affiliation{\vspace{.1cm} Centre for High Energy Physics, \\ Indian Institute of Science,\\
C. V. Raman Avenue, Bangalore 560012, India.}
\emailAdd{justin@iisc.ac.in, jyotirmoym@iisc.ac.in}

\abstract{
We study the time dependence of R\'{e}nyi/entanglement entropies of
 locally excited states  created by fields with integer spins $s \leq 2$ in $4$ dimensions. 
 For spins 0, 1  these states are characterised by  localised  energy densities of a given width which
 travel as a spherical
wave at the speed of light. 
For  the spin 2 case, in the absence of a local gauge invariant stress tensor, we probe these states
with the Kretschmann scalar and show they represent localised curvature densities 
which travel at the speed of light.  We consider the reduced density matrix 
of the half space with these excitations and develop methods which include a 
convenient gauge choice  to evaluate 
the time dependence of R\'{e}nyi/entanglement entropies as these quenches 
enter the half region.  In all cases, the entanglement entropy grows in time and saturates at 
$\log 2 $.  
In the limit, 
 the width of these excitations tends to zero, the growth 
 is determined by order $2s+1$  polynomials  in the ratio of 
 the  distance from the co-dimension-2 entangling surface and  time.
 The polynomials corresponding to  quenches 
created by  the  fields can be organised 
in terms of their  representations under  the $SO(2)_T\times SO(2)_L$  symmetry preserved by the presence of 
the co-dimension 2 entangling surface.
For fields transforming as scalars under this symmetry, the order $2s+1$ polynomial is 
completely determined by the spin. 

}

\begin{document} 
\maketitle
\flushbottom
\section{Introduction}

The study of systems driven out of equilibrium and is subsequent evolution is important to understand 
the phenomenon of thermalization in quantum systems. 
Holography relates thermalization  in  theories which admit gravitational duals 
  to the formation of black holes in the bulk which is another important  question in quantum gravity, see \cite{Liu:2018crr} for a review on the
   holographic aspects of systems away from equilibrium. 
A simple way to kick a system out of equilibrium is to change the Hamiltonian of the theory at  a given time and 
then study the dynamics as the system evolves unitarily with the new Hamiltonian \cite{Calabrese:2005in,Calabrese:2007rg}. 
These quenches can be of 2 types,  one in which the parameters of the Hamiltonian is changed uniformly 
in space which is called the `global quench'. 
The other in which the quench is generated by a localized change in the initial quantum state or the density matrix
so that the system is excited and departs from the ground state only locally at a specific instance of time. 
Then one studies the evolution of various observables subsequently, such quenches are termed 
`local quenches'  \cite{Calabrese:2007mtj,Eisler_2007,St_phan_2011,Asplund:2013zba,Calabrese:2016xau}.

One specific protocol to set up local quenches is to insert a local operator at some point, which then creates a 
local change in the ground state density matrix.  The time evolution of observables after local
such quenches have been extensively studied in  CFT and even in non-conformal field theories \cite{Ageev:2022kpm}.
In CFT's that admit a gravity dual, such quenches are dual to an in-falling particle in the bulk. 
Most of these detailed  studies  have been confined to 2 dimensions \cite{Asplund:2011cq,Nozaki:2013wia,Nozaki:2014hna,Caputa:2014vaa,Nozaki:2014uaa,He:2014mwa,Caputa:2015tua,Asplund:2014coa,Caputa:2014eta,David:2016pzn,David:2017eno,Kusuki:2017upd,Zhang:2019kwu,Kusuki:2019evw}. The quantum corrections to the entanglement entropy of local quenches have also been studied both in CFT and holography \cite{Agon:2020fqs}.  
But there have been  only a  few studies in higher dimensions, particularly in 4d  for quenches created by free fields 
and restricted to spins $\leq 1$ \cite{Nozaki:2014hna,Nozaki:2014uaa,Nozaki:2015mca,Nozaki:2016mcy}. 
In this paper, we  revisit the study of these quenches  and develop methods so that we can study 
quenches created by the linearized graviton. 
The observables that we wish to study after the quench  are the R\'{e}nyi entropies and 
entanglement entropy.  In this context it is important to point out that the entanglement properties
of the linearized graviton have only been recently studied. 
The coefficient of the logarithmic term for the entanglement entropy  of a spherical region in 
a theory of linearized gravitons has been evaluated in 
\cite{Benedetti:2019uej,David:2020mls}. The contribution to the entropy from gravitational 
 edge modes has been understood in \cite{David:2022jfd}. 
The graviton after all is a fundamental particle of nature,  and therefore it is useful and important 
to study its information theoretic 
properties.

Let us briefly describe the set up for the local quenches we  will consider  in this paper. 
We excite the ground state by an operator ${\cal O}$, 
then the density matrix at time $t$ is given by 
\begin{eqnarray} \label{quenchs}
\rho = {\cal N}   e^{- i Ht } e^{-\epsilon H}   {\cal O } ( y)   \big|0\big\rangle\big\langle 0\big|   {\cal O}^\dagger ( y) 
 e^{-\epsilon H}  e^{ i Ht }.
\end{eqnarray}
Here the excitation  is created by the operator ${\cal O}$ inserted at coordinates  $ y = (0, 0,  y_0, 0)$, 
the density matrix is then evolved by time $t - i \epsilon$. ${\cal N}$ is the factor used to normalised the state. 
The small imaginary part added to the 
time evolution,  gives a width to the excitation. 
An example for ${\cal O}$ is the conformal scalar field  $\phi$.  On 
probing this state by the stress tensor  we find that the 
the excitation creates a spherical pulse of energy which travels as a spherical wave at the speed of light
as shown in figure \ref{fig0}. 
Consider the reduced density matrix $\rho_A$ obtained by tracing out the region $y<0$,  the entangling surface
is  a plane of co-dimension $2$. 
We study the evolution of  both the R\'{e}nyi and entanglement entropies of this local quench. 
These entropies are defined by 
\begin{eqnarray}
S^{(n)}_A = \frac{\log {\rm Tr} (\rho_A)^n}{1- n}. \qquad 
\end{eqnarray}
As an example consider the  excitation corresponding to the scalar, the difference in entanglement entropy from that of the 
ground state is shown in the figure \ref{fig2}. 
As the pulse enters the entangling region, the entanglement increases and saturates to $\log(2)$. 

In this paper we  study the growth of entanglement for excitations for operators with 
spins $\leq 2$.  For local quenches in $2d$, the entanglement growth saturates to the logarithm of the 
quantum dimension of the corresponding operator \cite{He:2014mwa}. 
When  width $\epsilon$ approaches 
zero, the growth of  entanglement follows a step function \footnote{The leading behaviour 
of entanglement growth in $2d$ local quenches at 
finite width $\epsilon$ has been shown to be universal \cite{David:2016pzn}.  }
However in $d>2$,  even in the zero width limit, the entanglement growth  depends on a non-trivial function
and then saturates at $\log(2)$ for free fields.  In this paper we obtain this function for quenches created 
by all curvature components of the Riemann tensor. 
The saturation at $\log(2)$ has been interpreted 
in terms of entanglement pairs \cite{Nozaki:2014hna}. 

We first revisit the study of the quenches due to a scalar and   $U(1)$ field strengths 
and then extend the study to all the components of the curvature tensor of the linearised graviton. 
Note that for theories with local gauge invariance, the natural  local gauge invariant operators are 
field strengths.  Therefore we use components of the  Riemann as operators to create the 
quench. 
Furthermore,  since a  local stress tensor does not exist  for the graviton, we demonstrate the state 
that we consider behaves as a local quench by probing it with the Kretschmann scalar. 
The behaviour of the Kretschmann scalar for a graviton quench is shown in figure  \ref{kscalar}. 
Thus the quench created by Riemann tensors, represents a curvature density travelling at the 
speed of light.  
The study of R\'{e}nyi/entanglement entropies involve evaluating $2n$ point functions of the operators 
creating the quench on the replica surface. 
For the spin-1 and the graviton quenches, we need to choose a convenient gauge for
w the propagator of the gauge potential and the metric on the replica surface. 
We develop the gauge used in \cite{Candelas:1977zza,David:2020mls} for this purpose.  The presence of the 
co-dimension 2
entangling surface reduces the symmetry of the system to $SO(2)_T\times SO(2)_L$, where  the 
subscripts $T, L$ refers to 
transverse and longitudinal directions to the entangling surface. The gauge choice is compatible with this 
symmetry.

Since we are dealing with free fields, 
the $2n$ point function on the Replica surface  which captures the R\'{e}nyi entropies 
 of the operators creating the quench can be evaluated using Wick contractions. 
 In the limit of zero width of the quench, we can isolate the leading Wick contractions 
 and obtain the time dependence of the growth of R\'{e}nyi entropies when the quench enters
 the region $y>0$. 
 In all cases, including the quenches created by Riemann tensor components
  the entanglement grows and saturates at $\log(2)$. 
  We observe that the time dependence of R\'{e}nyi/entanglement entropies  for quenches created 
  by the spin $s$ field is determined by order $2s +1$ polynomials in $r = \frac{y_0}{t}$, where
  $y_0$ is the perpendicular distance at which the quench was released and $t$ is the time elapsed.
  We find the structure of the polynomials can be organised in terms of 
  the $SO(2)_T\times SO(2)_L$ representation of the operator creating the quench. 
  These polynomials can be read from the tables \ref{table1}, \ref{table3},  \ref{table5} for scalars, vectors and 
  symmetric tensors of  $SO(2)_T\times SO(2)_L$ respectively. 
  The growth of the entanglement entropy for the corresponding quenches are given in figures 
  \ref{fig2}, \ref{fig3}, \ref{fig4}. 
 The polynomial which determines the growth of R\'{e}nyi entropies for fields transforming as 
  scalar under $SO(2)_T\times SO(2)_L$ is completely determined by the spin. 
 The constraints on the polynomial also show that the long time behaviour of R\'{e}nyi entropies 
 can always be written as 
 \begin{equation}
 \lim_{t\rightarrow \infty}  \Delta S^{(n)}(t)  = \log 2 -  \frac{n}{2} \frac{ y_0^2 }{ t^2}   a_\infty^2 + \cdots
 \end{equation}
 where $a_\infty$ is a rational number and can be read out from tables \ref{table1}, \ref{table3},  \ref{table5}.

 One of our main motivations for this work is to explore the information theoretic properties of 
 the graviton.  As we mentioned earlier,  some of  these properties of the graviton
 have only been recently investigated.
  It has been argued that in a theory of quantum 
 gravity that  the wave function exterior to a given sub-region determines the wave function interior
 to it \cite{Laddha:2020kvp,Raju:2021lwh}. Therefore the `split property' of quantum local  field theories 
 does not hold in the theory of  quantum gravity.  
  From  this paper as well as the earlier work \cite{Benedetti:2019uej,David:2020mls,David:2022jfd}, 
 we see that  the theory of linearised gravitons behaves just as a local quantum field 
 theory with gauge symmetry. 
 It will be interesting to see  explicitly how  this behaviour departs from that of a local quantum field theory 
 once the theory becomes interacting.  The methods developed in this paper, especially the construction of the 
 graviton propagator on the  replica surface will be useful to study this question.

The organization of the paper is as follows. 
In section \ref{scalvecq} we discuss the set up for the quench in detail and revisit the study of quenches 
created by the scalar and the $U(1)$ field strengths. 
Here we introduce the gauge which is consistent with the  $SO(2)_T\times SO(2)_L$ symmetry of the problem. 
We also develop simple methods to obtain the leading contributions 
of the $2n$ point function which determines the R\'{e}nyi entropies. 
These methods make it convenient to study the graviton quenches. 
Then in section \ref{gravq} we study all the independent  quenches due to components of the curvature tensor. 
Our results are summarised in the tables and figures of this section. 
Section \ref{conclusions} contains our conclusions. 
The appendices \ref{dercorgrav}, \ref{appendixc} \ref{idproof}  contain details of the calculations to arrive at our results. 
Appendix \ref{appendixa} shows how the  scalar correlator on the replica surface can be interpreted 
as a 2 point function in a BCFT.

\section{Scalar and vector quenches} \label{scalvecq}

In this section we first describe the set up for local quenches.
The correlator which evaluated the change in R\'{e}nyi entropy/entanglement entropy 
as the quench enters the entangling region is the $2n$ point correlator on the replica geometry \cite{Nozaki:2014hna}. 
We  re-examine local quench represented by the state (\ref{quenchs}) where ${\cal O} =\phi (y) $ the free scalar field in 
$4d$.  We will see that to obtain the growth of entanglement when the width $\epsilon\rightarrow 0$ it is sufficient 
to study the $2$ point function of the scalar field in the replica geometry. 
We cast it as a correlator in a
boundary conformal field theory, this allows us to isolate the singularities of the correlator which are responsible for the 
growth of the entanglement entropy as the pulse or the local quench enters the entangling region. 
We carry this observation ahead for the Maxwell theory.  
For the Maxwell theory, the quenches we consider are where ${\cal O} = F_{\mu\nu} (y)$, the field strength, which is a local 
gauge invariant operator. 
One consistency check we perform for both the scalar and Maxwell theory 
relates to the leading correction to the shape of the 
growth of entanglement due to the   width $\epsilon$ before the pulse enters the 
entangling region. This correction is determined by the expectation value of the composite bilinear  $:{\cal O}^2:$  
in the replica geometry.

As we have stated in the introduction we wish to consider the quench represented by the state
\begin{eqnarray}
\rho &=& e^{\tau_e  H} {\cal O}( 0, -y_0) \big|0\big\rangle \big\langle 0\big| {\cal O} ( 0, -y_0) e^{-\tau_l H} , 
\\ \nonumber
&=&
{\cal O} ( \tau_e, -y_0,) \big|0\big\rangle \big\langle 0\big| {\cal O } ( \tau_l, -y_0) ,
\end{eqnarray}
where $\tau_e = -\epsilon  - i t, \tau_l = \epsilon - i t$ and we have assumed ${\cal O}$ is Hermitian. 
Let $\rho_A$ be the reduced density matrix obtained by tracing over the region $y<0$, therefore the $x,z$ plane divides 
space into two parts. 
To obtain $S^{(n)}_A$ we can appeal to the path integral  and write it as a path integral over replicas. 
The reduced density matrix  $\rho_A^n$  can be written as the following partition function
\begin{equation}
{\rm Tr}( \rho_A^n) = \frac{ Z_n}{( Z_1) ^n}.
\end{equation}
Here $Z_n$ is the path integral over the surface $\Sigma_n$ which is a 
$n$-branched cover  over $y>0$. On each sheet of the cover
there is  a pair of  operators  inserted at points $ r_e e^{i \theta_e^{( k ) }}, r_l e^{ i \theta_e^{(k) }} $ where 
$(k)$ labels the sheet.  These locations are specified by  setting
\begin{eqnarray}
r_e e^{ i \theta_e}  &=& - y_0  + i \tau_{e} , \qquad  r_l e^{ i \theta_l}  = - y_0  + i \tau_{l},  \\ \nonumber
\theta_e^{( k) } &=& \theta_e + 2\pi ( k-1) , \qquad  \theta_l^{( k) } = \theta_l + 2\pi ( k-1) , \quad k = 1, \cdots n.
\end{eqnarray}
The angle $\theta$ takes values from $0$ to $2\pi n$ on the branched cover $\Sigma_n$. 
The partition function $Z_1$ is just the normalization needed to ensure ${\rm Tr}\rho =1$ and therefore it is 
$2$ point function of $\langle {\cal O} ( \tau_e, -y_0)   {\cal O} ( \tau_l, -y_0)   \rangle_{\Sigma_1}$ in 
$4d$ flat Euclidean space. 
Similarly, when there is no insertions of operators, that is when we consider the vacuum we have the 
partition functions $Z_{0n}$ and $Z_{01}$. The former is the path integral of the theory on the $n$-branched replica surface and the latter is the path integral on $4d$ flat Euclidean space. 
The reduced density matrix of the vacuum is given by 
\begin{equation}
{\rm Tr} \rho_{0A}^n  = \frac{ Z_{0n} }{( Z_{01} ) ^n},
\end{equation}
where the subscript $`0'$ refers to the vacuum. 
Using these partition functions we can write the difference of the  R\'{e}nyi/entanglement entropy of the quench compared to the vacuum
\begin{eqnarray} \label{renform}
\Delta S_A^{(n)} &=& \frac{1}{1-n}\left(  \log{\rm Tr} \rho_A^n -  \log {\rm Tr}  \rho_{0A}^n \right) , \\ \nonumber
&=&\frac{1}{1-n}\left(  \log \frac{Z_n}{Z_{0n}} - n \log \frac{ Z_1}{Z_{01}} \right) , \\ \nonumber
&=& \frac{1}{1-n} \left( \log  \Big\langle \prod_{k = 1}^n {\cal O } ( r_l,  \theta_l^{(k)})  {\cal O } ( r_e, \theta_e^{(k)} ) \Big\rangle_{\Sigma_n}
 - n \log \langle {\cal O }( r_l , \theta_l) {\cal O }( r_e , \theta_e) \rangle_{\Sigma_1}  \right).
\end{eqnarray}
The study of the time evolution of the local quench  due to an operator
reduces to the evaluation of its $2n$ point function on the replica surface. For free fields, this $2n$ point function 
can be obtained using Wick contractions of the $2$ point function.

\subsection{Time evolution of the scalar quench}

\begin{figure}
\centering
\begin{subfigure}{.5\textwidth}
  \centering
  \includegraphics[width=1\linewidth]{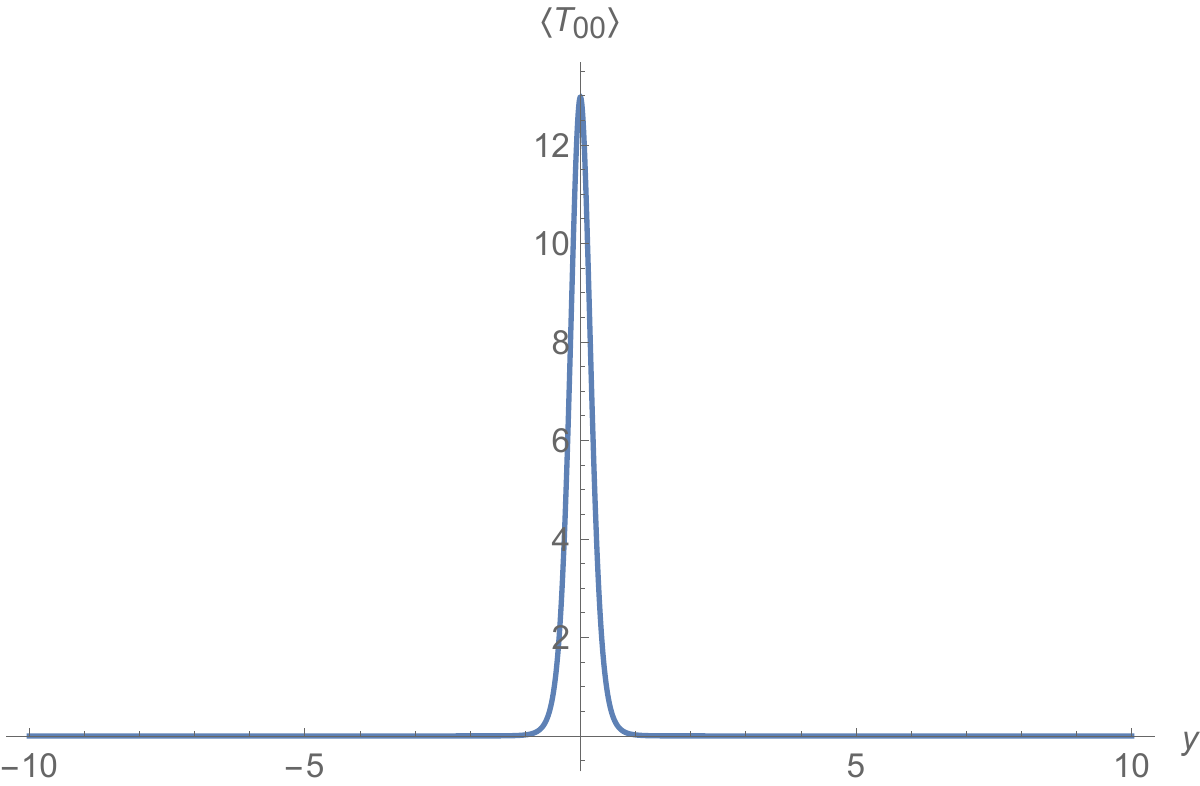}
  \caption{$t=0$ and $\epsilon=0.5$}
\end{subfigure}%
\begin{subfigure}{.5\textwidth}
  \centering
  \includegraphics[width=1\linewidth]{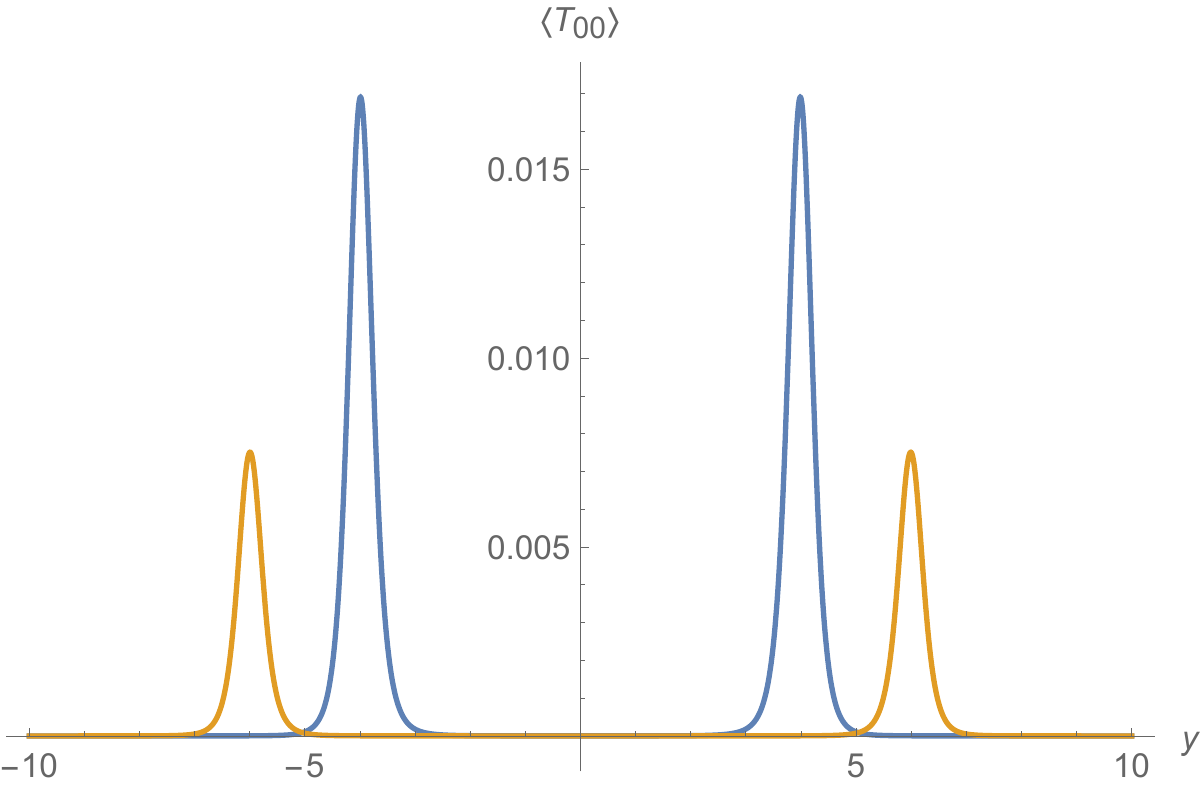}
  \caption{Blue: $t=4$, orange : $t=6$, $\epsilon=0.5$.}
\end{subfigure}
\caption{Energy density profile in the $y$ direction
 for the excitation created by a  quench due to the scalar $\phi$ placed at the origin at times $t=0, 4, 6$. The width $\epsilon = 0.5$. 
 The energy density is spherically symmetric and travels at the speed of light. Unlike in $d=2$, the amplitude 
 decreases in time as the density spreads over a sphere of increasing radius. }
\label{stress_scalar}
\end{figure}

Let us consider the case when the quench is due to the free scalar $\phi$ given by 
\begin{equation}\label{densca}
\rho (t, - y_0) = 
{\phi} ( \tau_e, -y_0,) \big|0\big\rangle \big\langle 0\big| {\phi } ( \tau_l, -y_0) .
\end{equation}
To show that this state corresponds to a pulse with width $\epsilon$ moving at the speed of light, we can evaluate the 
expectation value of the stress tensor on this state.  The stress tensor for the conformal scalar is given by 
\begin{eqnarray}
    T_{00}&=& \partial_0\phi\partial_0\phi-\frac{1}{12}\left(2\partial_0^2-\Box\right)\phi^2 ,\\ \nonumber
    &=& \frac{1}{2}\partial_0\phi\partial_0\phi-\frac{1}{2}\phi\partial_0^2\phi+\frac{1}{6}\partial_y\phi\partial_y\phi+\frac{1}{6}\phi\partial_y^2\phi+\frac{1}{6}\left(\partial_x\phi\partial_x\phi+\partial_z\phi\partial_z\phi\right)+\frac{1}{6}\left(\phi\partial_x^2\phi+\phi\partial_z^2\phi\right).
\end{eqnarray}
The expectation value of the stress tensor placed at the origin reduces to the evaluation of the following 3pt function
\begin{eqnarray}\label{expstresss}
{\rm Tr} \big[ T_{00}( 0, 0 )  \rho ( t, - y_0)  \big] &=& {\rm Tr} \big[ T_{00}  ( 0, y_0) \rho ( t, 0)  \big] ,  \\ \nonumber
&=&  \frac{\langle \phi(-\epsilon - i t ,0)T_{00}(0, y_0 )\phi(\epsilon - i t ,0) \rangle}{\langle \phi(-\epsilon - it, 0)
\phi( \epsilon - i t, 0)\rangle}.
\end{eqnarray}
Here we have specified only the temporal and the $y$ coordinate, it is understood that the fields are placed at 
$x_1=x_2,  z_1 = z_2$. 
In the first line of the above equation we have used translation invariance.  All the correlators are evaluated 
in $4d$  Euclidean space, the two-point function of the scalar is given by 
\begin {equation}\label{r4cor}
\langle \phi ( x) \phi ( y) \rangle_{\Sigma_1}= \frac{1}{4\pi^2 |x- y|^2 } .
\end{equation}
Evaluating both the $3$ point function and $2$ point in (\ref{expstresss}) using Wick's rule we obtain 
\begin{equation}
{\rm Tr} \big[ T_{00}( 0, y_0 )  \rho ( t, 0)  \big]= 
\frac{8\epsilon ^4 \left(3 t^4+2 t^2 \left(5 y_0^2+3 \epsilon ^2\right)+3 \left(y_0^2+\epsilon ^2\right)^2\right)}{3 \pi^2 \left(t^4+2 t^2 \left(\epsilon ^2-y_0^2\right)+\left(y_0^2+\epsilon ^2\right)^2\right)^3}.
\end{equation}
We have verified that this correlator agrees with the general expression for the $3$ point function of the 
stress tensor with $2$ scalar operators obtained in \cite{Osborn:1993cr} using the Ward identity. 
We interpret this result as the expectation value of the stress tensor placed at $y_0$ after the quench placed at the origin  has evolved for a 
time $t$.  The behaviour of the stress tensor is shown in figure \eqref{stress_scalar}. 
Since there is spherical symmetry in the set up, the  quench is a spherical pulse of energy of width 
$\epsilon$ travelling at the speed of 
light.  Note that the amplitude of the pulse decreases in time as the energy spreads over a sphere of increasing 
radius. 
 \label{fig0}
\subsubsection*{The two-point function on $\Sigma_n$  and its singularities}

\begin{figure}[htb] 
\centering
\begin{subfigure}{.5\textwidth}
  \centering
  \includegraphics[width=1\linewidth]{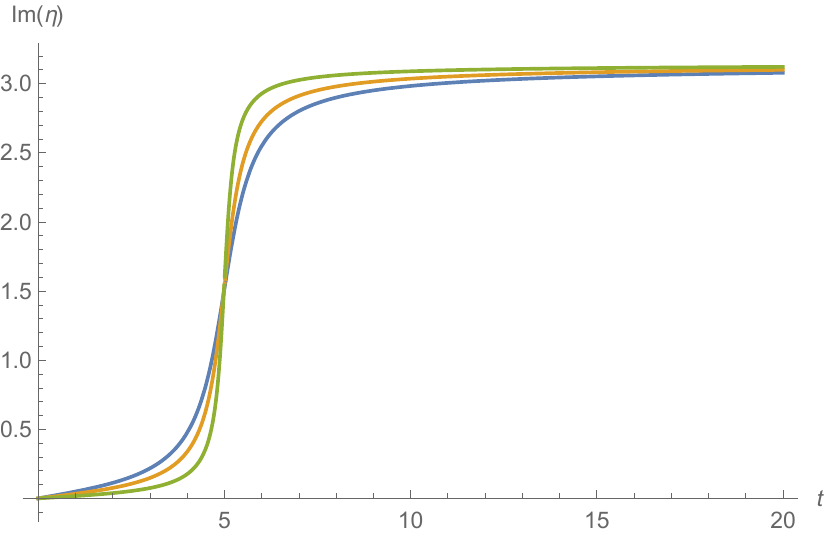}
  \caption{  Imaginary part of $\eta$ as a function of real-time.}
  \label{fig:sub12}
\end{subfigure}%
\begin{subfigure}{.5\textwidth}
  \centering
  \includegraphics[width=1\linewidth]{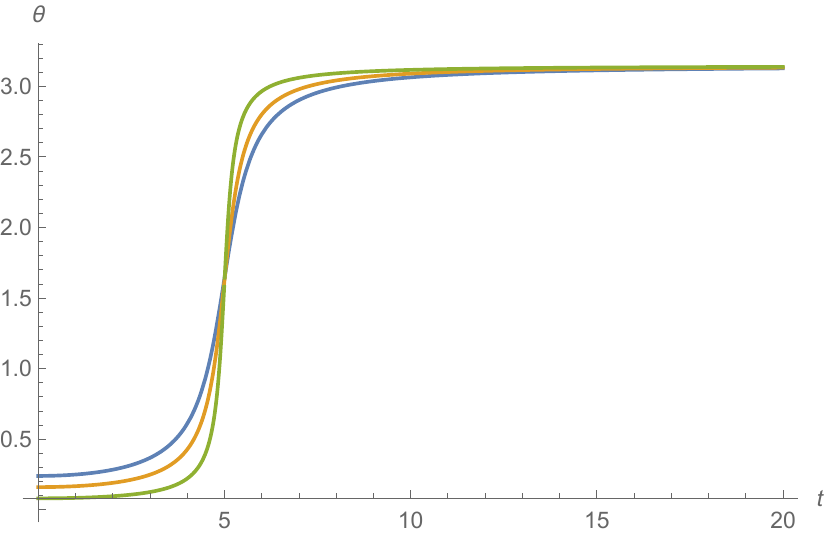}
  \caption{$\theta$ as a function of real-time.}
  \label{fig:sub22}
\end{subfigure}
\caption{We plot imaginary part of $\eta$ and $\theta$ as a function of real-time by varying $\epsilon$. Blue line corresponds to $\epsilon=0.6$, orange line corresponds to $\epsilon=0.4$, and green line corresponds to $\epsilon=0.2$. We keep $y_0=-5$ in all cases.} 
\label{figt1}
\end{figure}

As discussed in the paragraph leading to  (\ref{renform}), to 
evaluate the change in  R\'{e}nyi/entanglement entropy when the quench enters the region $y>0$, we need the 
two-point function of the scalar field on replica surface which is a cone in which the angle $\theta$ is identified with 
$\theta + 2\pi n $.  The Greens function is obtained by solving the differential equation
\begin{align}
    \left(\partial_r^2+\frac{1}{r}\partial_{r_1}+\frac{1}{r_1^2}\partial_{\theta_1}^2+\partial_{\textbf{x}_1}^2\right) G(r_1,r_2,\theta_1,\theta_2,\textbf{x}_1,\textbf{x}_2)&=-\frac{1}{r_1}\delta (r_1-r_2)\delta(\textbf{x}_1-\textbf{x}_2), \quad 
\end{align}
with the boundary condition  $G(r_1, r_2,\theta_1, \theta_2,  \textbf{x}_1, \textbf{x}_2)=
G(r_1, r_2,\theta_1+2\pi n,  \textbf{x}_1, \textbf{x}_2).$ The label $\textbf{x}$ refers to the cartesian $x, z$ directions. 
The solution of the Greens function is well known and has been studied in various contexts before, see 
\cite{Nozaki:2014hna} for a recent discussion. 
\begin{eqnarray}\label{sclcor}
 \langle \phi( r_1, \theta_1, \textbf{x}_1)  \phi ( r_2, \theta_2, \textbf{x}_2) \rangle_{\Sigma_n} &=&
  G(r_1,r_2,\theta_1,\theta_2,\textbf{x}_1,\textbf{x}_2), \\ \nonumber
 & =&  \frac{\sinh \frac{\eta}{n}}{8\pi^2 nr_1 r_2 \sinh \eta(\cosh \frac{\eta}{n}-\cos\frac{\theta_1-\theta_2}{n})},  \nonumber 
 \end{eqnarray}
 with
 \begin{eqnarray}\label{defcross}
    \cosh{\eta}=\frac{r_1^{2} +r_2^{2}+(x_1-x_2)^{2}+(z_1-z_2)^{2}}{2r_1 r_2},   & \qquad &
  \cos(\theta_1 - \theta_2) =  \frac{ y_1 y_2   + \tau_1\tau_2}{r_1 r_2} , \nonumber \\ \quad
 r_1 = y_1^2 +  \tau_1^2 ,  &\qquad& \qquad r_2 =  y_2^2 + \tau_2^2 .\nonumber
\end{eqnarray}
The expression of the Greens function suggests that the correlator is that of 2 conformal scalars of unit dimensions
in a BCFT with a co-dimension $2$ surface in $d=4$. This can be seen from comparing the expressions in 
\cite{Billo:2016cpy}. 
The behaviour correlator is determined by the   cross ratios $\eta$ and  $\theta = \theta_1 - \theta_2$. 
Let us study their behaviour as the quench starting from $y  =- y_0 <0$ and enters the region $y>0$. 
This is easily  seen in the plots shown in figure \ref{figt1}. 
To obtain these plots we have chosen $x_1 = x_2=z_1=z_2 = 0, y_0= -5$ and  three different values of $\epsilon$.
 Note that the angle $\eta$ is purely imaginary and 
and it begins with $\eta =0$ and for $t> y_0$ it reaches $\eta = -i \pi$, similarly $\theta$ begins with $\theta = 0$ and 
asymptotically tends to the value of $\theta = \pi$.  The transition is sharper as the width of the pulse becomes
narrower.  

 As seen from the plots, the cross ratios  would have different expansions for $t<y_0$ and for $t>y_0$. Let us first expand the parameters in small $\epsilon$ for $t<y_0$ case,
 \begin{align}\label{before}
            \begin{split}
            \eta_{\rm{b}} &= i\tan^{-1} \left( \frac{ \epsilon}{ y_0 - t}\right)   -i \tan^{-1}\left(  \frac{\epsilon}{ y_0+t} \right) , \\ 
          \theta_{\rm{b}} &= \tan^{-1} \left( \frac{\epsilon}{ y_0-t} \right) + \tan^{-1} \left( \frac{\epsilon}{ y_0 +t}  \right) .
           \\
            \end{split}
        \end{align}
        These expressions should be understood as a power series expansion in $\epsilon$. 
        Similarly the expansions of the cross ratios for $t>y_0$ are given by 
         \begin{align}\label{after}
            \begin{split}
            \eta_{\rm{a}} &= i \pi  - i \tan^{-1} \left( \frac{\epsilon}{ t -y_0} \right) - i\tan^{-1} \left( \frac{\epsilon}{ t+ y_0} \right) , 
            \\
            \theta_{\rm{a}} &= \pi  - \tan^{-1} \left( \frac{\epsilon}{ t- y_0} \right) + \tan^{-1} \left( \frac{\epsilon}{ t+ y_0} \right)  .
            \end{split}
        \end{align}

    We now examine the leading divergence of the correlator (\ref{sclcor}) 
     for $t<y_0$ in the $\epsilon\rightarrow 0$ limit. 
    This will be useful to isolate the leading behaviour of the  $2n$ point function in 
     (\ref{renform}) for $t< y_0$ and obtain the 
    R'{e}nyi entropy.  
    Note that in this regime both the cross ratios are proportional to $\epsilon$ and tend to zero. 
    Examining the correlator it is clear that the divergence arises due to the term 
    $\big( \cosh\frac{\eta}{n} - \cos\frac{\theta}{n}\big) $
    in the denominator which is proportional to $\epsilon^2$. 
        \begin{eqnarray}
             \lim_{\epsilon\rightarrow 0} 
             G(r_1,r_2,\theta_1,\theta_2 )_{(t<y_0)}&=&\frac{1}{4\pi^2(y^2_0-t^2)(\eta^2+\theta^2)}+ O(\epsilon^0), \\ \nonumber
             &=& \frac{1}{16\pi^2 \epsilon^2} +  O(\epsilon^0).
        \end{eqnarray}
    In the second line we have substituted the leading term in the expansions of the cross ratios  from (\ref{before}). 
    Thus this correlator in the small $\epsilon$ limit reduces to the correlator on $R^4$. 
    Essentially the 2 operators separated in Euclidean time on the same sheet approach each other closely, so the 
    fact that they are on the replica surface does not matter. 
    Note that we have suppressed  the dependence of the correlator on the $x, z$ coordinates since  the 
    Green's function is independent of these coordinates.
     It is useful to keep track of the $O(\epsilon^0)$ which is given by 
    \begin{equation} \label{bepcor}
     \lim_{\epsilon\rightarrow 0} 
             G(r_1,r_2,\theta_1,\theta_2 )_{(t<y_0)} = \frac{1}{16\pi^2\epsilon^2}\left(1+\epsilon^2 \frac{1-n^2}{3n^2 (y^2_0-t^2)}+\cdots\right).
    \end{equation}
    The Greeen's function with point $r_1, \theta_1$ on the $k_1$-sheet and 
    $r_2, \theta_2$ on the $k_2$-sheet is obtained
    by shifting $\theta = \theta_1 - \theta_2 $ to $\theta \rightarrow 2\pi ( k _1 - k_2) $.  From the expression 
    of the Green's function in (\ref{sclcor}) and the cross ratios for $t<y_0$ in (\ref{before}) , we see that there are no
    other singularities for $k_1 - k_2 \neq 0$. 
    Therefore, the leading contribution to the Green's function  in this regime arises when the operators are 
    present on the same sheet.

      Let us also understand the singularities of  correlator  in the $t>y_0$ regime.  From (\ref{after}) we see that the 
      cross ratios $\eta, \theta$ take values $i\pi, \pi$ respectively. 
  We  re-write the correlator as 
   \begin{align}\label{rewcor}
          G(r_1,r_2,\theta_1,\theta_2)&=-\frac{\sin \frac{i\eta}{n}}{16\pi^2 nr_1 r_2 \sin i \eta\Big(\sin \frac{i\eta+\theta}{2n} \sin\frac{i\eta-\theta}{2n} \Big)}.
      \end{align}
      From (\ref{after}) 
      we see that in this regime the $\sin( i \eta) $ goes to zero as $\epsilon$, and    $\sin \big( \frac{i\eta+\theta}{2n}\big)  $
      goes to zero as $\epsilon$. 
      The expression $\sin\big(\frac{i\eta-\theta}{2n} \big)$ remains finite and tends to $- \sin \frac{\pi}{n}$. 
      Therefore the correlator again diverges as $\epsilon^{-2}$. 
      Taking this limit we obtain 
      \begin{equation}\label{sing1}
      \lim_{\epsilon\rightarrow 0} G( r_1, r_2, \theta_1, \theta_2)_{ (t>y_0) } = \frac{t + y_0}{ 32\pi^2 t \epsilon^2} + O(\epsilon^0).
      \end{equation}
      This is the leading behaviour of the correlator  when the operators are placed on the same sheet. 
      Consider the situation when the operators are placed at $r_1, \theta_1$ on the $k_1$-sheet and $r_2, \theta_2$ 
   on the $k_2$-sheet. For this we need to substitute $\theta\rightarrow \theta + 2\pi ( k_1 - k_2) $ in (\ref{rewcor}) to 
   obtain
    \begin{align}\label{rewcor1}
          G(r_1,r_2,\theta_1^{(k_1)} ,\theta_2^{(k_2)} )&=-\frac{\sin \frac{i\eta}{n}}
          {16\pi^2 nr_1 r_2 \sin i \eta\Big(\sin \frac{i\eta+\theta +2\pi ( k_1-k_2) }{2n} \sin\frac{i\eta-\theta - 2\pi ( k_1- k_2) }{2n} \Big)}.
      \end{align}
   On using the expansions for the cross ratios in (\ref{rewcor1}) we see that the most singular behaviour of this 
   correlator arises when $k_1 - k_2 = - 1$. In this case $\sin i\eta$ goes to zero as $\epsilon$ and 
   $\sin ( \frac{i \eta - \theta - 2\pi ( k _1 - k_2)}{2n}  ) $ goes to zero as $\epsilon$, while $\sin (
   \frac{ i \eta + \theta  + 2\pi ( k _1 - k_2)}{2n}  )$ 
   remains finite. Again the correlator diverges as  $\epsilon^{-2}$, 
     \begin{equation}\label{sing2}
      \lim_{\epsilon\rightarrow 0} G( r_1, r_2, \theta_1^{(k_1)}, \theta_2^{(k_2)})_{ (t>y_0, k_1 - k_2 = -1 )} = \frac{t - y_0}{ 32\pi^2 t \epsilon^2} + O(\epsilon^0).
      \end{equation}
      Here the operators are placed so that they are on adjacent sheets and 
      the 2nd operator is on a sheet above the 1st operator. 
     It is important to note that the 2 singular behaviours for $t>y_0$ in (\ref{sing1}) and (\ref{sing2}) are related 
     by $y_0\rightarrow -y_0$.  
     This arises because the contribution to the singularity arises from different factors in the  product 
     $ \sin ( \frac{i \eta - \theta}{2n} ) \sin (\frac{ i \eta + \theta  }{2n} )$  present in the correlator (\ref{rewcor}). 
     There is one more observation regarding the singular behaviour of the correlators in (\ref{sing1}) and (\ref{sing2}). 
     We see that it apart from the factor of $1/\epsilon^2$, which sets the dimensions, the function of $y_0, t$ 
     is only  a function of the ratio $\frac{y_0}{t}$. 
     Note that there are no other singular contributions. One particular case worth mentioning is when one examines the 
     same point say $r_1 = r_2, \theta =0$, but they are on different sheets, that is $k_1 -k_2 \neq 0$.
     Then from (\ref{rewcor1}) we see that  the correlator is finite and not divergent. 
     
   \begin{figure} 
\begin{tikzpicture}
\draw[fill=gray] (0,0) -- (4,0) -- (6,1) -- (2,1
) -- (0,0);
\draw[fill=gray] (0,1.3) -- (4,1.3) -- (6,2.3) -- (2,2.3
) -- (0,1.3);
\draw[fill=gray] (0,2.6) -- (4,2.6) -- (6,3.6) -- (2,3.6
) -- (0,2.6);
\draw[fill=gray] (0,4.4) -- (4,4.4) -- (6,5.4) -- (2,5.4
) -- (0,4.4);
\draw[blue,thick] (3,0.5)--(5,0.5); 
\draw[blue,thick] (3,1.8)--(5,1.8);
\draw[blue,thick]  (3,4.95)--(4.95,4.95);
\draw[blue,thick] (3,3.15)--(5,3.15);
\draw[black,thick]  (5.6,4.7) node{$k=n$};
 \draw [black,thick](5.7,3.15) node{$k=3$};
 \draw [black,thick](5.7,1.8) node{$k=2$};
  \draw [black,thick](5.7,0.5) node{$k=1$};
\draw [->,>=stealth',thick,blue,thick](2.8,0.45) arc[start angle=180, end angle=360, radius=0.3cm];
\draw [->,>=stealth',thick,blue,thick](3.2,0.47) arc[start angle=0, end angle=185, radius=0.2cm];
\draw [->,>=stealth',thick,blue,thick](2.8,1.75) arc[start angle=180, end angle=360, radius=0.2cm];
\draw [->,>=stealth',thick,blue,thick](3.4,1.76) arc[start angle=0, end angle=180, radius=0.3cm];
\draw [->,>=stealth',thick,blue,thick](3.4,4.95) arc[start angle=0, end angle=180, radius=0.3cm];
\draw [->,>=stealth',thick,blue,thick](2.8,4.95) arc[start angle=180, end angle=360, radius=0.2cm];
\draw [->,>=stealth',thick,blue,dashed](3.4,0.45)--(3.4,1.76);
\draw [->,>=stealth',thick,blue,dashed](3.2,4.95)--(3.2,5.8);
\draw [->,>=stealth',thick,blue,dashed](3.2,5.8)--(6.8,5.8);\draw [->,>=stealth',thick,blue,dashed](3.2,5.8)--(6.8,5.8);
\draw [->,>=stealth',thick,blue,dashed](6.8,5.8)--(6.8,-0.2);
\draw [->,>=stealth',thick,blue,dashed](6.8,-0.2)--(3.2,-0.2);
\draw [->,>=stealth',thick,blue,dashed](3.2,-0.2)--(3.2,0.45);
\draw [->,>=stealth',thick,blue,dashed](3.4,3.15)--(3.4,4.95);
\draw [->,>=stealth',thick,blue,dashed](3.2,1.76)--(3.2,3.15);
\draw [->,>=stealth',thick,blue,thick](3.2,3.15) arc[start angle=0, end angle=180, radius=0.2cm];
\draw [->,>=stealth',thick,blue,thick](2.8,3.15) arc[start angle=180, end angle=360, radius=0.3cm];
\draw [-stealth](0,-0.15) -- (1,-0.15);
\draw [-stealth](-0.2,0.1) -- (0.5,0.5);
\draw [black,thick](1.2,-0.15) node{y};
\draw [black,thick](0.7,0.7) node{$\tau$};
\filldraw[red] 
                (2.7,0.8) circle[radius=2pt]
                (2.7,0.8) circle[radius=2pt];
                \filldraw[red] 
                (2.3,0.5) circle[radius=2pt]
                (2.3,0.5) circle[radius=2pt];
                \filldraw[red] 
                (2.7,2.1) circle[radius=2pt]
                (2.7,2.1) circle[radius=2pt];
                \filldraw[red] 
                (2.3,1.8) circle[radius=2pt]
                (2.3,1.8) circle[radius=2pt];
                 \filldraw[red] 
                (2.7,3.4) circle[radius=2pt]
                (2.7,3.4) circle[radius=2pt];
                \filldraw[red] 
                (2.3,3.1) circle[radius=2pt]
                (2.3,3.1) circle[radius=2pt];
                \filldraw[red] 
                (2.3,4.9) circle[radius=2pt]
                (2.3,4.9) circle[radius=2pt];
                  \filldraw[red] 
                (2.7,5.2) circle[radius=2pt]
                (2.7,5.2) circle[radius=2pt];
                \draw[red,thick](2.3,3.1)--(2.7,3.4);
                \draw[red,thick](2.3,0.5)--(2.7,0.8);
                \draw[red,thick](2.3,1.8)--(2.7,2.1);
                 \draw[red,thick](2.3,4.9)--(2.7,5.2);
                \draw [red,thick](2,1.65) node{e};
                  \draw [red,thick](2.5,2.2) node{l};
                   \draw [red,thick](2,2.9) node{e};
                  \draw [red,thick](2.5,3.45) node{l};
                  \draw [red,thick](2,0.4) node{e};
                  \draw [red,thick](2.5,0.9) node{l};
                   \draw [red,thick](2.5,5.25) node{l};
                  \draw [red,thick](2,4.7) node{e};
\end{tikzpicture}
\qquad
\begin{tikzpicture}
\draw[fill=gray] (0,0) -- (4,0) -- (6,1) -- (2,1
) -- (0,0);
\draw[fill=gray] (0,1.3) -- (4,1.3) -- (6,2.3) -- (2,2.3
) -- (0,1.3);
\draw[fill=gray] (0,2.6) -- (4,2.6) -- (6,3.6) -- (2,3.6
) -- (0,2.6);
\draw[fill=gray] (0,4.4) -- (4,4.4) -- (6,5.4) -- (2,5.4
) -- (0,4.4);
\draw[blue,thick] (3,0.5)--(5,0.5); 
\draw[blue,thick] (3,1.8)--(5,1.8);
\draw[blue,thick]  (3,4.95)--(4.95,4.95);
\draw[blue,thick] (3,3.15)--(5,3.15);
\draw[black,thick]  (5.6,4.7) node{$k=n$};
 \draw [black,thick](5.7,3.15) node{$k=3$};
 \draw [black,thick](5.7,1.8) node{$k=2$};
  \draw [black,thick](5.7,0.5) node{$k=1$};
\draw [-stealth](0,-0.15) -- (1,-0.15);
\draw [-stealth](-0.2,0.1) -- (0.5,0.5);
\draw [black,thick](1.2,-0.15) node{y};
\draw [black,thick](0.7,0.7) node{$\tau$};
\draw [->,>=stealth',thick,blue,thick](2.8,0.45) arc[start angle=180, end angle=360, radius=0.3cm];
\draw [->,>=stealth',thick,blue,thick](3.2,0.47) arc[start angle=0, end angle=185, radius=0.2cm];
\draw [->,>=stealth',thick,blue,thick](2.8,1.75) arc[start angle=180, end angle=360, radius=0.2cm];
\draw [->,>=stealth',thick,blue,thick](3.4,1.76) arc[start angle=0, end angle=180, radius=0.3cm];
\draw [->,>=stealth',thick,blue,thick](3.4,4.95) arc[start angle=0, end angle=180, radius=0.3cm];
\draw [->,>=stealth',thick,blue,thick](2.8,4.95) arc[start angle=180, end angle=360, radius=0.2cm];
\draw [->,>=stealth',thick,blue,dashed](3.4,0.45)--(3.4,1.76);
\draw [->,>=stealth',thick,blue,dashed](3.2,4.95)--(3.2,5.8);
\draw [->,>=stealth',thick,blue,dashed](3.2,5.8)--(6.8,5.8);\draw [->,>=stealth',thick,blue,dashed](3.2,5.8)--(6.8,5.8);
\draw [->,>=stealth',thick,blue,dashed](6.8,5.8)--(6.8,-0.2);
\draw [->,>=stealth',thick,blue,dashed](6.8,-0.2)--(3.2,-0.2);
\draw [->,>=stealth',thick,blue,dashed](3.2,-0.2)--(3.2,0.45);
\draw [->,>=stealth',thick,blue,dashed](3.4,3.15)--(3.4,4.95);
\draw [->,>=stealth',thick,blue,dashed](3.2,1.76)--(3.2,3.15);
\draw [->,>=stealth',thick,blue,thick](3.2,3.15) arc[start angle=0, end angle=180, radius=0.2cm];
\draw [->,>=stealth',thick,blue,thick](2.8,3.15) arc[start angle=180, end angle=360, radius=0.3cm];
\filldraw[red] 
                (2.7,0.8) circle[radius=2pt]
                (2.7,0.8) circle[radius=2pt];
                \filldraw[red] 
                (2.3,0.5) circle[radius=2pt]
                (2.3,0.5) circle[radius=2pt];
                \filldraw[red] 
                (2.7,2.1) circle[radius=2pt]
                (2.7,2.1) circle[radius=2pt];
                \filldraw[red] 
                (2.3,1.8) circle[radius=2pt]
                (2.3,1.8) circle[radius=2pt];
                 \filldraw[red] 
                (2.7,3.4) circle[radius=2pt]
                (2.7,3.4) circle[radius=2pt];
                \filldraw[red] 
                (2.3,3.1) circle[radius=2pt]
                (2.3,3.1) circle[radius=2pt];
                \filldraw[red] 
                (2.3,4.9) circle[radius=2pt]
                (2.3,4.9) circle[radius=2pt];
                  \filldraw[red] 
                (2.7,5.2) circle[radius=2pt]
                (2.7,5.2) circle[radius=2pt];
                \draw[red,thick](2.3,0.5)--(2.7,2.1);
                \draw[red,thick](2.3,1.8)--(2.7,3.4); \draw[red,thick,dashed](2.3,3.1)--(2.7,4.1);
              \draw[red,thick,dashed](2.7,5.2)--(2.3,4.3);
         
                \draw [red,thick](2,1.65) node{e};
                  \draw [red,thick](2.5,2.2) node{l};
                   \draw [red,thick](2,2.9) node{e};
                  \draw [red,thick](2.5,3.45) node{l};
                  \draw [red,thick](2,0.4) node{e};
                  \draw [red,thick](2.5,0.9) node{l};
                  \draw [red,thick](2.5,5.25) node{l};
                  \draw [red,thick](2,4.9) node{e};
\end{tikzpicture}
\caption{The leading Wick contractions in the  $2n$ point function on $\Sigma_n$. 
For $t<y_0$, only the  Wick contractions on the same sheet as shown in the figure on the left contributes.
For $t>y_0$,  both the Wick contractions, that is the one on the same sheet as well as 
the cyclic Wick contractions along adjacent sheet contributes equally. In the figure on the right, the
operator at `$e$' on the last sheet is contracted with the operator at `$l$ ' } \label{wickcf}
\end{figure}
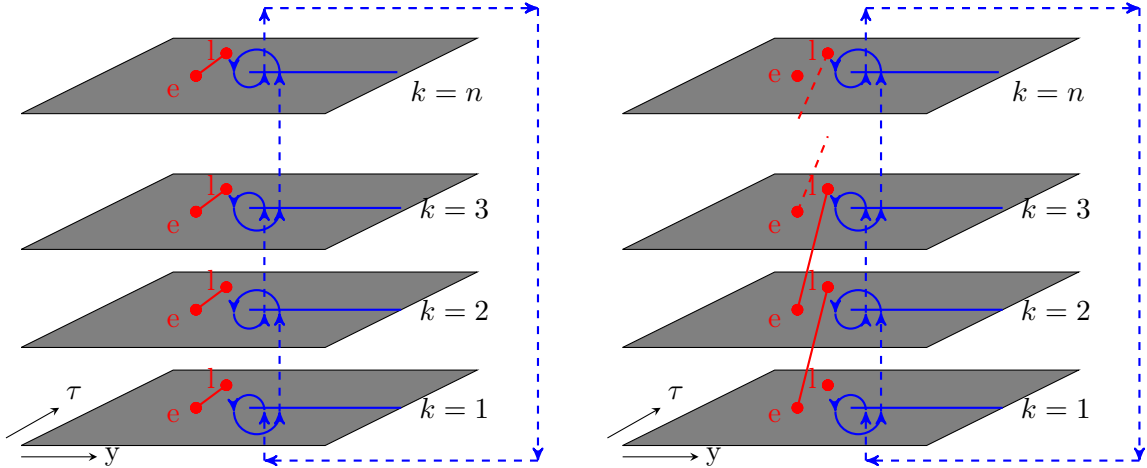

        \subsubsection*{Evaluation of $\Delta S^{(n)}_A$}
        
        With the information of the leading behaviour of the two-point function on the replica surface, we can proceed 
        to evaluate the leading contributions to the R\'{e}nyi entropies.
        For free fields the  correlator in (\ref{renform}) can be evaluated using Wick contraction. 
        For $t<y_0$ we see that the leading contribution  comes from the contraction in which the two
         operators are on the same sheet.  This is shown in the diagram on the left of figure \ref{wickcf}. 
        From (\ref{bepcor})  we see that this contribution is given by 
        \begin{eqnarray} \label{leadcontra}
       \lim_{\epsilon \rightarrow 0, t<y_0}
       & & \left.  \Big\langle \prod_{k = 1}^n {\cal O } ( r_l,  \theta_l^{(k)})  {\cal O } ( r_e, \theta_e^{(k)} ) \Big\rangle_{\Sigma_n} 
       \right|_{\rm same\;sheet\;contraction}  = \\ \nonumber 
       &&  \qquad\qquad\qquad\qquad\left[  \frac{1}{16\pi^2\epsilon^2}\left(1+\epsilon^2 \frac{1-n^2}{3n^2 (y^2_0-t^2)}+\cdots\right) \right]^{n}.
        \end{eqnarray}
         As discussed below (\ref{bepcor}),  the contractions of operators located on 2 different sheets are not singular they are $O(\epsilon^0)$. The  next leading contraction will be from $n-2$ pairs of operators on the same sheet and 
         $2$ pairs of operators contracted across sheets. This contribution is $O(\epsilon^{-2(n-2)} )$. 
         Therefore the leading and the next sub-leading contribution comes from the contractions on the same sheet
         given in (\ref{leadcontra}). Substituting this  in (\ref{renform}) 
         and using the expression for the correlator on  $\Sigma_1$  from (\ref{r4cor}) we obtain 
            \begin{align}\label{jbefore}
            \Delta S^{(n)}_{t<y_0}&=\frac{1}{1-n}\log\left(1+\epsilon^2 \frac{1-n^2}{3n^2 (y^2_0-t^2)}+\cdots\right)^n\nonumber\\
            &=\frac{n}{1-n}\frac{\left(1-n^2\right) \epsilon ^2}{3 n^2 \left(y^2_0-t^2\right)}+\mathcal{O}(\epsilon^4) \\
            &= \frac{ n+1}{ 3 n} \frac{\epsilon^2}{ y_0^2 - t^2}  +\mathcal{O}(\epsilon^4) .
        \end{align}
        The entanglement vanishes as $\epsilon^2$ before the pulse enters the entangling region. 
        The coefficient of the $\epsilon^2$ term has an interesting explanation. This is due to the fact that the leading contribution to the OPE of the operators on the same sheet  is identical to the OPE of 2 scalars on $R^4$
        \begin{equation}
        \phi(x_1) \phi(x_2)  \sim  \frac{1}{4\pi^2 |x_1 - x_2|^2}\Big( 1 +  C_{\phi\phi :\phi:^2} |x_1-x_2|^2  :\phi^2(x_2): + \cdots
        \Big) .
        \end{equation}
        The  constant then in (\ref{jbefore})  should
        be proportional to the  expectation value of the $:\phi^2:$ on the replica surface. This can be obtained from the 
        Green's function using the point split method. 
         \begin{align}
            \langle :\phi^2\rangle_{\Sigma_n} &=\lim_{\substack{r_1\rightarrow r_2\\ \theta_1\rightarrow\theta_2}} \left(G(r_1,r_2,\theta_1,\theta_2 )-\frac{1}{4\pi^2\left(r_1^2+r_2^2-2 r_1r_2\cos(\theta_1-\theta_2)\right)}\right).
        \end{align}
        We have set the parallel directions $x, y $ of the both points to be the same. 
        Now we can take the limit $r_1\rightarrow r_2 = \sqrt{ y^2 + \tau^2}$, from (\ref{defcross}) we see that 
        $\eta \rightarrow 0$, therefore we  obtain
        \begin{align}\label{composite}
           \langle :\phi^2:\rangle_{\Sigma_n} &=\lim_{\theta\rightarrow 0}\frac{1}{4\pi^2}\left(\frac{1}{2 n^2 (y^2 +\tau^2) \left(1-\cos \left(\frac{\theta }{n}\right)\right)}-\frac{1}{2 (y^2+\tau^2) (1-\cos (\theta ))}\right) , \nonumber\\
           &=\frac{1}{16\pi^2}\frac{1-n^2}{3n^2 (y^2+\tau^2)}.
        \end{align}
        Comparing (\ref{jbefore}) and (\ref{composite}), we see that indeed that the $\epsilon^2$ coefficient is proportional 
        to the expectation value of the bilinear  $:\phi^2$. 
        We will see that as expected a similar  feature persists in all the cases we study in this paper. This will serve as a simple consistency check of our calculations.

        In the domain $t >y_0$, from the discussion of the singularities of the 2 point function we see that 
        there are 2 set of contractions that contribute. 
        The contractions of pairs of operators on the same sheet as shown in the left diagram of  figure \ref{wickcf}. Using
        (\ref{sing1}) we  obtain the following contribtution
            \begin{eqnarray} \label{leadcontra1}
       \lim_{\epsilon \rightarrow 0, t>y_0}
       & & \left.  \Big\langle \prod_{k = 1}^n {\cal O } ( r_l,  \theta_l^{(k)})  {\cal O } ( r_e, \theta_e^{(k)} ) \Big\rangle_{\Sigma_n} 
       \right|_{\rm same\;sheet\;contraction}  = \\ \nonumber 
       &&  \qquad\qquad\qquad\qquad 
       \left( \frac{ t- y_0}{ 32\pi^2 t \epsilon^2} \right)^n +  O( \epsilon^{ -(2n -1) })   \cdots
       \end{eqnarray} 
       The contractions shown in the right diagram of 
       figure \ref{wickcf}, also leads to the same order of divergence in the $2n$ point function. 
       These contractions are due to contractions  across nearest neighbour sheets cyclically. 
       From (\ref{sing2}) we obtain 
           \begin{eqnarray} \label{leadcontra2}
       \lim_{\epsilon \rightarrow 0, t>y_0}
       & & \left.  \Big\langle \prod_{k = 1}^n {\cal O } ( r_l,  \theta_l^{(k)})  {\cal O } ( r_e, \theta_e^{(k)} ) \Big\rangle_{\Sigma_n} 
       \right|_{ {\rm cyclic\; contraction} } = \\ \nonumber 
       &&  \qquad\qquad\qquad\qquad 
       \left( \frac{ t- y_0}{ 32\pi^2 t \epsilon^2} \right)^n +  O( \epsilon^{ -(2n -1) })   \cdots
       \end{eqnarray} 
       All other contractions are sub-leading since they involve at least one contraction which is not nearest neighbour 
       contraction or  they involve contractions of the same point say $r_e, \theta_e$ but across 2 different sheets.
       Combining the contributions (\ref{leadcontra1}) and ( \ref{leadcontra2}) we see that we obtain the following 
       contribution to the change in entanglement entropy for $t>y_0$
       \begin{eqnarray} \label{jumpa}
           \lim_{\epsilon\rightarrow 0, t>y_0}
            \Delta S^{(n)}_A&=\frac{1}{1-n}\log \left[ \frac{ \left(\frac{t+y_0}{32 \pi ^2 t}\right)^n+\left(\frac{t-y_0}{32 \pi ^2 t}\right)^n } {\left(\frac{1}{16 \pi ^2}\right)^n}  \right].
       \end{eqnarray}
       At this point, it is useful to observe that the term in the square bracket is the leading contribution of the 
       ratio of the $2n$ point function on the surface $\Sigma_n$ to  $n$ products of the 
       $2$ point function on $\Sigma_1$.  By definition this ratio should be one at $n=1$, which is indeed true
       as can be seen in (\ref{jumpa}). Also observe that the expression is  a function of a linear polynomial  in 
       the ratio of $y_0/t$. 
We can take the $n\rightarrow 1$ limit to obtain the entanglement entropy 
\begin{eqnarray}
    \Delta S_A&=& \lim_{n\rightarrow 1}\Delta S^{(n)}_A ,\\ \nonumber
    &=& \log(2) + \frac{y_0}{2t}   \log\left[ \frac{ 1 -\frac{y_0}{t} }{ 1 + \frac{y_0}{t} } \right]
    -\frac{1}{2} \log \left[ 1- \big( \frac{y_0}{t} \big)^2 \right].
\end{eqnarray} 
Another important behaviour to observe is the long time behaviour of the growth of R\'{e}nyi/entanglement entropies, 
we see that it is given by 
\begin{equation}
 \lim_{t\rightarrow \infty}\Delta S^{(n)}_A = \log 2-\frac{n y^2_0}{2 t^2}+ {O}(\frac{y^4_0}{t^4})+\cdots
\end{equation}

We have seen that the growth profile of the entanglement  as the quench enters the interval is entirely determined 
from the singular behaviour of  $2$ point function on the replica surface  (\ref{sclcor}).
 This correlator is of the form of 2 conformal primaries 
     of weight $1$ in a BCFT with a defect of co-dimension $2$.  Therefore it is interesting to cast the behaviour at the 
     singularities to that seen for the case of quenches in $2$ dimensions earlier in \cite{David:2016pzn}. 
     In the appendix we cast the correlator for $n=2$ in terms of cross ratios similar to that in $2$ dimensions and 
     we will show that the growth of entanglement entropy is due to a similar phenomenon seen in $2$ dimensions, that is the  holomorphic cross ratio crosses a branch cut, while the anti-holomorphic does not. 

\subsection{Time evolution of the vector quench}

In this section we study local quantum quenches in free Maxwell field theory in $d=4$ dimension. The theory is free and conformal in four dimension. We study quenches  obtained by exciting  the ground state by field strengths.
Therefore we would need the two-point function of the gauge field on the replica surface to 
evaluate the change in entanglement entropy. 
This two-point function can be constructed from the two-point function of the scalar  given in (\ref{sclcor}) by taking 
suitable derivatives. 
Just as in the 
case of a co-dimension $2$ defect
 correlator in (\ref{sclcor})  preserves $SO(2)_T \times SO(2)_L$ symmetry of $R^4$. 
 It is convenient to fix a gauge which
 gauge which is compatible with this symmetry. 
 The  $U(1)$ theory is gauge invariant under the transformation
      \begin{equation}
      A_\mu \rightarrow A_\mu + \partial_\mu \epsilon,
      \end{equation}
  where $\epsilon$ is the gauge parameter.    We use this gauge symmetry to first fix the gauge
      \begin{equation}\label{covcond}
      \partial^\mu A_\mu =0.
      \end{equation}
       The equations of motion in the covariant gauge  reduces to 
      \begin{equation}\label{maxeom}
      \nabla^2 A_\mu =0.
      \end{equation}
      Now  there are still allowed gauge transformations which preserve the covariant gauge, these are 
     of the form
      \begin{equation}
      A_\mu' = A_\mu + \partial_\mu \epsilon, \qquad  {\rm with }\quad  \Box\epsilon =0.
      \end{equation}
      Given a gauge potential which satisfies (\ref{covcond}) and (\ref{maxeom}) we make 
      a further gauge transformation  so that  the gauge potential satisfies 
      \begin{equation}\label{maxgcond}
      \partial^a A_a' =0, \qquad \partial^i A_{i}' =0  \quad a \in \{ t, y\} , i \in \{x, z\}.
      \end{equation}
      The gauge potential satisfies the transversality condition independently in the directions perpendicular to the 
      defect as well as the directions parallel to the defect.
      These  conditions can be satisfied 
         by choosing the gauge  transformation  to be
      \begin{equation}
      \epsilon  = - \frac{\partial^i A_i }{\nabla^2} , \qquad \nabla^2 = \partial_x^2 + \partial_z^2.
      \end{equation}
      To show \eqref{maxgcond} is satisfied, we need to use (\ref{maxeom}) and (\ref{covcond}).
       The gauge transformation also satisfies $\Box \epsilon=0$, which ensures that it is a 
     it is a valid choice of gauge.  Such a gauge was first used to construct the vector propagator for a conical 
     defect in \cite{Candelas:1977zza} and evaluate the expectation value of the stress tensor. 
     We can follow the same construction for the replica surface $\Sigma_n$,  which leads to  \cite{David:2020mls}, 
      \begin{eqnarray}\label{corm1}
      G_{\mu\nu'} ( x,x')&\equiv& \langle A_\mu ( x) A_{\nu'} ( x') \rangle.   \\ \nonumber
       G_{ab'} (x,x') &=&   \frac{P_a P_{b'} }{ \nabla^2}  G (x, x'),  \qquad
      G_{i j'}(x, x')    = \left[ \delta_{ij}  - \frac{\partial_i\partial_j}{ \nabla^2} \right] \tilde G(x, x') ,
      \\ \nonumber
     G_{a i'} (x,x') &=&  G_{i a'} ( x, x') =0.
      \end{eqnarray}
      where $G(x, x') $ is the scalar propagator on the cone which is given in \eqref{sclcor}.
      Note that by construction, the two-point function satisfies the  condition
      \begin{equation}
      \partial^\mu G_{\mu\nu'} ( x, x') = 0 .
      \end{equation}
      Here we label the 2 points as $x$ and $x'$,  and $P_a$ are defined as
\begin{eqnarray}\label{defpa}
P_a = \epsilon_{a b} g^{bc} \nabla_c , \qquad\qquad
\epsilon_{t y} = -\epsilon_{yt} = 1, \quad \epsilon_{tt} = \epsilon_{yy} =0.
\end{eqnarray}
One useful fact of writing the correlator in this way is that it is easy to write the components $G_{ab'}$ in 
polar coordinates if necessary.  We just need to write the covariant form of $P_a$ and $P_b'$.
Earlier work on quenches for the $U(1)$ field used the Feynman gauge. We will  see, the 
result for the entanglement jump  in the $\epsilon\rightarrow 0$ limit
 is the same as that obtained in \cite{Nozaki:2016mcy} using the gauge used here. 

\subsection*{Two-point function of field strengths}

There are $6$ components of the field strengths and in principle one would have thought one needs to 
evaluate the $6$  correlators of these field strengths to study their respective quenches. 
However due to the rotational symmetry in $x$ and $z$ directions and 
electro-magnetic duality,  there are only $2$ independent quenches. 
We will show this by explicit evaluation of the correlators, the fact the due to duality the entanglement growth 
of quenches of electric fields is the same as magnetic fields was first observed in \cite{Nozaki:2016mcy}. 
Here we will see that the correlator on the replica surface, $\Sigma_n$  of electric fields is same as that of 
magnetic fields.

First consider the two-point function of $F_{ty}$. 
\begin{align} \label{fty}
 & \langle F_{ty}(x )F_{ty}(x')\rangle_{\Sigma_n}  \nonumber \\
  &=\partial_{t_1}\partial_{t_2}\langle A_{y}(x)A_{y}(x')\rangle+  \partial_{y_1}\partial_{y_2}\langle A_{t}(x)A_{t}(x')\rangle 
-\partial_{t_1}\partial_{y_2}\langle A_{y}(x)A_{t}(x')\rangle-\partial_{t_2}\partial_{y_1}\langle A_{t}(x )A_{y}(x')\rangle,\nonumber\\
  &=\Big[\partial_{t_1}\partial_{t_2}(\frac{\partial_{t_1} \partial_{t_2}}{\nabla^2})+\partial_{y_1}\partial_{y_2}(\frac{\partial_{y_1}\partial_{y_2}}{\nabla^2}) 
-\partial_{t_1}\partial_{y_2}(\frac{-\partial_{t_1}\partial_{y_2} }{\nabla^2})
  -\partial_{t_2}\partial_{y_1}\frac{(-\partial_{t_2}\partial_{y_1)}}{\nabla^2})\Big] G(x ,x '),\nonumber\\
  &=-(\partial_{t_1}^2+\partial_{y_1}^2)G(x, x'),\nonumber\\
  &=-\frac{ (\coth (\eta )+1) \left(\cosh \left(\frac{\eta }{n}\right) \cos \left(\frac{\theta }{n}\right)+n \coth (\eta ) \sinh \left(\frac{\eta }{n}\right) \left(\cosh \left(\frac{\eta }{n}\right)-\cos \left(\frac{\theta }{n}\right)\right)-1\right)}{4 \pi ^2 n^2 r_1^3r_2\sinh\eta\left(\cos \left(\frac{\theta }{n}\right)-\cosh \left(\frac{\eta }{n}\right)\right)^2}.
\end{align}
In this and subsequent equations, it is understood that $x, x'$ refers to the first and the second coordinate. 
To derive the last line, we use the on-shell condition on the Green's function for separated points
\begin{align}\label{onshell}
    (\partial_{t_2}^2+\partial_{y_2}^2+\partial_{x_2}^2+\partial_{z_2}^2)G(x, x')=0.
\end{align}
We then  used the fact there is translation invariance in the $x, z$, which results in the following relations
\begin{align} \label{transinv}
    \begin{split}
        \partial_{x_1}G(x ,x ')&=-\partial_{x_2}G(x, ,x '),\\
        \partial_{z_1}G(x ,x')&=-\partial_{z_2}G(x,x').
    \end{split}
\end{align}
Note that the final result is (pseudo) scalar in the $t, y$ directions, this is expected since we are working with the field strength $F_{ty}$. 

To evaluate the time dependence of the entanglement entropy after the quench we need the singular behaviour of 
the correlator for $t<y_0$ and $t>y_0$. 
This is can be obtained by examining (\ref{fty}) just as in the case of the scalar correlator. 
Again the leading  singularity for $t<y_0$ occurs when the $2$ points of the correlator are on the same sheet, 
        \begin{align} \label{ftyb}
             \lim_{\epsilon\rightarrow 0}
              \big\langle F_{ty}(r_1,\theta_1^{k} ) F_{ty}( r_2,  \theta_2^{(k)})\big\rangle_{t<y_0}&=-\frac{1}{16 \pi ^2 \epsilon ^4}-\frac{-11 n^4+10 n^2+1}{720 \pi ^2 n^4 \left(t^2-y^2_0\right)^2}+{O}(\epsilon^2).
        \end{align}
        As expected, the leading singularity is the same as that of the correlator on $\Sigma_1$. 
       The correlator is  not singular when the $2$ points lie on different sheets. The leading correction at order $O(\epsilon^0)$  in (\ref{ftyb}) 
       can be understood using the same reasoning as in the case for the scalar correlator. It is due to the expectation value 
       of the composite $:F_{ty}^2:$ on  $\Sigma_n$. We will demonstrate this in the next sub-section. 
       Let us now obtain the  leading behaviour in the $\epsilon\rightarrow 0$ limit for $t>y_0$. 
       For this we can again examine  (\ref{fty})   when   $\theta, \eta $ is around $\pi$ and $i\pi$   as discussed for the 
       scalar case.   When the operators are on the same sheet $k_1 = k_2 = k$,  we obtain 
       \begin{align}\label{ftya1}
             \lim_{\epsilon\rightarrow 0}
              \langle F_{ty}(r_1, \theta_1^{(k)} ) F_{ty}( r_2, \theta_2^{(k)} )\rangle_{t>y_0}&=-\frac{(2 t-y_0) (t+y_0)^2}{64 \pi ^2 t^3 \epsilon ^4}+\mathcal{O}(\frac{1}{\epsilon^3})+\cdots
        \end{align}
        Similarly when the operators are placed on adjacent sheets we obtain 
         \begin{align} \label{ftya2}
         \langle F_{ty}( r_1, \theta_1^{(k_1)} ) F_{ty}(r_2, \theta_2^{(k_2)} )\rangle_{t>y_0, k_1- k_2 = -1}&=
         -\frac{(t-y_0)^2 (2 t+y_0)}{64 \pi ^2 t^3 \epsilon ^4}+\mathcal{O}(\frac{1}{\epsilon^3})+\cdots
        \end{align}
        As we discussed in detail for the scalar,  the reason the correlators in (\ref{ftya1}), (\ref{ftya2}) are related 
        by $y\rightarrow y_0$ is due to the fact in one case $\sin\big( \frac{i\eta + \theta}{2n} \big)$ tends to zero 
        and in the other  $\sin\big(\frac{ i \eta- \theta - 2\pi ( k_1 - k_2) }{2n}\big) $ tends to zero. 
        Further more, the power of $\epsilon^{-4}$ is determined by the dimension of the operator $F_{ty}$.
 The time dependence of the leading term is  determined by a cubic polynomial through the ratio $y_0/t$. 
 The order of the polynomial is one less  the power of the $\epsilon$.  The leading behaviours
 of these correlators  in (\ref{ftyb}), (\ref{ftya1}) and (\ref{ftya2}) obtained using the gauge in this paper agree with
 that obtain in \cite{Nozaki:2016mcy} using the Feynman gauge.

Let us proceed and evaluate the two-point function of $F_{xz}$. 
\begin{align}\label{fxz}
    & \langle F_{xz}(x)F_{xz}(x')\rangle_{\Sigma_n} \nonumber \\ \nonumber
 &    =\partial_{x_1}\partial_{x_2}\big\langle A_{z}(x )A_{z}(x')\big\rangle+  
 \partial_{z_1}\partial_{z_2}\big\langle A_{x}(x)A_{x}(x')\big\rangle 
-\partial_{x_1}\partial_{z_2}\big\langle A_{z}(x )A_{x}(x')\big\rangle-\partial_{x_2}\partial_{z_1}\big\langle A_{x}(x )A_{z}(x')\big\rangle\nonumber\\
  &=\left(\partial_{x_1}\partial_{x_2}\frac{\partial_{x_1}^2}{\nabla^2}+  \partial_{z_1}\partial_{z_2}\frac{\partial_{z_1}^2}{\nabla^2}-\partial_{x_1}\partial_{z_2}( \frac{-\partial_{x_1}\partial_{z_1}}{\nabla^2})-\partial_{x_2}\partial_{z_1}( \frac{-\partial_{x_1}\partial_{z_1}}{\nabla^2} )\right)G(x ,x'),\nonumber\\
           &=-(\partial_{x_1}^2+\partial_{z_1}^2)G(x ,x')
           =(\partial_{t_1}^2+\partial_{y_1}^2)G(x ,x').
\end{align}
To arrive at the last line, we have again used translation invariance (\ref{transinv}) and the on shell condition (\ref{onshell}). 
Comparing (\ref{fty}) and (\ref{fxz}), we see that they are identical up to a sign. 
This is because the source free Maxwell theory is self dual under electro-magnetic duality and  in Euclidean space
the equations relating the dual components is given by 
\begin{equation}
\tilde F_{ty} = -i \epsilon_{tyxz} F_{xz}.
\end{equation}
Let us now  evaluate  the two-point function of  $F_{tx}$
\begin{align}\label{ftx}
  &   \langle F_{tx}(x)F_{tx}(x')\rangle_{\Sigma_n}= \nonumber\\
 & \partial_{t_1}\partial_{t_2}\langle A_{x}(x )A_{x}(x')\rangle+  \partial_{x_1}\partial_{x_2}\langle A_{t}(x )A_{t}(x')\rangle
 -\partial_{t_1}\partial_{x_2}\langle A_{x}(x )A_{t}(x')\rangle-\partial_{t_2}\partial_{x_1}\langle A_{t}(x )A_{x}(x')\rangle,\nonumber\\
  &=  \partial_{t_1}\partial_{t_2}\langle A_{x}(x )A_{x}(x ')\rangle+  \partial_{x_1}\partial_{x_2}\langle A_{t}(x )A_{t}(x')
  \rangle = \left( \partial_{t_1}\partial_{t_2}  \frac{\partial_{y_1}^2}{ \nabla^2}  - \partial_{x_1}^2  
  \frac{\partial_{y_1} \partial_{y_2}}{\nabla^2}   \right) G(x, x') 
  ,\nonumber\\
  &=\frac{1}{2}\left(\partial_{t_1}\partial_{t_2}-\partial_{y_1}\partial_{y_2}\right)G(x, x').
\end{align}
In the last but one line we have used translation invariance (\ref{transinv}) 
and then we have used the following isotropy property
of the Greens function
\begin{eqnarray} \label{isotropy}
\left. \partial_{x_1}^2 G(x, x')\right|_{x_1\rightarrow x_2, z_1\rightarrow z_2}  = \left. \frac{1}{2} \nabla^2 G(x, x')\right|_{x_1\rightarrow x_2, z_1\rightarrow z_2}, \\ ,\nonumber
\left. \partial_{z_1}^2 G(x, x')\right|_{x_1\rightarrow x_2, z_1\rightarrow z_2}  = \left. \frac{1}{2} \nabla^2 G(x, x')\right|_{x_1\rightarrow x_2, z_1\rightarrow z_2}.\\ \nonumber
\end{eqnarray}
We can take the coincident limit in the $x, z$ directions, since all the operators involved in the quench are placed
at $x=z=0$.  In appendix \ref{idproof},   
we have derived these and similar relations which helps to simplify the computation 
of the correlators.  Repeating the   steps involved in the evaluation of the $F_{tx}$  in (\ref{ftx}),  it is easy to see 
the two-point function of $F_{tz}$ would be identical. 
\begin{eqnarray}
\left. \langle F_{tz} (x) F_{tz} (x') \rangle_{\Sigma_n}\right|_{x_1\rightarrow x_2, z_1\rightarrow z_2}  =  
\left. \langle F_{tx} (x) F_{tx} (x') \rangle_{\Sigma_n} \right|_{x_1\rightarrow x_2, z_1\rightarrow z_2} .
\end{eqnarray}
Finally we have the two-point function of $F_{yx}$
\begin{align}\label{fyx}
    & \langle F_{yx}(x)F_{yx}(x')\rangle_{\Sigma_n} ,\nonumber\\
     &=\partial_{y_1}\partial_{y_2}\langle A_{x}(x )A_{x}(x' )\rangle+  
     \partial_{x_1}\partial_{x_2}\langle A_{y}(x)A_{y}(x')\rangle 
 -\partial_{y_1}\partial_{x_2}\langle A_{x}(x)A_{y}(x')\rangle-\partial_{y_2}\partial_{x_1}\langle A_{x}(x )A_{y}(x')\rangle,\nonumber\\
  &=\partial_{y_1}\partial_{y_2}\langle A_{x}(x )A_{x}(x')\rangle+  \partial_{x_1}\partial_{x_2}\langle A_{y}(x)A_{y}(x')\rangle 
  ,\nonumber \\
  &= \left( \partial_{y_1} \partial_{y_2} ( \frac{ \partial_{z_1^2} }{\nabla^2 })  + \partial_{x_1}\partial_{x_2} 
  ( \frac{ \partial_{t_1}  \partial_{t_2} }{\nabla^2 })  \right) G(x, x') 
   ,\nonumber\\
  &=\frac{1}{2}\left(-\partial_{t_1}\partial_{t_2}+\partial_{y_1}\partial_{y_2}\right)G(x ,x ').
\end{align}
To arrive at the last line we have used translation invariance  (\ref{transinv}) as well as (\ref{isotropy}). 
It is again clear from this calculation that if we repeat the same steps for we have the relation 
\begin{eqnarray}
\left. \langle F_{yx} (x) F_{yx} (x') \rangle_{\Sigma_n}\right|_{x_1\rightarrow x_2, z_1\rightarrow z_2}  =  
\left. \langle F_{yz} (x) F_{yz} (x') \rangle_{\Sigma_n} \right|_{x_1\rightarrow x_2, z_1\rightarrow z_2} .
\end{eqnarray}
Comparing (\ref{ftx})  and (\ref{fyx}) we see that as expected they are identical up to a sign due to the 
electro magnetic duality of the Maxwell theory. 
In conclusion due to isotropy and duality we have only one independent quench among those created by 
field strengths $F_{tx}, F_{tz}, F_{yx}, F_{yz}$. 
The explicit form of two-point function  of say $F_{tx}$ on $\Sigma_n$  is quite long and cumbersome. 
But for our purpose we just need the singular behaviour in the $\epsilon\rightarrow 0$ limit for $t<y_0$ and 
$t>y_0$. A simple method to evaluate this is to first convert the derivatives in $t_1, t_2, y_1, y_2$ in terms of 
$\theta, \eta, r_1, r_2$ using the chain rule of differentiation, 
substitute the  singular behaviour of the scalar correlator and the coefficients occurring form the chain rule and 
isolate the leading singularity 
\footnote{This has been done using Mathematica.}.
For $t<y_0$, the leading behaviour again is when the operators are placed on the same sheet and is given by 
  \begin{align}\label{ftxb}
             \lim_{\epsilon\rightarrow 0}
              \big\langle F_{tx}(r_1,  \theta_1^{k} ) F_{tx}( r_2, \theta_2^{(k)})\big\rangle_{(t<y_0)}&=-\frac{1}{16 \pi ^2 \epsilon ^4} +O(\epsilon^0).
        \end{align}
    For operators placed on different sheets, the correlator is finite.
        Again, for $t>y_0$, the leading singular behaviour for operators on the same sheet is 
          \begin{align}\label{ftxa1}
              \lim_{\epsilon\rightarrow 0}
              \big\langle F_{tx}( r_1,  \theta_1^{k}  ) F_{tx}( r_2, \theta_2^{(k)} )\big\rangle_{( t>y_0)}&=-\frac{(t+y_0) \left(4 t^2-t y_0+y^2_0\right)}{128 \pi ^2 t^3 \epsilon ^4}+\cdots.
        \end{align}
        Then there is an equally singular contribution from operators located on adjacent sheets which is given by 
             \begin{align}\label{ftxa2}
              \langle F_{tx}( r_1, \theta_1^{(k_1)} ) F_{tx}( r_2, \theta_2^{(k_2))} )\rangle_{ (t>y_0, k_1-k_2 =-1)}&=-\frac{(t-y_0) \left(4 t^2+t y_0+y^2_0\right)}{128 \pi ^2 t^3 \epsilon ^4}+\cdots.
        \end{align}
        Observe, as discussed in detail for the scalar case the correlator in (\ref{ftxa1}) with operators on the same sheet 
        is related to (\ref{ftxa2}) that with operators on adjacent sheets by $y_0\rightarrow - y_0$. 
        Also note that power of 
        $\epsilon$ is determined by the dimension of the field strength, so it is $\epsilon^4$ and the dependence  in time 
        is through a polynomial of order  $3$ in the ratio $y_0/t$.   Again,  the singular behaviour of the 
        correlators in (\ref{ftxb}), (\ref{ftxa1}) and (\ref{ftxa2}) agree with that 
        obtained in \cite{Nozaki:2016mcy} using the Feynman gauge.

\subsubsection*{Evaluation of $\Delta S_A^{(n)}$}

\begin{figure}[htb]
\centering
\begin{subfigure}{.5\textwidth}
  \centering
  \includegraphics[width=1\linewidth]{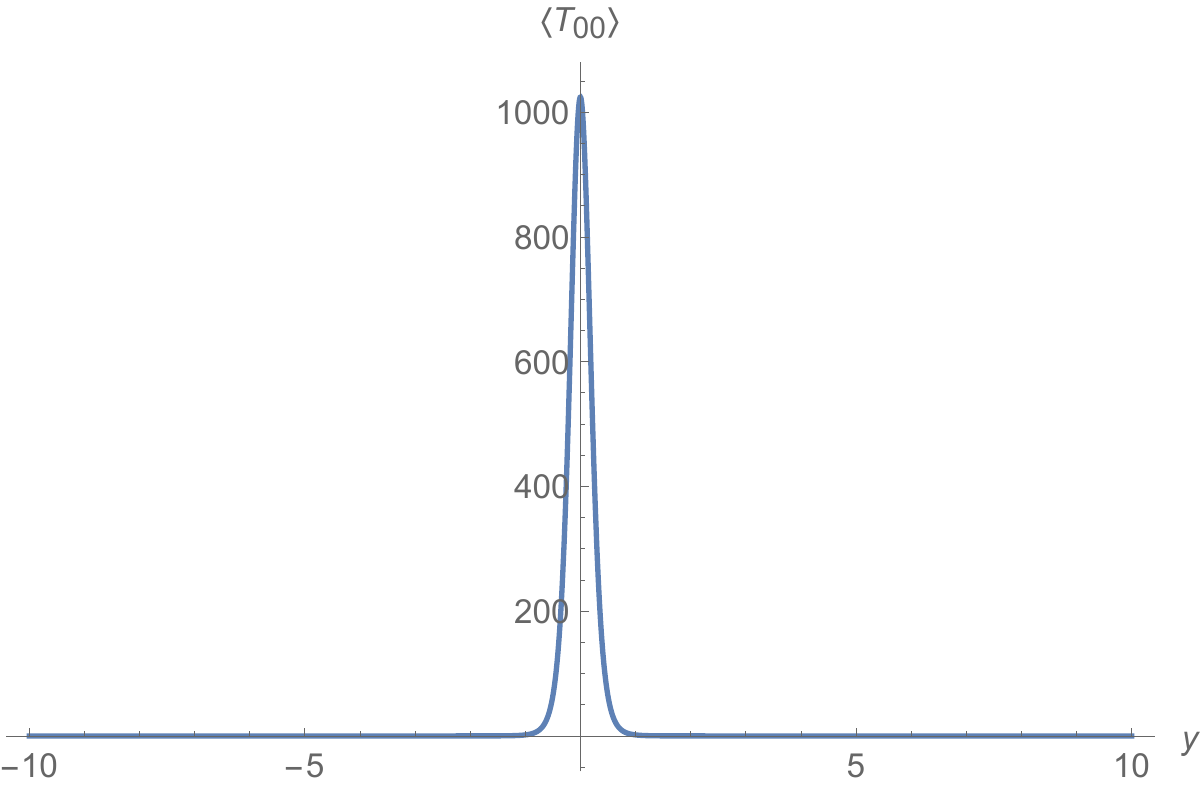}
  \caption{$t=0$ and $\epsilon=0.5$}
\end{subfigure}%
\begin{subfigure}{.5\textwidth}
  \centering
  \includegraphics[width=1\linewidth]{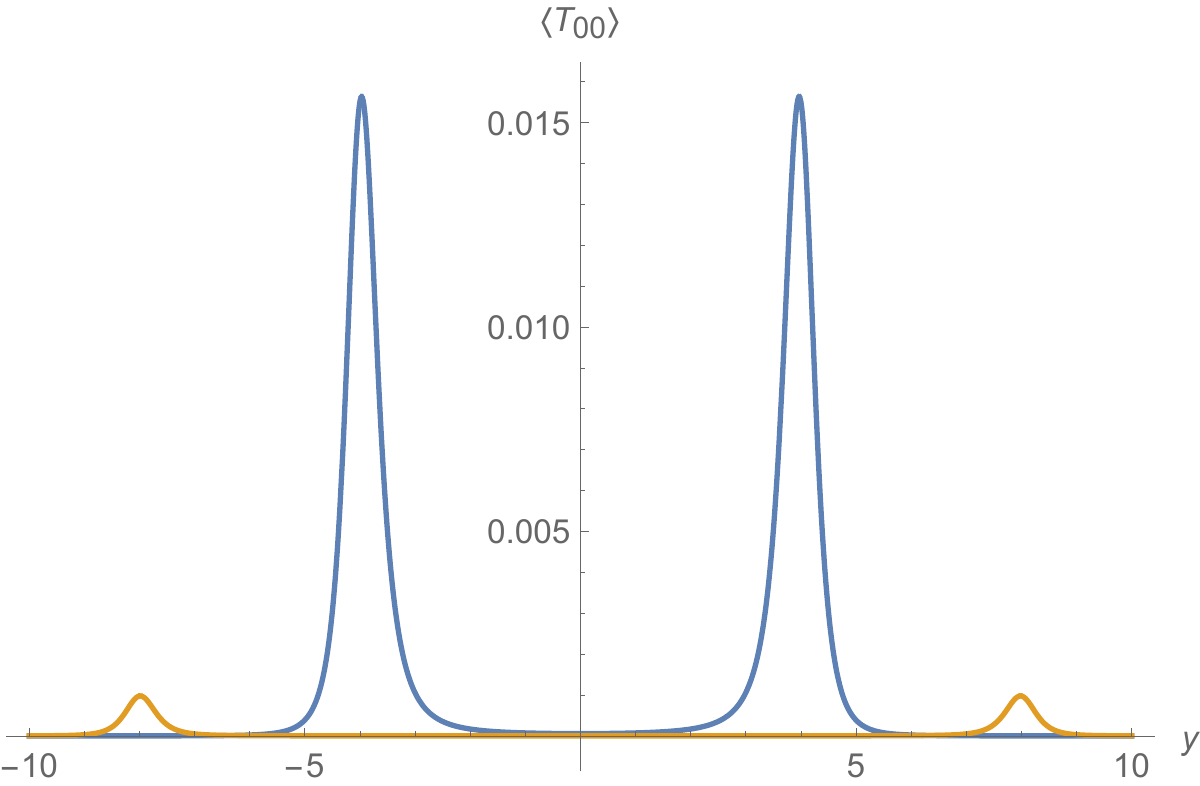}
  \caption{Blue curve: $t=4$, orange curve : $t=8$ .}
\end{subfigure}
\caption{Energy density profile for the local quench  created by field strength $F_{ty}$ at times $t=0, 4, 8$ with 
width $\epsilon=0.5$.} \label{fig:my_label}
\end{figure}

Let us consider the state in which the ground state is excited by the operator $F_{ty}$ given by 
\begin{equation}\label{exvec1}
\rho ( t, - y_0) = F_{ty} ( \tau_e, - y_0) \big|0\big\rangle \big\langle 0 \big| F_{ty} ( \tau_l, - y_0) .
\end{equation}
Just as in the scalar case we wish to probe this state  by the stress tensor
and show that  it corresponds to a  spherical pulse moving at the speed of light. 
We evaluate the expectation value of then energy density 
\begin{eqnarray}\label{enmax}
{\rm Tr}[ T_{00} ( 0, 0) \rho ( t, -y_0)  ]  &=& {\rm Tr} [T_{00} ( 0, y_0) \rho( t, 0) ], \\ \nonumber
&=& \frac{\Big\langle F_{ty} ( - \epsilon - i t, 0) T_{tt} (0, y_0) F_{ty}( \epsilon - i t, 0) \Big\rangle}{
\Big\langle  F_{ty} ( - \epsilon - i t, 0) F_{ty} (\epsilon - i t, 0)  \Big\rangle }.
\end{eqnarray}
The  energy density  of the Maxwell field is given by 
\begin{equation}
     T_{00}=F_{0\lambda}F_{0}^{\;\lambda}-\frac{1}{4} F_{\alpha\beta}F^{\alpha\beta}.
\end{equation}
The Wick contractions in (\ref{enmax}) are on flat $R^4$ or $\Sigma_1$, therefore we use the same $2$ point functions 
given in (\ref{corm1}) with $n=1$. 
Expanding the terms in the stress tensor, one encounters  Wick contractions of the following type 
  \begin{align}
       \begin{split}
      \lim_{\substack{x_1\rightarrow x_2\\ z_1\rightarrow z_2}}    \langle F_{ty}(r_1,\theta_1,x_1,z_1)F_{ti}(r_2,\theta_2,x_2,z_2)\rangle&=0,\\
          \lim_{\substack{x_1\rightarrow x_2\\ z_1\rightarrow z_2}}  \langle F_{ty}(r_1,\theta_1,x,z)F_{yi}(r_2,\theta_2,x_2,z_2)\rangle&=0,\\
       \lim_{\substack{x_1\rightarrow x_2\\ z_1\rightarrow z_2}}    \langle F_{ty}(r_1,\theta_1,x,z)F_{ij}(r_2,\theta_2,x_2,z_2)\rangle&=0,\qquad\qquad \{i,j\}\in \{x,z\}.
       \end{split}
       \end{align}
       Using these simplifications, we find  the evaluation of the expectation value of the energy density 
       of the  excited state (\ref{exvec1}) involves only  product of   two-point functions of  $F_{ty}$. 
       Evaluating this product at the coordinates in (\ref{enmax}) we obtain
          \begin{align}
          {\rm Tr}[ T_{00} ( 0, 0) \rho ( t, -y_0)  ] 
           &=\frac{64 \epsilon ^4}{\left(t^4+2 t^2 \left(\epsilon ^2-y^2\right)+\left(y^2+\epsilon ^2\right)^2\right)^2}.
       \end{align}
       We can also interpret this  expectation value can also be thought of as the value of the 
       energy density at time $t$ at position $y$.  Figure \ref{fig:my_label} shows the profile of energy density 
       at times $t=0, 4, 6$ for quenches created by $F_{ty}$. Again the profile indicates a spherical wave of 
       energy density 
       which travels at the speed of light. 

Let us evaluate the time dependence of the entanglement due to the excitation in (\ref{exvec1}).  
The analysis proceeds just as in the case of the scalar excitation. 
We  need to evaluate the $2n$-point function   in (\ref{renform}) with ${\cal O} = F_{ty}$. 
For $t<y_0$ the leading contribution comes from Wick contractions on the same sheet  given in (\ref{ftyb}). 
Substituting this in (\ref{renform}) we obtain 
   \begin{align}\label{eeftyb}
            \Delta S^{(n)}_{t<y} [F_{ty}] &=\frac{1}{1-n}\log\left(1+\frac{\left(-11 n^4+10 n^2+1\right) \epsilon ^4}{45 n^4 \left(-t^2+y_0^2\right)^2}+\cdots\right)^n,\nonumber\\
            &=\frac{n}{1-n}\frac{\left(-11 n^4+10 n^2+1\right) \epsilon ^4}{45 n^4 \left(t^2-y_0^2\right)^2}+\mathcal{O}(\epsilon^4).
        \end{align}
        To evaluate the denominator in (\ref{renform}) we have used the two-point 
        function of $F_{ty}$ on $\Sigma_1$ which is given by 
        \begin{equation}
        \langle F_{ty} ( \tau_1, y_1 ) F_{ty} (\tau_2, y_2) ) \rangle_{\Sigma_1}  
        = -\frac{1}{\pi^2  [ ( \tau_1 - \tau_2) ^2 + ( y_1 - y_2) ^2 ]^2}.
        \end{equation}
        Therefore for $t<y_0$, the change in R\'{e}nyi entropy vanishes as $\epsilon^4$. 
        This  term is due to the presence of the  composite operator  $:F_{ty}^2:$ in the OPE of the two excitations on the same sheet which implies that the coefficient of the $\epsilon^4$ term should be proportional to the 
        expectation value of the composite operator on  $\Sigma_n$. 
        This expectation value can be obtained by  the point split method
        \begin{align} \label{compositemax}
            \langle :F_{ty}^2:\rangle_{\Sigma_n}  &=-\lim_{\substack{r_2\rightarrow r_1\\ \theta_1\rightarrow\theta_2}}\tilde{\nabla}^2 \left(G(r_1,r_2,\theta_1,\theta_2,\textbf{x},\textbf{x})-\frac{1}{4\pi^2\left(r_1^2+r_2^2-2 r_1r_2\cos(\theta_1-\theta_2)\right)}\right),\nonumber\\
             &=-\lim_{\theta\rightarrow 0}\left(\frac{\csc ^2\left(\frac{\theta }{2 n}\right) \left(2 n^2+3 \csc ^2\left(\frac{\theta }{2 n}\right)-2\right)}{48 \pi ^2 n^4 r_1^4}-\frac{1}{4 \pi ^2 r_1^4 (\cos (\theta )-1)^2}\right) ,\nonumber \\
            &=-\frac{\left(-11 n^4+10 n^2+1\right) }{720 \pi^2  n^4 \left(\tau^2 +y^2\right)^2}.
        \end{align}
        Here $\tilde{\nabla}^2=(\partial_{r_1}^2+\frac{1}{r_1}\partial_{r_1}+\frac{1}{r_1^2}\partial_{\theta_1}^2)$ is the Laplacian in $r-\theta$ plane. Therefore the
       $\epsilon^4$ term in \eqref{eeftyb} is due to the  presence of the composite operator $:F_{ty}^2:$. 
       
 In the regime $t>y_0$, the leading contributions in the $\epsilon\rightarrow 0$ limit arises from the 
      Wick contractions on the same sheet given in (\ref{ftya1}) as well as cyclic contractions  on adjacent sheets  
      given in (\ref{ftya1}). 
     Substituting these contributions in (\ref{renform}) and normalising with the correlator on $\Sigma_1$ we obtain 
      \begin{align} \label{rfty}
           \lim_{\epsilon\rightarrow 0} \Delta S^{(n)}_A [F_{ty}] &=\frac{1}{1-n}\log\left[ \frac{\left(\frac{(2 t-y_0) (t+y_0)^2}{64 \pi ^2 t^3 \epsilon ^4}\right)^n+\left(\frac{(t-y_0)^2 (2 t+y_0)}{64 \pi ^2 t^3 \epsilon ^4}\right)^n}{\left(\frac{1}{16\pi^2\epsilon^4}\right)^n} \right].
        \end{align}
       At the leading order in $\epsilon$, $\Delta S^{(n)}_A$ is independent of $\epsilon$.  The asymptotic behaviour at large times is given by
        \begin{align}
        \lim_{t\rightarrow \infty}    \Delta S^{(n)}_A [F_{ty}] &=\log (2)-\frac{9 n y_0^2}{8 t^2}+ O(\frac{y_0^4}{t^4})+\cdots
        \end{align}
          The coefficient of $y_0^2/t^2$ is important, in fact we will see that the knowledge of this coefficient is 
        sufficient to determine the entire time dependent profile given in (\ref{rfty}). 

        Taking the $n\rightarrow 1$ limit we obtain the change in the entanglement entropy 
          \begin{align}
         \lim_{n\rightarrow 1}   \Delta S^{(n)} [F_{ty}] &=\Delta S_{\rm{EE}} [F_{ty}] ,\nonumber\\
         &=\log2-\frac{1}{2}\log\left(1-\frac{y_0^2}{t^2}\right)- \frac{ y_0}{t}  \tanh ^{-1}\left(\frac{y_0}{t}\right).
        \end{align}
    
        We  repeat the same exercise  for the excitation created by the field strength $F_{tx}$. 
        In the $\epsilon\rightarrow 0$ limit, the change in entanglement entropy vanishes for $t<y_0$. 
        For this case we are not keeping track of the $\epsilon^4$ contribution. 
        The growth in the R\'{e}nyi entropy for $t>y_0$  in the zero width limit is given by 
             \begin{align} \label{rftx}
         \lim_{\epsilon\rightarrow 0} \Delta S^{(n)}_A [F_{tx}] &=\frac{1}{1-n}\log \left[ \frac{\left(\frac{(t+y_0) \left(4 t^2-t y_0+y_0^2\right)}{128 \pi ^2 t^3 \epsilon ^4}\right)^n+\left(\frac{(t-y_0) \left(4 t^2+t y_0+y_0^2\right)}{128 \pi ^2 t^3 \epsilon ^4}\right)^n}{\left(\frac{1}{16\pi^2\epsilon^4}\right)^n} \right].
        \end{align}
        The asymptotic behaviour of the R\'{e}nyi entropy at large times is given by 
                \begin{align}
        \lim_{t\rightarrow \infty}    \Delta S^{(n)}_A [F_{tx}] &=\log (2)-\frac{9 n y_0^2}{32 t^2}+ O(\frac{y_0^3}{t^3})+\cdots.
        \end{align}
        Finally the entanglement entropy is given by 
         \begin{align}
         \lim_{n\rightarrow 1}   \Delta S^{(n)}[F_{tx}] &=\Delta S_{\rm{EE}}[F_{tx}] ,\nonumber\\
         &=\frac{1}{8} \Big[8\log 8-\left(\frac{y_0^3}{t^3}+\frac{3 y_0}{t}+4\right) \log \left(\frac{y_0^3}{t^3}+\frac{3 y_0}{t}+4\right),\nonumber\\
         &-\left(4-\frac{y_0^3}{t^3}-\frac{3 y_0}{t}\right) \log \left(4-\frac{y_0}{t}  \left(\frac{y_0^2}{t^2}+3\right)\right)\Big].
        \end{align}

We  make some observations regarding the R\'{e}nyi/entanglement entropy growth after $t>y_0$ for 
the quenches created by the field strengths $F_{ty}$ in (\ref{rfty}) and $F_{tx}$ in (\ref{rftx}). 
\begin{enumerate}
\item
The  2 leading Wick contractions, those on the same sheet and that on adjacent sheets are related by $y_0 \rightarrow - y_0$. 
\item  The time dependence  of the leading Wick contractions is a polynomial in the  ratio $y_0/t$ of order 3, which is
equal to $2s+1$ for $s=1$.   
For example, the $F_{tx}$, the leading contribution to the $F_{tx}$ correlator can be written as 
   \begin{align}\label{ftxa11}
              \lim_{\epsilon\rightarrow 0}
              \big\langle F_{tx}( r_1,  \theta_1^{k}  ) F_{tx}( r_2, \theta_2^{(k)} )\big\rangle_{( t>y_0)}&=-\frac{(1+r) \left(4 -r +r^2\right)}{128 \pi ^2  \epsilon ^4}+\cdots, \qquad r = \frac{y_0}{t}.
        \end{align}
        The scaling dimension is set by the power of $\epsilon$ in the denominator which is $4$. 
        The polynomial is of the order $3$.  The correlator of the operator
         on the  adjacent sheets is obtained by $r\rightarrow - r$.
\item The ratio of the $2n$ point function on $\Sigma_n$ to the $n$-th power of the 2-point function on $\Sigma_1$ 
at $n=1$,  by definition is unity. This, together with the fact that two leading contributions for the $2n$ point function are related by $r\rightarrow - r$ 
 implies that at $n=1$, the polynomials that occur in the logarithm of the R\'{e}nyi entropy must satisfy 
 \begin{equation}
\qquad P(r) + P( -r) = 1. 
 \end{equation} 
 Note that due to this condition the polynomial should be of the form
 \begin{equation}
 P(r) = \frac{1}{2} ( 1+ a_1 r + a_3 r^2) .
 \end{equation}
 Note that the even power  $r^2$ is missing in the polynomial. 
 By examining (\ref{rfty}) and (\ref{rftx}) we see that both the contributions in the argument of the logarithm is in this form. 
 \item 
 The  leading contribution to the $F_{ty}$
  correlator  on the same sheet 
  has a factor of $(1+r)^2$, while the  corresponding contribution from the $F_{tx}$ correlator has a factor $(1+r)$. 
  $F_{ty}$ is    transforms as a (pseudo)scalar under $SO(2)_T$ and a scalar 
 under  $SO(2)_L$. 
  $F_{tx}$  transforms as  vector under $SO(2)_T$ as well as $SO(2)_L$.
   The pseudo-scalar has higher power of the factor $(1+r)^2$ which is  equal  to $s+1$. 
  \item 
  The polynomial corresponding to the scalar  of $SO(2)_T\times SO(2)_L$, $F_{ty}$ is completely
  determined by the above conditions.
  It should be of the form
  \begin{equation}
  P_{F_{ty} } (r)  = \frac{1}{2} ( 1+r)^2 ( 1 + a_1 r)  = \frac{1}{2} [  1 + ( a_1 + 2 ) r +   ( 2 a_1  + 1) r^2  + a_1 r^3].
  \end{equation}
  This implies we must have $a_1 = -1/2$, therefore we have 
  \begin{equation}
   P_{F_{ty} } (r)  = \frac{1}{2} ( 1+r)^2 ( 1  -\frac{1}{2}  r) 
  \end{equation}
  This reasoning agrees with the result in (\ref{ftya1}).
  
  \item The polynomial corresponding to the vector of $SO(2)_T\times SO(2)_L$ is determined 
  to a single number.  It  is of the form
  \begin{eqnarray}
  P_{F_{ty}} ( r) &=& \frac{1}{2} ( 1+r) (  1+ a_1 r + a_2 r^2) , \\ \nonumber
  &=& \frac{1}{2} ( 1 + ( a_1 + 1) r + ( a_1 + a_2) r^2 +  a_2 r^3 ) .
  \end{eqnarray}
  Since we have only odd terms in $r$ we must have $a_2 = - a_1$. 
  There the polynomial is of the form
  \begin{equation}
  P_{F_{ty}} ( r)  = \frac{1}{2} ( 1+ r) ( 1 + a_1 r - a_1 r^2) 
  \end{equation}
  Comparison with (\ref{ftxa1}) we see that indeed the polynomial is of this form with $a_1 = -1/4$. 
  \item 
  Finally if the polynomial is given by 
  \begin{equation}
  P(r) = \frac{1}{2} ( 1+ a_1 r +   a_2 r^3 ). 
  \end{equation}
  The leading terms in the R\'{e}nyi entropy is given by 
  \begin{eqnarray}
  S^{(n)}_A( P) &=& \frac{1}{ (1- n) } \log\left[  \Big(  \frac{1}{2} \Big)^n \Big( 1+ n a_1 r +  \frac{n(n-1)}{2}  a_1^ 2 r^2 +\cdots\Big) 
  + (r\rightarrow - r)  \right]   ,\nonumber \\
  &=& \log ( 2) - \frac{n}{2} a_1^2 r^2 + \cdots.
  \end{eqnarray}
  The reason the sub-leading correction to the long time behaviour of the R\'{e}nyi entropies starts ar $r^2$ is clearly
  due to the fact that there are no even powers of $r$ is the polynomial corresponding to the 2-point function. 
The coefficient which determines the  asymptotic behaviour at large times is entirely determined by 
  the linear term  $a_1$ of the polynomial corresponding to the correlator.  Thus the polynomial of 
  the field strengths in the vector representation of $SO(2)_T\times SO(2)_L$ is completely determined by the 
  asymptotic behaviour of the R\'{e}nyi entropies. 
\end{enumerate}

It is easy to see the linear polynomial $P_{\phi} = \frac{1}{2}( 1+ r)$ obtained for the scalar 
$s=0$, quench is consistent with these observations.
In this next section we will see that these observations generalise 
 to quenches induced by the curvature of the linearised graviton. 

        \section{Local gravitational quenches} \label{gravq}
        
        In this section we apply the methods developed in previous section 
         to quenches created by the Riemann curvature tensor 
        in the theory of linearised gravity in $d=4$.  
        As we have mentioned in the introduction, the 
        information theoretic properties of the graviton has only been recently studied. 
        Quenches with the spin-2 field have not been studied earlier. In \cite{Laddha:2020kvp,Raju:2021lwh} 
        it has been argued that the absence of the 
        `split' property of local quantum theories in gravity implies that sub-regions in gravity are ill defined. 
      We will see in this section, that the methods of the previous section can be extended to
       the theory of linearised gravity and the results for the entanglement of quenches is qualitatively similar to that obtained for the vector and scalar quenches.
      
        The Lagrangian for the theory of linearised gravity is given by 
               \begin{eqnarray} \label{laggrav}
{\cal L } = - \partial_\mu h^{\mu\nu} \partial_\alpha h^\alpha_{\, \nu} 
+ \frac{1}{2} \partial^\alpha h_{\mu\nu} \partial_\alpha h^{\mu\nu} 
+ \partial_\mu h^{\mu \nu} \partial_\nu h^\alpha_{\, \alpha}
- \frac{1}{2} \partial_\alpha h^{\mu}_{\, \mu} \partial^\alpha h^\nu_{\, \nu}.
\end{eqnarray}
It admits the following gauge symmetry
\begin{equation}\label{gsym}
\delta h_{\mu\nu}   = \partial_\mu \xi_\nu + \partial_\nu \xi_\mu.
\end{equation}
We need to consider operators which are gauge invariant to create the quenches. 
In the theory of linearised gravity, it is easy to see that the Riemann curvature is gauge invariant. 
It is defined by 
\begin{equation} \label{reim}
R_{\mu\nu\rho\sigma} = \frac{1}{2} ( \partial_\nu \partial_\rho h_{\mu \sigma}
-\partial_\mu \partial_\rho h_{\nu \sigma}
+\partial_\mu \partial_\sigma h_{\nu \rho} - \partial_\nu \partial_\sigma h_{\mu \rho} ). 
\end{equation}
We will study quenches created by all the components of the Riemann tensor. 
By explicit computation we will show that there are only 7 distinct quenches created by the 
20 independent components of the Riemann tensor. 
 
 \subsection{Gauge fixing and the graviton propagator}
 
Just as in the case of the  $U(1)$ theory, we need to obtain the 2 point function of the graviton on the
replica surface $\Sigma_n$. 
For this it is convenient to fix the gauge which preserves the $SO(2)_T\times SO(2)_L$  symmetry of the 
defect geometry.  
Using the gauge invariance in (\ref{gsym}),  we can ensure that the graviton is transverse and traceless
\begin{equation} \label{ggcond}
\partial^\mu h_{\mu\nu} = 0, \qquad\qquad h^\mu_{\;\mu} = 0.
\end{equation}
The Einstein's  equations then implies that the  graviton satisfies the equations
\begin{equation} \label{eineq}
\Box h_{\mu\nu} =0.
\end{equation}
There is still a residual gauge invariance which preserves the transverse, traceless conditions, 
these transformations are of the form
\begin{equation} \label{hshift}
h'_{\mu\nu}  = h_{\mu\nu} + \partial_\mu \epsilon_\nu + \partial_\nu \epsilon_\mu , \qquad 
\end{equation}
where the gauge parameter satisfies the conditions
\begin{equation} \label{newgc}
\Box \epsilon_\mu = 0, \qquad\qquad \partial^\mu \epsilon_\mu =0.
\end{equation}
We can use this degree of freedom to ensure that the graviton satisfies the  transverse conditions   both perpendicular 
and parallel to the defect. That is obtain a gauge parameter,  so that we have
\begin{equation} \label{newgg}
\partial^a h'_{a\mu} = 0, \qquad \partial^i h'_{i \mu} = 0, \qquad a\in \{ t , y\}, i \in \{x, z\}, 
\qquad h^{\prime \mu}_{\mu} =0.
\end{equation}
The gauge parameter which does this job  is given by 
\begin{eqnarray} \label{gravgau}
\epsilon_a &=& -\frac{1}{\nabla^2} \left( \partial^i h_{ia}  - \frac{  \partial_a \partial^i\partial^j h_{ij} }{ 2  \nabla^2} \right), 
\\ \nonumber
\qquad \epsilon_i &=&  - \frac{1}{\nabla^2} \left( \partial^j h_{i j} - \frac{ \partial_i \partial^j \partial^k h_{jk}  }{ 2 \nabla^2 } 
\right) , \\ \nonumber
\nabla^2  &=& \partial^i\partial_i .
\end{eqnarray}
On substituting this gauge parameter in (\ref{hshift}) it is easy to see that the conditions (\ref{newgg}) is satisfied. 
To show this we need the transverse condition in (\ref{ggcond})  and the equations of motion (\ref{eineq}). 
It is also  useful to realise that the choice in (\ref{gravgau}) satisfies
\begin{equation}
\partial^a \epsilon_a =  - \partial^i \epsilon_i = +  \frac{ \partial^i \partial^j}{2\nabla^2}  h_{ij}, 
\end{equation}
To arrive at these relations we have again used the gauge condition $\partial^ i h_{i\mu } = 0$ and the on shell condition
$\Box h_{\mu\nu } = 0$. Therefore the gauge transformation in (\ref{gravgau})  satisfies the transverse condition in  (\ref{newgc}).   It is also easy to see that the gauge parameter satisfies the condition $\Box \epsilon_\mu =0$ using the  on shell condition  of $h_{\mu\nu}$.

\subsubsection*{Graviton Propagator}

Since it is possible to choose the gauge conditions in (\ref{newgg} ), we can can construct a propagator 
that is consistent with these conditions. 
This  propagator is  given by  
\begin{eqnarray}\label{defcor1}
            \langle h_{ab}(x) h_{cd} (x') \rangle &=&\left(\frac{P_aP_c' P_bP_d'}{\nabla^4}+\frac{P_aP_d'P_bP_c'}{\nabla^4}\right)G (x, x'),\nonumber \\
           \langle h_{ij}(x) h_{kl} (x') \rangle &=&\Bigg[\left(\delta_{ik}-\frac{\partial_i\partial_k}{\nabla^2}\right)\left(\delta_{jl}-\frac{\partial_j\partial_l}{\nabla^2}\right)+\left(\delta_{il}-\frac{\partial_i\partial_l}{\nabla^2}\right) \left(\delta_{jk}-\frac{\partial_j\partial_k}{\nabla^2}\right)\Bigg]G  (x, x') , \nonumber \\
         \langle h_{ai}(x)  h_{b j} (x') \rangle &=& \frac{P_a P_b'  }{\nabla^2}
          \left( \delta_{ij} - \frac{\partial_i \partial_j}{\nabla^2} \right)  G(x, x') , \nonumber \\ 
        \langle h_{ab} (x) h_{ij} ( x') \rangle &=&  2 \frac{ P_a P_b }{\nabla^2} 
        \left( \delta_{ij} -  \frac{\partial_i \partial_j}{\nabla^2} \right) G (x, x') , \\ \nonumber
           \langle h_{ij} (x) h_{ab} ( x') \rangle &=&  2 \frac{ P'_a P'_b }{\nabla^2} \left( \delta_{ij} - \frac{\partial_i \partial_j}{\nabla^2} \right) G (x, x'). \\ \nonumber
\langle h_{ai} (x) h_{ij} (x') \rangle &=&  \langle h_{ai} (x)  h_{cd} (x')  \rangle =0,
\end{eqnarray}
Here $G(x, x')$ is the scalar propagator on the replica surface given in (\ref{sclcor}). 
It can be easily verified that the propagator satisfies the conditions
\begin{eqnarray}
\nabla^a_x\langle h_{a \mu } (x)  h_{\rho \sigma} (x') \rangle = 0, \qquad 
\partial^i \langle h_{i \mu } (x) h_{j k} (x') \rangle =0, \\ \nonumber
\langle h^\mu_\mu ( x) h_{\rho \sigma} (x') \rangle =0,  \qquad 
\Box_x \langle h_{\mu\nu} (x) h_{\rho\sigma} (x') \rangle = 0. 
\end{eqnarray}
A similar set of equations in which the derivatives acts on $x'$ 
and the trace is taken over $h_{\rho\sigma} (x') $ also 
holds.

        \subsection{Two-point function of Riemann curvatures} \label{2ptrcurv}
        
        In this section we evaluate the 2-point functions of all the 20 components of the Riemann curvatures. 
        We will see by explicit calculation
         that among the 20   correlators we can use to create the quenchs,  there are only 7 independent
        quenches. In each case we evaluate the behaviour of the correlator  before and after it reaches the 
        entangling surface in the small width limit. 
        
        \subsection*{Class 1: $\{R_{tyty}$,  $R_{xzxz}$,  $R_{tyxz}$\}} 
        
        We begin by evaluating two-point functions of $R_{tyty}$ on replica surface.
      From (\ref{reim}),  $R_{tyty}$ is given by
       \begin{align}
            R_{tyty}&=\frac{1}{2}\Big[\partial_y\partial_t h_{ty}-\partial_t^2h_{yy}+\partial_t\partial_y h_{yt}-\partial_y^2h_{tt}\Big],\nonumber\\
            &=\frac{1}{2}\Big[2\partial_y\partial_t h_{ty}-\partial_t^2h_{yy}-\partial_y^2h_{tt}\Big].
        \end{align}
        We now use the graviton corellator on the replica surface
         \eqref{defcor1} to evaluate the two-point function of $R_{tyty}$
          \begin{align} \label{rtyty}
            \langle R_{tyty}(x ) R_{tyty}(x' )\rangle&=\frac{1}{4}\Big[4\partial_{y_1}\partial_{y_2}\partial_{t_1}\partial_{t_2}\langle h_{ty}h_{ty}\rangle-2\partial_{y_1}\partial_{t_1}\partial_{t_2}^2\langle h_{ty}h_{yy}\rangle-2\partial_{y_1}\partial_{t_1}\partial_{y_2}^2\langle h_{ty}h_{tt}\rangle\nonumber\\
            &-2\partial_{t_1}^2\partial_{y_2}\partial_{t_2}\langle h_{yy}h_{ty}\rangle+\partial_{t_1}^2\partial_{t_2}^2\langle h_{yy}h_{yy}\rangle +\partial_{t_1}^2\partial_{y_2}^2\langle h_{yy}h_{tt}\rangle\nonumber\\
           & -2\partial_{y_1}^2\partial_{y_2}\partial_{t_2}\langle h_{tt}h_{ty}\rangle+\partial_{y_1}^2\partial_{t_2}^2\langle h_{tt}h_{yy}\rangle+\partial_{y_1}^2\partial_{y_2}^2\langle h_{tt}h_{tt}\rangle\Big],\nonumber\\
           &=\frac{1}{2}\frac{(\partial_{t_1}^2+\partial_{y_1}^2)^2(\partial_{t_2}^2+\partial_{y_2}^2)^2}{\nabla^4}G(x , x' ),\nonumber\\
           &=\frac{1}{2}(\partial_{t_1}^2+\partial_{y_1}^2)^2G(x ,x').
        \end{align}
        To derive the last line we use the on-shell condition of the scalar Green's function  in the second coordinate $x'$
        given in (\ref{onshell})
        and translation invariance 
        in $x, z$ direction (\ref{transinv}), to convert the derivatives to the first coordinate. 
        
Similarly, we evaluate $\langle R_{xzxz} R_{xzxz}\rangle.$ The expression of $R_{xzxz}$ is given by (\ref{reim})
        \begin{align}
           R_{xzxz}&=
            \frac{1}{2}\Big[2\partial_x\partial_z h_{xz}-\partial_x^2h_{zz}-\partial_z^2h_{xx}\Big].
        \end{align}
   Using the correlators from  \eqref{defcor1},
        we get
        \begin{align}\label{rxzxz}
            \langle R_{xzxz}(x ) R_{xzxz}(x ')\rangle&=\frac{1}{4}\Big[4\partial_{x_1}\partial_{x_2}\partial_{z_1}\partial_{z_2}\langle h_{xz}h_{xz}\rangle-2\partial_{x_1}\partial_{z_1}\partial_{x_2}^2\langle h_{xz}h_{zz}\rangle-2\partial_{x_1}\partial_{z_1}\partial_{z_2}^2\langle h_{xz}h_{xx}\rangle\nonumber\\
            &-2\partial_{x_1}^2\partial_{z_2}\partial_{x_2}\langle h_{zz}h_{xz}\rangle+\partial_{x_1}^2\partial_{x_2}^2\langle h_{zz}h_{zz}\rangle +\partial_{x_1}^2\partial_{z_2}^2\langle h_{zz}h_{xx}\rangle\nonumber\\
           & -2\partial_{z_1}^2\partial_{z_2}\partial_{x_2}\langle h_{xx}h_{xz}\rangle+\partial_{z_1}^2\partial_{x_2}^2\langle h_{xx}h_{zz}\rangle+\partial_{z_1}^2\partial_{z_2}^2\langle h_{xx}h_{xx}\rangle\Big],\nonumber\\
           &=\frac{(\partial_{x_1}^2+\partial_{z_1}^2)^4}{2\nabla^4}G(x ,x'),\nonumber\\
           &=\frac{1}{2}(\nabla^2)^2G(x, x') = \frac{1}{2} ( \partial_{t_1}^2 + \partial_{t_2} )^2 G(x, x') .
        \end{align}
        In the last line we have used the on-shell condition satisfied by $G(x, x')$. 
        Therefore this correlator is identical  to $\langle R_{tyty} (x) R_{tyty} (x') \rangle$ given in (\ref{rtyty}). 
        This implies quenches created by inserting the component $R_{tyty}$ would behave identically  to the
        component $R_{xzxz}$.

     Consider the correlator 
        $\langle R_{tyty}(x)R_{tyty}(x')\rangle$,  the expression of $R_{tyxz}$ is given by
         \begin{align}
            R_{tyxz}&= \frac{1}{2}\Big[ \partial_y\partial_x h_{tz} - \partial_t\partial_x h_{yz} + \partial_t\partial_z h_{yx} - \partial_y\partial_z h_{tx} \Big],
        \end{align}
        There are 16 contractions using the graviton correlator in (\ref{defcor1}), which on adding results in 
        \begin{align} \label{rtyxz}
            \langle R_{tyxz}(x) R_{tyxz}(x')\rangle&=-\frac{1}{4}(\partial_{t_1}^2+\partial_{y_1}^2)(\partial_{t_2}^2+\partial_{z_2}^2)G(x,x'),\nonumber\\
            &=-\frac{1}{4}(\partial_{t_1}^2+\partial_{y_1}^2)^2G(x ,x').
        \end{align}
        Comparing with  (\ref{rtyty}), (\ref{rxzxz}) , we see that the two-point function of $R_{tyxz}$ differs from the 
        from two correlators by a factor of $2$. 
      The overall factor of $2$ does not affect the calculation of 
        The R\'{e}nyi/entanglement entropy corresponding to quenches just depends on the ratio of the  correlators as can be seen from (\ref{renform}). Therefore the time dependence of the entanglement entropy of quenches created 
        by all components $R_{tyty}, R_{xzxz}, R_{tyxz}$ will be the same.  For these quenches we will take the 
        excited state by $R_{tyty}$ to be the representative quench.           
        
        The explicit expression of the $R_{tyty}$ correlator on $\Sigma_n$ is given in (\ref{exprtyty}). But, 
         all we need is the leading singular behaviour of the 2-points functions to obtain the 
        time dependence of the entanglement entropy in the $\epsilon\rightarrow 0$ limit. 
        For $t<y_0$, the leading singularity arises when the operators are on the same sheet and is given by 
        \begin{align}\label{tlessy_grav}
             \lim_{\epsilon\rightarrow 0}
            &  \langle R_{tyty}(r_1, \theta_1^{(k)} ) R_{tyty}(r_2, \theta_2^{(k)} )\rangle_{t<y_0} \\ \nonumber
            & \qquad\qquad  =\frac{1}{2}\left(\frac{1}{4 \pi ^2 \epsilon ^6}+\frac{(n-1) (n+1) \left(191 n^4+23 n^2+2\right)}{3780 \pi ^2 n^6 \left(t^2-y_0^2\right)^3}+\cdots\right).
        \end{align}
        The leading singularity is same as that when the 2 points are on $R^4$ or $\Sigma_1$. 
        The leading  correction at $O(\epsilon^0)$ is due to expectation value of the composite $:R_{tyty}^2:$ on 
        $\Sigma_n$.  Correlators with points on different sheets are not singular in this regime. 
        
      In $t>y_0$ regime,  there are 2 leading singular contributions. When the operators are placed on the same sheet
      we obtain 
            \begin{align}\label{rtytya1}
             \lim_{\epsilon\rightarrow 0}
              \langle R_{tyty}(r_1, \theta_1^{(k)} ) R_{tyty}( r_2, \theta_2^{(k)}  )\rangle_{t>y_0}&=\frac{1}{2}\left(\frac{(t+y_0)^3 \left(8 t^2-9 t y_0+3 y_0^2\right)}{64 \pi ^2 t^5 \epsilon ^6}+\mathcal{O}(\frac{1}{\epsilon^5})+\cdots\right).
        \end{align}
        The other leading singular contribution occurs when operators are placed on adjacent sheet and is given by 
        \begin{align}\label{rtytya2}
         \lim_{\epsilon\rightarrow 0}
             & \langle  R_{tyty}( r_1, \theta_1^{(k_1} ) R_{tyty}( r_2, \theta_2^{(k_2)} )\rangle_{t>y_0, k_1- k_2 =-1} \nonumber \\ \qquad\qquad 
             &=\frac{1}{2}\left(\frac{(t-y_0)^3 \left(8 t^2+9 t y_0+3 y^2_0 \right)}{64 \pi ^2 t^5 \epsilon ^6}+\mathcal{O}(\frac{1}{\epsilon^5})+\cdots\right).
        \end{align}
        Observe that the power of $\epsilon$ in the denominator is determined by the dimension of the field and the 
        singular behaviours in this regime on the same sheet and on adjacent sheets are related by $y_0\rightarrow - y_0$. 
        The time dependence is through the ratio $y_0/t$ which is a polynomial of order $5 = 2s+1, s= 2$. 
        Finally  the correlator on $\Sigma_1$ is given by 
         \begin{align} \label{rtytyf}
              \langle R_{tyty}(\tau_1, y_1) R_{tyty}(\tau_2, y_2 )\rangle_{n=1}&=
              \frac{8}{ \pi ^2 \big[ ( \tau_1 - \tau_2)^2 + ( y_1 - y_2)^2 \big]^3 }.
        \end{align}
        
        \subsection*{Class 2:  $\{ R_{tytx},  R_{tytz},  R_{yzxz}, R_{yxxz} \} $}
        
       In  equations (\ref{Rtytx}), (\ref{Rtytz}), (\ref{Ryzxz}), (\ref{Ryxxz})
          of the appendix   we show that the curvature 
        components  $\{ R_{tytx},  R_{tytz},  R_{yzxz}, R_{yxxz} \} $ have identical $2$ point functions when 
        one takes the coincident limit in the directions parallel  to the defect, that is $x_1= x_2, z_1 = z_2$. 
                 \begin{align}
           \langle R_{tytx}(x)R_{tytx}(x')\rangle&=  \left.  \partial_{x_1}^2( -2 \partial_{y_1}\partial_{y_2}+ \partial_{t_1}\partial_{t_2})G(x,x')  \right|_{x_1= x_2, z_1 = z_2}.
         \end{align}
         After taking the derivative with respect to $x_1$ coordinate, we evaluate the two-point function in the limit $x_1\rightarrow x_2$ and $z_1\rightarrow z_2$.
         Using this expression the leading singularity of the correlator when the two-points are on the same 
         sheet in the regime $t>y_0$ is given by 
         \begin{align}
        \lim_{\epsilon\rightarrow 0}
          \langle R_{tytx}(r_1, \theta_1^{(k)} )R_{tytx}(r_2, \theta_2^{(k)} )\rangle_{t>y_0}  &= \frac{(t+y_0)^2 \left(14 t^3-13 t^2 y_0+12 t y_0^2-6 y_0^3\right)}{512 \pi ^2 t^5 \epsilon ^6}+\cdots
         \end{align}
         In this regime, the other leading singularity arises from correlators with points on adjacent sheets which is
         is given by 
         \begin{align}
          \langle R_{tytx}(r_2, \theta_2^{(k_1)} )R_{tytx}(r_2, \theta_2^{(k_2)} )\rangle_{t>y_0, k_1 -k_2 =-1} &= 
         \frac{(t-y_0)^2 \left(14 t^3+13 t^2 y_0+12 t y_0^2+6 y_0^3\right)}{512 \pi ^2 t^5 \epsilon ^6}+\cdots
         \end{align}
         Finally the two-point function on $\Sigma_1$ is given by 
         \begin{align}
          \langle R_{tytx}(\tau + \epsilon )R_{tytx}( \tau - \epsilon ) \rangle_{\Sigma_1} &=  \frac{7}{128 \pi ^2 \epsilon ^6}.
         \end{align}
         Here it is understood, all other coordinates  are taken to  be coincident. 
         As expected, the above result is also the 
          leading singularity when points are on the same sheet in the regime $t<y_0$.

               \subsection*{Class 3:  $\{R_{txxz}$ or $R_{tzxz}$ or $R_{tyyx}$ or $R_{tyyz} \} $}
        
        All the curvature components in this class have the same two-point function in the limit the points along the 
        defect are coincident. 
        The details of demonstrating this is given  in equations (\ref{Rtxxz}), (\ref{Rtzxz}), (\ref{Rtyyx}), (\ref{Rtyyz})
         of the appendix. 
        Let us take $R_{txxz}$ as the representative in this class, the correlator is given by 
        \begin{equation}
        \langle R_{txxz} (x) R_{txxz}(x') \rangle = 
        \left. \partial_{x_1}^2(\partial_{y_1}\partial_{y_2}-2 \partial_{t_1}\partial_{t_2})G(x ,x')\right|_{x_1= x_2, z_1 = z_2}.
        \end{equation}
        In the regime $t>y_0$, the leading singularity in the $\epsilon\rightarrow 0$ limit from the correaltors on the 
        same sheet are given by 
        \begin{equation}
        \lim_{\epsilon\rightarrow 0} \langle R_{txxz} ( r_1, \theta_1^{(k)}) R_{txxz}(  r_2,  \theta_2^{(k)} )   \rangle_{t>y_0} 
= -\frac{(t+y_0)^2 \left(22 t^3-14 t^2 y_0+6 t y_0^2-3 y_0^3\right)}{512 \pi ^2 t^5 \epsilon ^6}+\cdots
\end{equation}
Similarly the leading singularity form operators  on adjacent sheets is given by 
\begin{equation}
   \lim_{\epsilon\rightarrow 0} \langle R_{txxz} ( r_1, \theta_1^{(k_1)}) R_{txxz}(  r_2 ,\theta_2^{(k_2)} )   \rangle_{t>y_0, k_1 -k_2 = -1} =  -\frac{(t-y_0)^2 \left(22 t^3+14 t^2 y_0+6 t y_0^2+3 y_0^3\right)}{512 \pi ^2 t^5 \epsilon ^6}+\cdots
\end{equation}
For $t<y_0$ as well as the correlator on the $\Sigma_1$, we obtain the following behaviour
\begin{align}
          \langle R_{txxz}(\tau + \epsilon )R_{txxz}(\tau - \epsilon)\rangle_{\Sigma_1} &=   -\frac{11}{128 \pi ^2 \epsilon ^6}.
         \end{align}
         where all the other coordinates are understood to be coincident for the two operators. 

     \subsection*{ Class 4: $\{ R_{txtx}, R_{yzyz}, R_{tztz}, R_{yxyx} \}$}
           
          All the curvature components in this class have the same two-point function in the limit the points along the 
        defect are coincident. 
        The details of this is given in equations (\ref{Rtxtx}), (\ref{Ryzyz}), (\ref{Rtztz})
        Let us take $R_{txtx}$ as the representative in this class, the correlator is given by 
             \begin{equation}
        \langle R_{txtx} (x) R_{txtx}(x') \rangle = 
      \Big[  \frac{3}{16}\left(\partial_{t_1}^2\partial_{t_2}^2+ \partial_{y_1}^2\partial_{y_2}^2\right)+\frac{1}{16}\left(\partial_{t_1}\partial_{y_2}-\partial_{y_1}\partial_{t_2}\right)^2)\Big]G(x ,x')|_{x_1= x_2, z_1 = z_2}.
        \end{equation}
         In the regime $t>y_0$, the leading singularity in the $\epsilon\rightarrow 0$ limit from the correaltors on the 
        same sheet are given by 
        \begin{equation}
        \lim_{\epsilon\rightarrow 0} \langle R_{txtx} ( r_1, \theta_1^{(k)}) R_{txtx}(  r_2 , \theta_2^{(k)} )   \rangle_{t>y_0} 
= \frac{9 (t+y_0) \left(6 t^4-t^3 y_0+t^2 y_0^2-t y_0^3+y_0^4\right)}{1024 \pi ^2 t^5 \epsilon ^6}+\cdots.
\end{equation}
Similarly the leading singularity form operators on adjacent sheet is given by 
\begin{eqnarray}
   &&\lim_{\epsilon\rightarrow 0} \langle R_{txtx} ( r_1, \theta_1^{(k_1)}) R_{txtx}(  r_2,  \theta_2^{(k_2)} )   \rangle_{t>y_0, k_1 -k_2 = -1} =  \\ \nonumber
   & & \qquad\qquad\qquad \qquad \frac{9 (t-y_0) \left(6 t^4+t^3 y_0+t^2 y_0^2+t y_0^3+y_0^4\right)}{256 \pi ^2 t^5 \epsilon ^6}+\cdots
\end{eqnarray}
For $t<y_0$ as well as the correlator on the $\Sigma_1$, we obtain the following behaviour
\begin{align}
         \lim_{\epsilon\rightarrow 0}
          \langle R_{txxz}(\tau + \epsilon )R_{txxz}(\tau - \epsilon)\rangle_{\Sigma_1} &=   \frac{27}{64 \pi ^2 \epsilon ^6}+\cdots
         \end{align}
         where all the other coordinates are understood to be coincident for the two operators.

        \subsection*{Class 5:  $\{ R_{txtz}, R_{yxyz} \} $}
  
          All the curvature components in this class have the same two-point function in the limit the points along the 
        defect are coincident
     as shown  in equations (\ref{Rtxtz}), (\ref{Ryxyz}).
        Let us take $R_{txtz}$ as the representative in this class, the correlator is given by 
             \begin{eqnarray}
       && \langle R_{txtz} (x) R_{txtz}(x') \rangle =  \\ \nonumber
     && \Big[ \frac{1}{8}\partial_{y_1}\partial_{y_2}\partial_{t_1}\partial_{t_2}+\frac{1}{16}\left(\partial_{t_1}\partial_{t_2}-\partial_{y_1}\partial_{y_2}\right)^2
            -\frac{1}{16}\left(\partial_{t_1}\partial_{y_2}-\partial_{y_1}\partial_{t_2}\right)^2\Big]G(x ,x')|_{x_1= x_2, z_1 = z_2}.
        \end{eqnarray}
      
         In the regime $t>y_0$, the leading singularity in the $\epsilon\rightarrow 0$ limit from the correaltors on the 
        same sheet are given by 
        \begin{eqnarray}
       &&   \lim_{\epsilon\rightarrow 0} \langle R_{txtz} ( r_1, \theta_1^{(k)}) R_{txtz}(  r_2, \theta_2^{(k)} )   \rangle_{t>y_0} 
       \\ \nonumber
&&  \qquad\qquad\qquad\qquad
=\frac{(t+y_0) \left(38 t^4-23 t^3 y_0+23 t^2 y_0^2-3 t y_0^3+3 y_0^4\right)}{1024 \pi ^2 t^5 \epsilon ^6}+\cdots
\end{eqnarray}
Similarly the leading singularity form operators  on adjacent sheets is given by 
\begin{eqnarray}
   && \lim_{\epsilon\rightarrow 0} \langle R_{txtz} ( r_1, \theta_1^{(k_1)}) R_{txtz}(  r_2 ,\theta_2^{(k_2)} )   \rangle_{t>y_0, k_1 -k_2 = -1}  \\ \nonumber
   && \qquad\qquad\qquad\qquad=\frac{(t-y_0) \left(38 t^4+23 t^3 y_0+23 t^2 y_0^2+3 t y_0^3+3 y_0^4\right)}{1024 \pi ^2 t^5 \epsilon ^6}+\cdots
\end{eqnarray}
For $t<y_0$ as well as the correlator on the $\Sigma_1$, we obtain the following behaviour
\begin{align}
         \lim_{\epsilon\rightarrow 0}
          \langle R_{txtz}(\tau + \epsilon )R_{txtz}(\tau - \epsilon)\rangle_{\Sigma_1} &=   \frac{19}{256 \pi ^2 \epsilon ^6}+\cdots
         \end{align}
      
      \subsection*{Class 6: $ \{ R_{txyx}, R_{tzyz} \}$ }

      The two curvature components in this class have identical two-point functions in the limit the points along the 
        defect  coincide as shown  in equation (\ref{Rtxyx}), (\ref{Rtzyz}).
        Let us take $R_{txyx}$ as the representative in this class, the correlator is given by 
             \begin{equation}
        \langle R_{txyx} (x) R_{txyx}(x') \rangle = 
      \Big[\frac{1}{32}\left(\partial_{t_1}\partial_{y_2}+\partial_{y_1}\partial_{t_2}\right)^2-\frac{1}{32}\left(\partial_{t_1}\partial_{t_2}-\partial_{y_1}\partial_{y_2}\right)^2\Big]G(x ,x')|_{x_1= x_2, z_1 = z_2}.
        \end{equation}
      
         In the regime $t>y_0$, the leading singularity in the $\epsilon\rightarrow 0$ limit from the correaltors on the 
        same sheet are given by 
        \begin{equation}
        \lim_{\epsilon\rightarrow 0} \langle R_{txyx} ( r_1, \theta_1^{(k)}) R_{txyx}(  r_2,  \theta_2^{(k)} )   \rangle_{t>y_0} 
=-\frac{3 (t+y_0) \left(16 t^4-11 t^3 y_0+11 t^2 y_0^2-t y_0^3+y_0^4\right)}{2048 \pi ^2 t^5 \epsilon ^6}+\cdots.
\end{equation}
Similarly the leading singularity from operators  on adjacent sheet is given by 
\begin{eqnarray}
  && \lim_{\epsilon\rightarrow 0} \langle R_{txyx} ( r_1, \theta_1^{(k_1)}) R_{txyx}(  r_2 , \theta_2^{(k_2)} )   \rangle_{t>y_0, k_1 -k_2 = -1}  \\ \nonumber
  && \qquad\qquad\qquad\qquad
   =-\frac{3 (t-y_0) \left(16 t^4+11 t^3 y_0+11 t^2 y_0^2+t y_0^3+y_0^4\right)}{2048 \pi ^2 t^5 \epsilon ^6}+\cdots
\end{eqnarray}
For $t<y_0$ as well as the correlator on the $\Sigma_1$, we obtain the following behaviour
\begin{align}
         \lim_{\epsilon\rightarrow 0}
          \langle R_{txyx}(\tau + \epsilon )R_{txyx}(\tau - \epsilon)\rangle_{\Sigma_1} &=   -\frac{3}{16 \pi ^2 \epsilon ^6}+\cdots
         \end{align}
       
               \subsection*{Class 7: $\{R_{txyz}\}$} 
               
               We are now left with one curvature component $R_{txyz}$.  The correlator is given by  (\ref{Rtxyz})
             \begin{equation}
        \langle R_{txyz} (x) R_{txyz}(x') \rangle = -
      \Big[\frac{3}{32}\left(\partial_{t_1}\partial_{t_2}-\partial_{y_1}\partial_{y_2}\right)^2+\frac{1}{32}\left(\partial_{y_1}\partial_{t_2}-\partial_{t_1}\partial_{y_2}\right)^2\Big]G(x ,x')|_{x_1= x_2, z_1 = z_2}.
        \end{equation}
        In the regime $t>y_0$, the leading singularity in the $\epsilon\rightarrow 0$ limit from the correaltors on the 
        same sheet are given by 
        \begin{equation}
        \lim_{\epsilon\rightarrow 0} \langle R_{txyx} ( r_1, \theta_1^{(k)}) R_{txyx}(  r_2 , \theta_2^{(k)} )   \rangle_{t>y_0} 
=-\frac{3 (t+y) \left(28 t^4-13 t^3 y_0+13 t^2 y_0^2-3 t y_0^3+3 y_0^4\right)}{2048 \pi ^2 t^5 \epsilon ^6}+\cdots
\end{equation}
Similarly the leading singularity from operators  on adjacent sheet is given by 
\begin{eqnarray}
  && \lim_{\epsilon\rightarrow 0} \langle R_{txyz} ( r_1, \theta_1^{(k_1)}) R_{txyz}(  r_2 , \theta_2^{(k_2)} )   \rangle_{t>y_0, k_1 -k_2 = -1}  \\ \nonumber
 && \qquad\qquad\qquad\qquad 
   =-\frac{3 (t-y_0) \left(28 t^4+13 t^3 y_0+13 t^2 y_0^2+3 t y_0^3+3 y_)^4\right)}{2048 \pi ^2 t^5 \epsilon ^6}+\cdots
\end{eqnarray}
For $t<y_0$ as well as the correlator on the $\Sigma_1$, we obtain the following behaviour
\begin{align}
          \lim_{\epsilon\rightarrow 0}
          \langle R_{txyz}(\tau + \epsilon )R_{txyz}(\tau - \epsilon)\rangle_{\Sigma_1} &=   -\frac{21}{256 \pi ^2 \epsilon ^6}+\cdots
         \end{align}

        \subsection{Growth of entanglement after  curvature quenches}
        
        In this section we will use the leading singularities of the  two-point functions evaluated evaluated 
        in section  \ref{2ptrcurv}
        to obtain the time dependence of the growth of entanglement after the quench enters the 
        region
        $y>0$.  We will  discuss in detail the evaluation for  curvature tensors belonging to class 1 and 
        present the results for the other classes in  the tables   and  figures at the end of this section. 
        We see that,  the time dependent polynomials which determine the growth of R\'{e}nyi/entanglement
     have similar features when one organises the curvature components in terms of the representations 
     of $SO(2)_T\times SO(2)_L$. 
        
               \subsubsection*{Scalars under $SO(2)_T \times SO(2)_L$,  Class 1: $\{R_{tyty}$,  $R_{xzxz}$,  $R_{tyxz}$\}} 
         
\begin{figure}[htb]
\centering
\begin{subfigure}{.5\textwidth}
  \centering
  \includegraphics[width=1\linewidth]{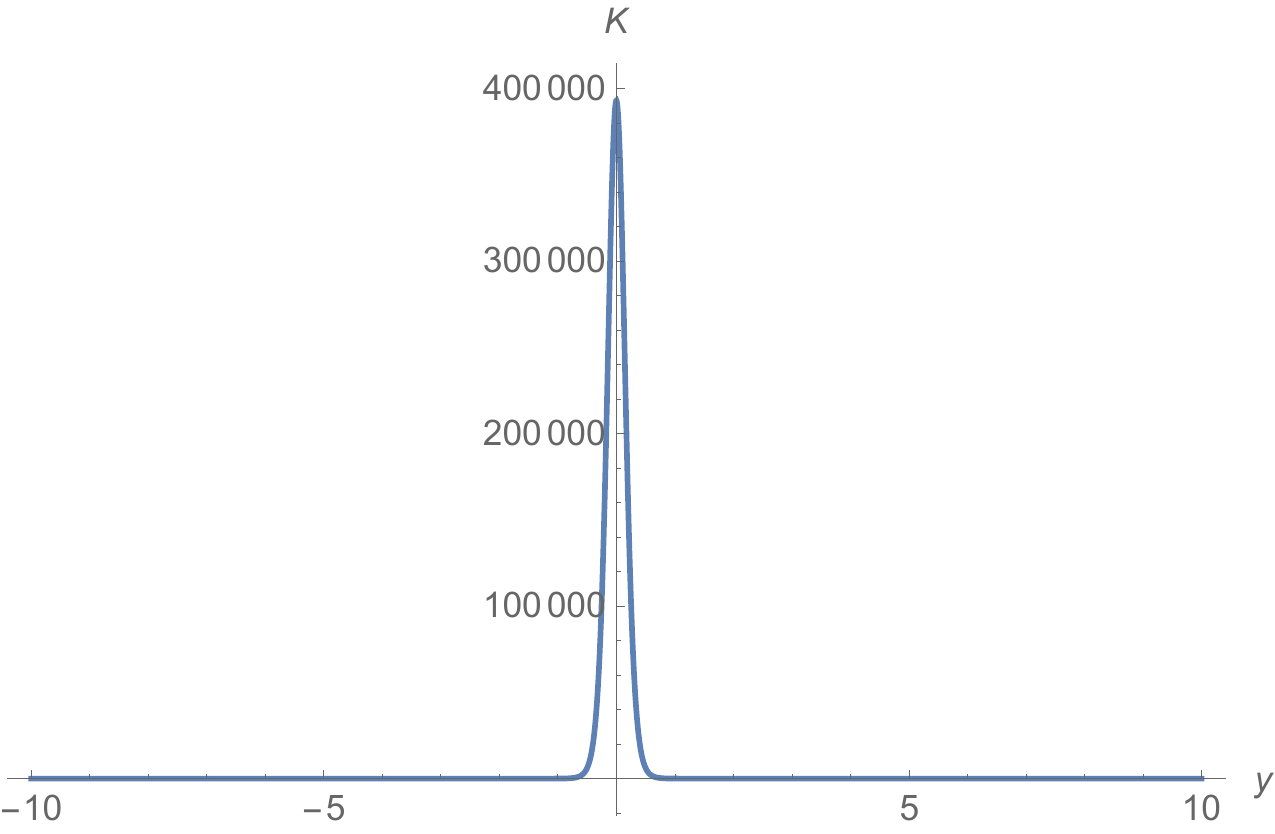}
  \caption{$t=0$ and $\epsilon=0.5$}
\end{subfigure}%
\begin{subfigure}{.5\textwidth}
  \centering
  \includegraphics[width=1\linewidth]{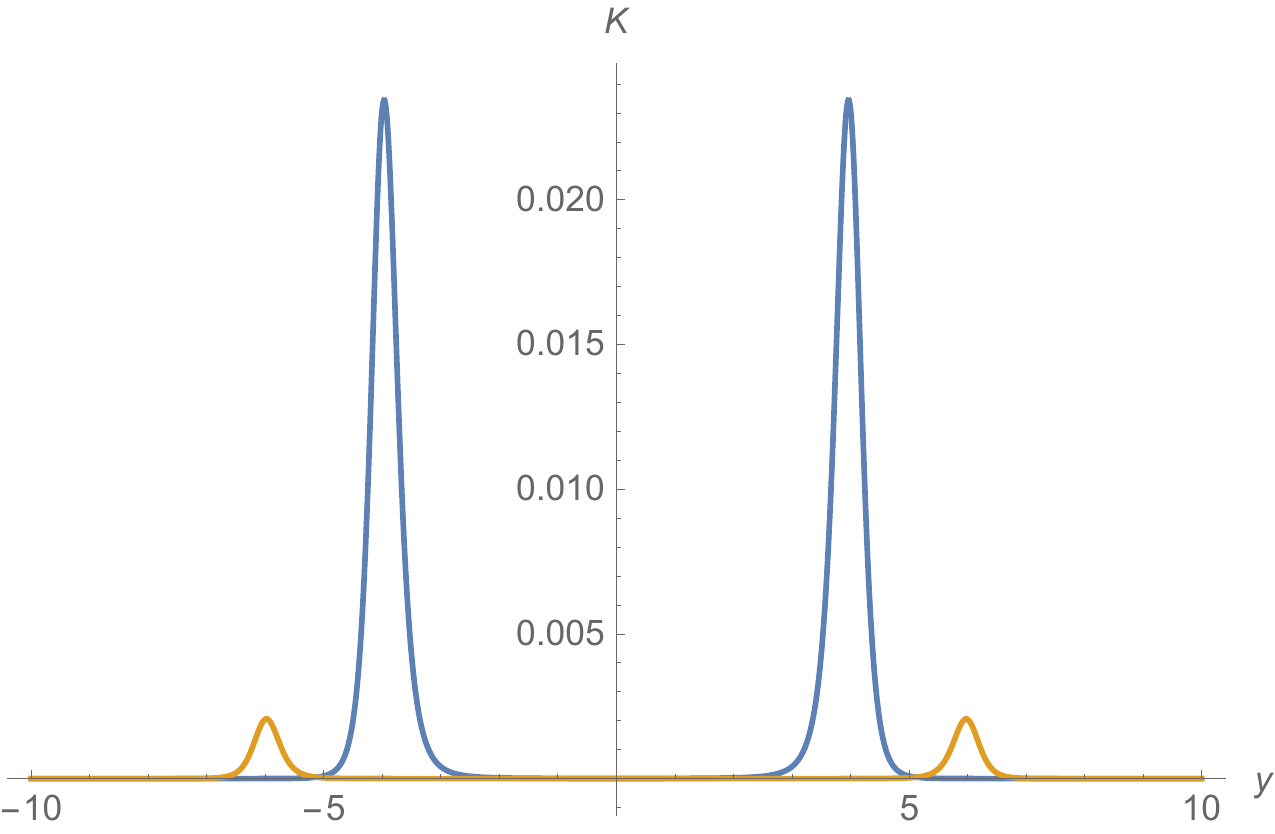}
  \caption{Blue curve: $t=4$, orange curve : $t=6$ .}
\end{subfigure}
\caption{Kretschmann curvature density for  the quench created by Riemann curvature $R_{tyty}$ placed at the origin
at times $t=0, t=4$ and $t=6$. We choose  $\epsilon=0.5$. The Kretschmann curvature density travels at the speed of 
light as spherical wave. }  \label{kscalar}
\end{figure}

\begin{figure}[htb]
\centering
  \includegraphics[width=1\linewidth]{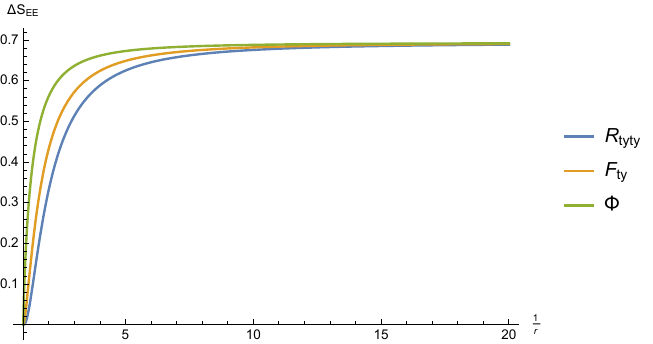}
  \caption{ $\Delta S_{\rm{EE}}$ as a function of $1/r = t/y_0$ in the zero width limit for scalars under 
  $SO(2)_T\times SO(2)_L$. Note that the entanglement grows slower as  the spin of the field creating 
  the quench increases. the plot for $R_{tyty}$  represents the growth for all components in class 1.} \label{fig2}
\end{figure}

         The time dependence of quenches  for 
          components in this class behave identically. The components in this class are 
          either scalars are pseudo-scalars under both  $SO(2)_T$ and $SO(2)_L$. 
          To represent this class we
         consider the state 
         \begin{equation} \label{rtytystate}
         \rho( t, - y_0)  = R_{tyty} ( \tau_e, - y_0) |0\rangle \langle 0| R_{tyty} ( \tau_l , - y_0) .
         \end{equation}
         Since there no local gauge invariant stress tensor for the theory of gravitons, we use the Kretschmann scalar
         as probe of the quench.  
         The expectation value of the Kretschmann scalar  in the above state will show that it is a state 
         which corresponds to a spherical pulse with non-zero value of the Kretschmann
          scalar density travelling at the speed of light. 
          The expectation value of the Kretschmann scalar in the state (\ref{rtytystate}) is given by 
          \begin{eqnarray}
          {\rm Tr} [ K (0, 0) \rho(t, -y_0) ]&=&
          {\rm Tr} [K (0, y_0)  \rho (t, 0)] , \\ \nonumber
          &=& \frac{\langle R_{tyty} ( - \epsilon - i t, 0 ) K (0, y_0)
           R_{tyty} (  \epsilon - i t, 0 ) \rangle}{ \langle R_{tyty} ( - \epsilon - i t, 0 ) R_{tyty} (  \epsilon - i t, 0 ) \rangle}.
          \end{eqnarray}
          where
          \begin{equation}
          K = R_{\mu\nu\rho\sigma} R^{\mu\nu\rho\sigma}.
          \end{equation}
          The appendix \ref{appendixc}, contains  the details of evaluating this correlator. This 
          involves the use of the graviton propagator given in (\ref{defcor1}) systematically. 
          The result is given by (\ref{kexpect})
          \begin{equation}
          {\rm Tr} [K(0, y_0)  \rho (t, 0) ]
          =\frac{6144 \epsilon ^6}{\left(t^4+2 t^2 \left(\epsilon ^2-y_0^2\right)+\left(y_0^2+\epsilon ^2\right)^2\right)^3}.
          \end{equation}
          Figure \ref{kscalar} plots the profile of the Kretschmann scalar as a function of the $y_0$ at $t= 0, 4, 6$ for 
          width $\epsilon = 0.5$. The figure clearly shows that the curvature density due to quench travels as a spherical
          wave at the speed of light. 

Now let us proceed to obtain the change in the R\'{e}nyi/entanglement in the regime $t<y_0$ before the  pulse hits the entangling surface $y_0=0$.  In this regime and in the $\epsilon \rightarrow 0$ limit, it is only Wick contractions 
on the same sheet that contribute in the $2n$ point function (\ref{renform}).  
From (\ref{tlessy_grav}), we see that this contribution goes as $\epsilon^{-6(n)}$ and the leading correction to this
Wick contraction is  proportional to $\epsilon^{-6(n-1)}$.
Just as in the case for the scalar, the next leading Wick contraction arises from $n-2$ operators located on the 
same sheet and $2$ pairs of operators contracted across the sheets, this contribution is proportional 
to $\epsilon^{-6(n-2)}$. So, we conclude that the leading correction to the entanglement in the regime 
$t<y_0$ arises from the correction in the two-point function on the same sheet. 
Using (\ref{tlessy_grav}) we get 
     \begin{align}\label{ope2}
            \Delta S^{(n)}_{t<y_0}&=\frac{1}{1-n}\log\left(1+\epsilon^6\frac{(n-1) (n+1) \left(191 n^4+23 n^2+2\right)}{945 n^6 \left(t^2-y_0^2\right)^3}+\cdots\right)^n,\nonumber\\
            &=\frac{n}{1-n}\frac{(n-1) (n+1) \left(191 n^4+23 n^2+2\right)}{945 n^6 \left(t^2-y_0^2\right)^3}\epsilon^6+\mathcal{O}(\epsilon^8).
        \end{align}
        The denominator in the ratio (\ref{renform}) is evaluated using the two-point function of $R_{tyty}$ on $\Sigma_1$, which is given  in (\ref{rtytyf}). Using the same arguments as in the case of the scalar and the $U(1)$ field 
        the leading term proportional to $\epsilon^6$ should arise from expectation value of the composite 
        $:R_{tyty}^2:$ on the replica surface. Let us evaluate this expectation value directly for a  simple cross check. 
               \begin{align}\label{compositemax2}
            \langle :R_{tyty}R_{tyty}:\rangle_{\Sigma_n}&=-\lim_{\substack{r_1\rightarrow r_2\\ \theta_1\rightarrow\theta_2}}\tilde{\nabla}^4 \left(G(r_1,r_2,\theta_1,\theta_2, {\bf x}, {\bf x} )-\frac{1}{4\pi^2\left(r_1^2+r_2^2-2 r_1r_2\cos(\theta_1-\theta_2)\right)}\right),\nonumber\\
          &=\lim_{\theta\rightarrow 0}  \Big[\frac{15 \left(n^2-1\right) \csc ^4\left(\frac{\theta }{2 n}\right)+2 \left(4 n^4-5 n^2+1\right) \csc ^2\left(\frac{\theta }{2 n}\right)+15 \csc ^6\left(\frac{\theta }{2 n}\right)}{60 \pi ^2 n^6 r_1^6}\nonumber\\
          &\qquad\qquad\qquad+\frac{2}{\pi ^2 r_1^6 (\cos (\theta )-1)^3}\Big],\nonumber\\
            &=\frac{(n-1) (n+1) \left(191 n^4+23 n^2+2\right)}{3780 \pi ^2 n^6 (y^2-t^2)^3}.
        \end{align}
        Here $\tilde{\nabla}^2=(\partial_{r_1}^2+\frac{1}{r_1}\partial_{r_1}+\frac{1}{r_1^2}\partial_{\theta_1}^2)$ is the Laplacian in $r-\theta$ plane.  Comparing (\ref{ope2}) and (\ref{compositemax2}), we see that indeed the 
        the leading contribution in the $\epsilon\rightarrow 0$ limit to the entanglement is due to the expectation value of the 
        composite of the curvature field. 
        One curious observation is that the $n$ dependence in the  expectation value in (\ref{compositemax2}) coincides 
        with the expectation value of the stress tensor for the conformal $2$ form in $d=6$ on $\Sigma_n$, 
        which was evaluated in \cite{David:2020mls} \footnote{See equation 3.58 of \cite{David:2020mls}}. 
        
The growth in entanglement in the regime $t>y_0$ in the $\epsilon\rightarrow 0$ limit
 is evaluated using the leading Wick contractions 
given in (\ref{rtytya1}) and (\ref{rtytya2}).  
This results in 
        \begin{align} \label{renrtyty}
          &  \lim_{\epsilon\rightarrow 0}\Delta S^{(n)}_A[R_{tyty}]=\frac{1}{1-n}\log\frac{\left(\frac{(t+y_0)^3 \left(8 t^2-9 t y_0+3 y^2\right)}{32 \pi ^2 t^5 \epsilon ^6}\right)^n+\left(\frac{(t-y_0)^3 \left(8 t^2+9 t y_0+3 y_0^2\right)}{32 \pi ^2 t^5 \epsilon ^6}\right)^n}{\left(\frac{1}{2 \pi ^2 \epsilon ^6}\right)^n},\nonumber\\
          &  =\frac{1}{1-n}\log \left[ \left(\frac{(t+y)^3 \left(8 t^2-9 t y_0+3 y_0^2\right)}{16  t^5}\right)^n+\left(\frac{(t-y_0)^3 \left(8 t^2+9 t y_0+3 y_0^2\right)}{16  t^5 }\right)^n \right].
        \end{align}
         As discussed earlier, the two Wick contractions are related to each other by $y_0\rightarrow - y_0$. 
        Observe that the time dependence of the growth of R\'{e}nyi entropy is through a polynomial 
        of order $5 = 2s+1, s= 2$ through  the ratio $y_0/t$.  
        We will see that all time dependences for the curvature quenches are determined by  order $5$ polynomials. 
        The asymptotic behaviour of the R\'{e}nyi entropy in the large time limit is given by 
             \begin{align}
        \lim_{t\rightarrow \infty}    \Delta S^{(n)}_A[R_{tyty}]
        &=\log (2)-\frac{n}{2}\left(\frac{15}{8}\right)^2\frac{ y^2}{ t^2}+\mathcal{O}(\frac{y^4}{t^4})+\cdots
        \end{align}
        Finally the entanglement entropy can also be written as a function of $y_0/t$ and is given by 
         \begin{eqnarray}
            \lim_{n\rightarrow 1}\Delta S^{(n)}_A[R_{tyty}] &=& \Delta S_{\rm{EE}}[R_{tyty}] ,\nonumber\\
            &=&\frac{1}{16}\Big[64 \log 2-(r+1)^3 (3 (r-3) r+8) \log \left((r+1)^3 (3 (r-3) r+8)\right)\nonumber\\
            & &+(r-1)^3 (3 r (r+3)+8) \log \left(-(r-1)^3 (3 r (r+3)+8)\right)\Big],  \nonumber \\ && \qquad\qquad r =\frac{y_0}{t}
        \end{eqnarray}
       Before we conclude the analysis of this class of curvature components,  let us generalise the observations of the 
        seen for the $U(1)$ quench.  For this class of curvature components
       the   polynomial which determines the growth of R\'{e}nyi entropy can be completely determined.       
       The components in this class of curvatures are scalars under  both $SO(2)_T$ and $SO(2)_L$, 
       they are in fact pseudo-scalars. 
       Therefore from the earlier observations, we expect the polynomial 
       to be 
       \begin{equation}\label{polrtyty}
       P_{R_{tyty}}(r)  = \frac{1}{2} ( 1+ r)^3 (1 + a_1 r + a_2 r^2) ,
       \end{equation}
       where order of the factor $(1+r)$ is  $3= s+1, s=2$. 
       The coefficients $a_1, a_2$ should be such that
       \begin{equation} \label{polcond}
       P_{R_{tyty}}(r) +  P_{R_{tyty}}(-r) = 1.
       \end{equation}
       This is because the argument of the logarithm in (\ref{renrtyty})
        is the ratio of $2n$ point function to the $n$ powers of the 
       two-point function and at $n=1$, this ratio is unity.  This implies that  the even powers $r^2, r^4$ of the polynomial 
       in (\ref{polrtyty}) must vanish. 
       The condition in (\ref{polcond}) is enough to determine the coefficients $a_1, a_2$ uniquely, they are given by 
       \begin{equation}
       a_1 = -\frac{9}{8}, \qquad a_2 = \frac{3}{8}.
       \end{equation}
       The polynomial is therefore
       \begin{equation}
       P_{R_{tyty} }(r) = \frac{1}{2} ( 1 + \frac{15}{8} r  - \frac{5}{4} r^3 + \frac{3}{8} r^8) .
       \end{equation}
       Comparing this with the polynomial in  (\ref{renrtyty}) we see that they agree. 
       The R\'{e}nyi/entanglement entropy growth in the $\epsilon\rightarrow 0$ limit is uniquely  determined 
       for the components in class 1 by demanding that the polynomial of the order of $5$ is of the form (\ref{polrtyty}). 
       The power of the factor $(1\pm r)$ is $3$, which also coincides with the scaling dimension of curvature.
       In tables (\ref{table1}), (\ref{table2}) we list the behaviour of the 
       R\'{e}nyi/entanglement entropy of quenches created by components 
       belonging to class 1 as well as scalars under $SO(2)_T\times SO(2)_L$ of the vector quench and the scalar. 
       Figure (\ref{fig2}) plots the behaviour of entanglement entropy.

\subsubsection*{Vectors under either $SO(2)_T \times SO(2)_L$, Class 2, 3}

\begin{figure}[htb]
\centering
  \includegraphics[width=1\linewidth]{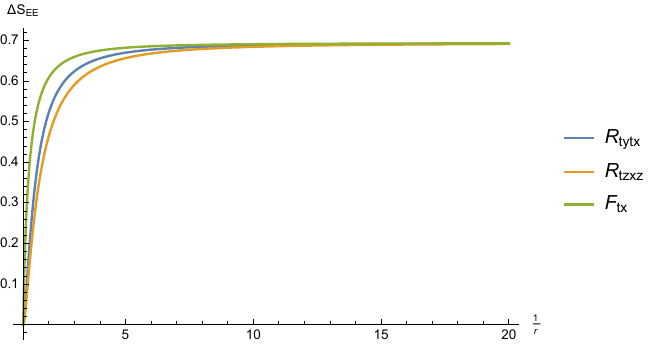}
  \caption{ $\Delta S_{\rm{EE}}$ as a function of $1/r = t/y_0$  in the zero width limit for vectors under 
  $SO(2)_T\times SO(2)_L$. Class 2, 3  are represented by $R_{tytx}, R_{tzxz}$ respectively. 
  The growth due to the quench due to the vectors from  the $U(1)$ field is higher than those due to the spin-2 fields. } \label{fig3}
\end{figure}

Consider the components in class 2, 3  for example
$R_{tytx}, R_{tytz}$ transform as a  (pseudo) vector of  $SO(2)_T$ and a vector $SO(2)_L$.
 Similarly components  $R_{yxxz}, R_{txxz}$ transform as (pseudo)  vector of $SO(2)_L$ and 
a vector of  $SO(2)_T$. There are $8$ components in class 2 and 3 combined. 
The components of class $2$ and class $3$ each  have the identical behaviour.
The expressions for the R\'{e}nyi entropy for  quenches created by curvature components belonging to 
class 2 and class 3 are given in tables. 
The polynomials which determine the growth of R\'{e}nyi entropy in both these classes are of the form
\begin{equation} \label{polclass23}
P_{{\rm class \; 2, 3}} (r) = \frac{1}{2} ( 1+ r)^2 ( 1+ a_1 r + a_2 r + a_3 r^2).
\end{equation}
Now the condition 
\begin{equation}
P(r) + P( -r) = 1.
\end{equation}
determines the polynomial uniquely up to one coefficient, This is because there are $2$ equations we obtain when 
we demand that the coefficients of even terms in $r$ must vanish and there are $3$ unknown coefficients in 
(\ref{polclass23}). 
Therefore say given $a_\infty$, the entire polynomial can be determined uniquely.   Table   (\ref{table3}
lists the R\'{e}nyi entropies and its asymptotic behaviour at large times, table \ref{table4}) lists the corresponding 
entanglement growth for quenches created by components in this class. 
These polynomials differ from that of the class 1 given in (\ref{polrtyty}), in that the  power of the 
factor  $(1\pm r)$ reduces  by one.  The figure \ref{fig3} plots the growth of the entanglement entropy  against $t/y_0$ 
for the vectors under $SO(2)_T\times SO(2)_L$ both from the $U(1)$ field as well as from spin-2 field. 

\subsubsection*{ Symmetric tensor of $SO(2)_T\times SO(2)_L$,  Class 4, 5, 6, 7}

\begin{figure}[htb]
\centering
  \includegraphics[width=1\linewidth]{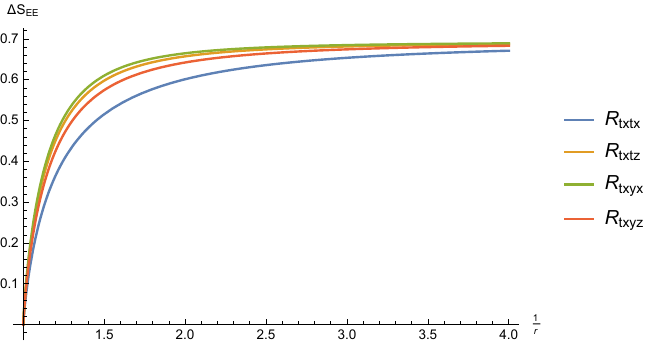}
  \caption{ $\Delta S_{\rm{EE} }$ as a function of $\frac{1}{r} = \frac{t}{y_0}  $  in the zero width limit for the symmetric  
  tensor under 
  $SO(2)_T\times SO(2)_L $. Class 4, 5, 6, 7 contain all the 9 curvature components of these  classes.
  We have plotted a representative from each of the classes} 
  \label{fig4}
\end{figure}

All the components in these classes
 transform as a 2nd rank  symmetric tensor in both $SO(2)_T$ as well as  $SO(2)_L$, therefore 
there are $9$ components in all. 
The expressions for the R\'{e}nyi entropy for  quenches created by curvature components belonging to 
class 4, 5, 6 are given in tables. 
For this class, the polynomial which determines the growth in R\'{e}nyi entropy is of the form
\begin{equation} \label{polclass456}
P_{{\rm class \; 4, 5, 6}} (r) = \frac{1}{2} ( 1+ r) ( 1+ a_1 r + a_2 r + a_3 r^2 +a_4 r^4).
\end{equation}
The equation $P(r) + P(-r) =1$ results in $2$ equations for the vanishing of the even powers of $r$.
Therefore, these polynomials are uniquely determined once $2$ coefficients are given. 
The power of the factor for $(1\pm r)$ further reduces by $1$ compared to the case of vectors in (\ref{polclass23}). 
The R\'{e}nyi/entanglement entropies for quenches in these classes are listed in tables \ref{table5}, \ref{table6} respectively. Figure \ref{fig4} plots the growth of entanglement.

        \begin{table}[ht]
\centering { \footnotesize{
\begin{tabular}{|c|l|c|}
\hline
Field &\qquad\qquad\qquad $\Delta S^{(n)}_A$ & $\lim_{t\rightarrow \infty}\Delta S^{(n)}_A$ \\
\hline 
& &  \\
$\phi$ & \quad$\frac{1}{1-n}\log\Big[\left(\frac{t+y_0}{2  t}\right)^n+\left(\frac{t-y_0}{2 t}\right)^n\Big] $ & $\log 2-\frac{n}{2}\frac{y_0^2}{t^2}$
   \\   
&  & \\
$F_{ty}, F_{xz}$ & \quad $\frac{1}{1-n}\log\Big[\left(\frac{(2 t-y_0) (t+y_0)^2}{4 t^3 }\right)^n+\left(\frac{(t-y_0)^2 (2 t+y_0)}{4 t^3 }\right)^n\Big]$ & $ \log 2-\frac{n}{2}\left(\frac{3y_0}{2t}\right)^2$  \\
& &  \\
$\rm{Class} \hspace{0.2mm}1 $& \quad$\frac{1}{1-n}\log\left[  \left(\frac{(t+y_0)^3 \left(8 t^2-9 t y_0+3 y^2\right)}{16  t^5}\right)^n+\left(\frac{(t-y_0)^3 \left(8 t^2+9 t y_0+3 y_0^2\right) }{16  t^5 }\right)^n  \right]$ &
$ \log 2-\frac{n}{2}\left(\frac{15y_0}{8t}\right)^2$  \\
& &  \\
\hline
\end{tabular}
\caption{ \Re entropy growth   of scalars under $SO(2)_T\times SO(2)_L$. The last column contains the asymptotic behaviour of \Re entropy after the quench enters the entangling region.  We have combined the scalars under 
$SO(2)_T\times SO(2)_L$ created by quenches for fields $\leq 2$. Note that the order of the polynomial in 
$r = y_0/t$, 
which determines the R\'{e}nyi entropy is $2s+1$ and the power of the factor $(1+r)$ is $s+1$ or the dimension of the operator creating the quench. Class 1 contains the curvature components $R_{tyty}, R_{zxzx}, R_{tyzx}$.  
This polynomial  for the scalars $SO(2)_T\times SO(2)_L$ is completely determined by the spin of the field. }
\label{table1}
}}
\end{table} 
\renewcommand{\arraystretch}{3.5}
\begin{table}[ht]
\centering { \footnotesize{
\begin{tabular}{|c|l|}
\hline
Field &\qquad\qquad\qquad $\Delta S_{\rm{EE}}$ \\
\hline 
$\Phi$ & \quad$\frac{1}{2} \left(\log \left(\frac{4}{1-r^2}\right)-2 r \tanh ^{-1}(r)\right) $  
 \\
 $F_{ty},F_{xz}$ & \quad$\frac{1}{4} \left((r-2) (r+1)^2 \log \left(-r^3+3 r+2\right)-\left(r^3-3 r+2\right) \log \left(r^3-3 r+2\right)+\log (256)\right) $  
 \\
 $\rm{Class}\hspace{1mm}1$ & $\makecell{\frac{1}{16}\Big[64 \log 2-(r+1)^3 (3 (r-3) r+8) \log \left((r+1)^3 (3 (r-3) r+8)\right)\\
            +(r-1)^3 (3 r (r+3)+8) \log \left(-(r-1)^3 (3 r (r+3)+8)\right)\Big]},   $ 
            \\
\hline
\end{tabular}}
\caption{Growth of entanglement entropy for scalars under $SO(2)_T\times SO(2)_L.$, here $r= y_0/t$. }
\label{table2}
}
\end{table}

         \begin{table}[ht]
\centering { \footnotesize{
\begin{tabular}{|c|l|c|}
\hline
Field &\qquad\qquad\qquad $\Delta S^{(n)}_A$ & $\lim_{t\rightarrow \infty}\Delta S^{(n)}_A$ \\
\hline 
& &  \\
$F_{tx}, F_{tz}, F_{yx}, F_{yz}$ & \quad $\frac{1}{1-n}\log\Big[\left(\frac{(t+y_0) \left(4 t^2-t y_0+y_0^2\right)}{8  t^3 }\right)^n+\left(\frac{(t-y_0) \left(4 t^2+t y_0+y_0^2\right)}{8  t^3 }\right)^n\Big]  $ &  $\log2-\frac{n}{2}\left(\frac{3y_0}{4t}\right)^2$  \\
& &  \\
$\rm{Class} \hspace{0.2mm}2 $& \quad$\frac{1}{1-n}\log \left[\frac{\left(\frac{(t-y_0)^2 \left(14 t^3+13 t^2 y_0+12 t y_0^2+6 y_0^3\right)}{128 \pi ^2 t^5 \epsilon ^6}\right)^n+\left(\frac{(t+y_0)^2 \left(14 t^3-13 t^2 y_0+12 t y_0^2-6 y_0^3\right)}{128 \pi ^2 t^5 \epsilon ^6}\right)^n}{\left(\frac{7}{32 \pi ^2 \epsilon ^6}\right)^n}\right] $ &
$ \log 2-\frac{n}{2}\left(\frac{15y_0}{14t}\right)^2$  \\
& &  \\
$\rm{Class} \hspace{0.2mm}3 $& \quad$\frac{1}{1-n}\log \left[\frac{\left(\frac{(t-y_0)^2 \left(22 t^3+14 t^2 y_0+6 t y_0^2+3 y_0^3\right)}{128 \pi ^2 t^5 \epsilon ^6}\right)^n+\left(\frac{(t+y_0)^2 \left(22 t^3-14 t^2 y_0+6 t y_0^2-3 y_0^3\right)}{128 \pi ^2 t^5 \epsilon ^6}\right)^n}{\left(\frac{11}{32 \pi ^2 \epsilon ^6}\right)^n}\right]$ &
$ \log 2-\frac{n}{2}\left(\frac{15y_0}{11t}\right)^2$  \\
& &  \\
\hline
\end{tabular}
\caption{ \Re entropy  growth  of vectors under $SO(2)_T\times SO(2)_L$.  The order of the  polynomial 
in $r = y_0/t$ is $2s+1$. The power of the pre-factor  $(1+r)$ is one less that the scaling dimension of the field. 
Class $2, 3$ together contain $8$ curvature components which can be organised
as vectors of  $SO(2)_T\times SO(2)_L$.  }
\label{table3}
}}
\end{table}
\renewcommand{\arraystretch}{3.5}
\begin{table}[ht]
\centering { \footnotesize{
\begin{tabular}{|c|l|}
\hline
Field &\qquad\qquad\qquad $\Delta S_{\rm{EE}}$ \\
\hline 
$F_{tx}, F_{tz}, F_{yx}, F_{yz}$ & \quad$\frac{1}{8} \left(-\left(r^3+3 r+4\right) \log \left(r^3+3 r+4\right)+\left(r^3+3 r-4\right) \log \left(4-r \left(r^2+3\right)\right)+8 \log (8)\right) $  
 \\
 $\rm{Class}\hspace{1mm}2$ & $\makecell{\frac{1}{28} \Big[\left(-6 r^5+5 r^3+15 r-14\right) \log \left(6 r^5-5 r^3-15 r+14\right)+\\
 \left(6 r^5-5 r^3-15 r-14\right) \log \left(-6 r^5+5 r^3+15 r+14\right)+28 \log (28)\Big]}   $ 
            \\
            $\rm{Class}\hspace{1mm}3$ & $\makecell{\frac{1}{44} \Big[-\left(3 r^5+5 r^3-30 r+22\right) \log \left(3 r^5+5 r^3-30 r+22\right)\\+\left(3 r^5+5 r^3-30 r-22\right) \log \left(-3 r^5-5 r^3+30 r+22\right)+44 \log (44)\Big]}   $ 
            \\
\hline
\end{tabular}}
\caption{Entanglement growth of quenches  due to vectors under $SO(2)_T\times SO(2)_L.$ }
\label{table4}
}
\end{table}

       \begin{table}[ht]
\centering { \footnotesize{
\begin{tabular}{|c|l|c|}
\hline
Field &\qquad\qquad\qquad $\Delta S^{(n)}_A$ & $\lim_{t\rightarrow \infty}\Delta S^{(n)}_A$ \\
\hline 
& &  \\
$\rm{Class}\hspace{0.2mm}4$ & \quad$\frac{1}{1-n}\log\Big[\frac{\left(\frac{9 (t-y_0) \left(6 t^4+t^3 y_0+t^2 y_0^2+t y_0^3+y_0^4\right)}{256 \pi ^2 t^5 \epsilon ^6}\right)^n+\left(\frac{9 (t+y_0) \left(6 t^4-t^3 y_0+t^2 y_0^2-t y_0^3+y_0^4\right)}{256 \pi ^2 t^5 \epsilon ^6}\right)^n}{\left(\frac{27}{64 \pi ^2 \epsilon ^6}\right)^n}\Big] $ & $\log 2-\frac{n}{2}\left(\frac{5y_0}6{t}\right)^2$
   \\   
&  & \\
$\rm{Class}\hspace{0.2mm}5$ & \quad $\frac{1}{1-n}\log\Big[ \frac{\left(\frac{(t-y_0) \left(38 t^4+23 t^3 y_0+23 t^2 y_0^2+3 t y_0^3+3 y_0^4\right)}{1024 \pi ^2 t^5 \epsilon ^6}\right)^n+\left(\frac{(t+y_0) \left(38 t^4-23 t^3 y_0+23 t^2 y_0^2-3 t y_0^3+3 y_0^4\right)}{1024 \pi ^2 t^5 \epsilon ^6}\right)^n}{\left(\frac{19}{256 \pi ^2 \epsilon ^6}\right)^n}\Big]$ & $ \log 2-\frac{n}{2}\left(\frac{15y_0}{38t}\right)^2$  \\
& &  \\
$\rm{Class}\hspace{0.2mm}6 $& \quad$\frac{1}{1-n}\log\Big[\frac{\left(\frac{3 (t-y) \left(16 t^4+11 t^3 y_0+11 t^2 y_0^2+t y_0^3+y_0^4\right)}{2048 \pi ^2 t^5 \epsilon ^6}\right)^n+\left(\frac{3 (t+y_0) \left(16 t^4-11 t^3 y_0+11 t^2 y_0^2-t y_0^3+y_0^4\right)}{2048 \pi ^2 t^5 \epsilon ^6}\right)^n}{\left(\frac{3}{16 \pi ^2 \epsilon ^6}\right)^n}\Big] $ &
$ \log 2-\frac{n}{2}\left(\frac{5y_0}{16t}\right)^2$  \\
& &  \\
$\rm{Class}\hspace{0.2mm}7 $& \quad$\frac{1}{1-n}\log\Big[\frac{\left(\frac{3 (t-y_0) \left(28 t^4+13 t^3 y_0+13 t^2 y_0^2+3 t y_0^3+3 y_0^4\right)}{512 \pi ^2 t^5 \epsilon ^6}\right)^n+\left(\frac{3 (t+y_0) \left(28 t^4-13 t^3 y_0+13 t^2 y_0^2-3 t y_0^3+3 y_0^4\right)}{512 \pi ^2 t^5 \epsilon ^6}\right)^n}{\left(\frac{21}{64 \pi ^2 \epsilon ^6}\right)^n}\Big] $ &
$ \log 2-\frac{n}{2}\left(\frac{15y_0}{28t}\right)^2$  \\
& &  \\
\hline
\end{tabular}
\caption{ \Re entropy  growth for  of symmetric  tensors under  both $SO(2)_T\times SO(2)_L$, Class 4, 5, 6, 7 together 
represent the 9 components of the symmetric tensors.  The polynomial in $r$  determining the 
R\'{e}nyi entropy growth is a 5th order polynomial and the power of the  $(1+r)$ is unity. }
\label{table5}
}}
\end{table}

   \begin{table}[ht]
\centering { \footnotesize{
\begin{tabular}{|c|l|}
\hline
Field &\qquad\qquad\qquad $\Delta S_{\rm{EE}}$ \\
\hline 
$\rm{Class}\hspace{1mm}4$ & \quad$\frac{1}{12} \left(-\left(r^5+5 r+6\right) \log \left(r^5+5 r+6\right)+\left(r^5+5 r-6\right) \log \left(6-r \left(r^4+5\right)\right)+12 \log (12)\right)$  
 \\
 $\rm{Class}\hspace{1mm}5$ & $\makecell{\frac{1}{76} \Big[-\left(3 r^5+20 r^3+15 r+38\right) \log \left(3 r^5+20 r^3+15 r+38\right)\\+\left(3 r^5+20 r^3+15 r-38\right) \log \left(-3 r^5-20 r^3-15 r+38\right)+76 \log (76)\Big]},   $ 
            \\
            $\rm{Class}\hspace{1mm}6$ & $\makecell{\frac{1}{32} \Big[-\left(r^5+10 r^3+5 r+16\right) \log \left(r^5+10 r^3+5 r+16\right)\\+\left(r^5+10 r^3+5 r-16\right) \log \left(16-r \left(r^4+10 r^2+5\right)\right)+32 \log (32)\Big]},   $ 
            \\
              $\rm{Class}\hspace{1mm}7$ & $\makecell{\frac{1}{56} \Big[-\left(3 r^5+10 r^3+15 r+28\right) \log \left(3 r^5+10 r^3+15 r+28\right)\\+\left(3 r^5+10 r^3+15 r-28\right) \log \left(-3 r^5-10 r^3-15 r+28\right)+56 \log (56)\Big]},   $ 
            \\
\hline
\end{tabular}
\caption{Entanglement entropy of components transforming as tensors under $SO(2)_T\times SO(2)_L$.}
\label{table6}
}}
\end{table}   

\subsubsection*{Summary:}

From all our  explicit computations we find that the R\'{e}nyi/entanglement entropies  for 
quenches created by components of a  spin $s$ field are determined 
by polynomials $P(r)$  of the degree $2s+1$ where $r = y_0/t$. 
For components which are scalars under $SO(2)_T\times SO(2)_L$,  the polynomial is  of the form
\begin{equation}
P(r) = \frac{1}{2} ( 1+ r)^{s+1} ( 1+ a_1 r + a_2 r^2 + \cdots a_s r^s) 
\end{equation}
These polynomials have the property that $P(r) + P(-r) = 1$ which implies that the even powers of $r$ apart 
from $r^0$ should vanish.
This represents $s$ conditions from which the polynomial can be completely determined. 
For components which below to vectors of $SO(2)_T\times SO(2)_L$, the form of the polynomial is  again 
of degree $2s+1$, but the power of the prefactor $(1+r)$ shifts from $s \rightarrow s-1$. 
A similar shift occurs for tensors under $SO(2)_T\times SO(2)_L$.  Given the fact that the even power $r^2$ does 
not occur in the polynomial, the long time behaviour of the R\'{e}nyi entropies is given by 
\begin{equation}
\lim_{t\rightarrow 0} \Delta S^{(n)}_A = \log (2) - \frac{n}{2} a_{\infty}^2 \frac{y_0^2}{t^2} + \cdots
\end{equation}
where $a_\infty$ is the coefficient of the linear term in the polynomial $P(r)$. 
        \section{Conclusions} \label{conclusions}
        
        We have studied local quenches created by fields with integer spins $s\leq 2$. 
        When the width of the quenches tends to zero, 
        the growth of  R\'{e}nyi/entanglement  entropies is determined by a $2s+1$ order polynomial 
        in $r = \frac{y_0}{t}$ where $y_0$ is the location the quench is initiated and $t$ is the time elapsed.
        The behaviour of the quenches can be organised in terms of the $SO(2)_T\times SO(2)_L$ representation 
        of the field creating the quench. 
        The polynomial determining the R\'{e}nyi entropy growth for $SO(2)_T\times SO(2)_L$ scalars is completely 
        determined by the spin of the field creating the quench. 
        
        An obvious generalisation of this work is to study quenches created by fields with spins $1/2, 3/2$. 
        Quenches created by   Dirac fermions 
        were studied in \cite{Nozaki:2015mca}. On examining the singularities of the propagator  for $t>y_0$ , 
        \footnote{See equations (48), (49) of  \cite{Nozaki:2015mca}} we see that they are indeed  proportional to polynomials of degree $2 = 2s+1, s= 1/2$ in the ratio $r$. 
        It will interesting to revisit this work and also study the case of spin-3/2. 
        One needs to choose a suitable gauge for  studying quenches due to  the gravitino. 
        
        One of the goals of our work is to explore the information theoretic properties of the graviton. 
       This work has shown that the linearised graviton behaves just as a local field with gauge symmetry. 
        It will be interesting to turn on the leading cubic interaction  in Einstein's theory 
        and see how ory departs from local quantum field theories which have
        the `split property' as argued in \cite{Laddha:2020kvp,Raju:2021lwh}.  The methods developed in this paper and 
         \cite{Benedetti:2019uej,David:2020mls,David:2022jfd} 
        as well as the new approach to 
        evaluating entanglement entropy  in \cite{Anegawa:2021osi,Anegawa:2022pce} 
        hopefully will help in providing a direct answer to this question.

        \appendix

         \section{BCFT interpretation of scalar 2-point function on $\Sigma_2$. } \label{appendixa}
         
           In this appendix we study the properties of the scalar correlator given in (\ref{sclcor}) on the replica surface. 
          We see that it can be thought of as a correlator in a boundary conformal field theory with a co-dimension $2$ defect.
          We demonstrate it explicitly for the case of $n=2$. 
        Let us begin with the two-point function on the replica surface for the replica parameter $n=2$. 
        \begin{align}\label{2ptn2}
            G(x, x')_{\Sigma_2}&=\frac{\sinh \frac{\eta}{2}}{16\pi^2 r_1 r_2\sinh\eta\left(\cosh\frac{\eta}{2}-\cos\frac{\theta}{2}\right)}.
        \end{align}
        It is convenient to use the cross ratio's $z, \bar z$ introduced in \cite{Lemos:2017vnx} to rewrite this correlator. 
        These variables are related to 
         $\eta$ and $\theta$ by the following 
        \begin{align}\label{reltn}
            \cosh\eta=\frac{1+z\bar{z}}{2(z\bar{z})^{\frac{1}{2}}},\qquad\qquad \cos\theta=\frac{z+\bar{z}}{2(z\bar{z})^{\frac{1}{2}}}.
        \end{align}
      The  two-point function in (\ref{2ptn2}) can be re-written as 
               \begin{align} \label{2ptn2z}
             G(z, \bar{z})_{\Sigma_2}&= \frac{1}{16\pi^2 r_1 r_2}
             \frac{2\sqrt{z\bar{z}}}{(1+\sqrt{z\bar{z}})(1-\sqrt{z})(1-\sqrt{\bar{z}})}.
        \end{align}
        Now if this is a correlator of $2$ scalar primaries in a BCFT with a codimension
        $2$ defect  it should admit an expansion of the form
           \begin{align} \label{bcftform}
            G(z, \bar{z})_{\Sigma_2}&=\frac{g(z,\bar{z})}{r_1 r_2},\nonumber\\
            &= \frac{1}{ r_1 r_2} \sum_{\mathcal{O}}(b_{\phi\mathcal{O}})^2\hat{f}_{\hat{\tau},s}(z,\bar{z}).
        \end{align}
         $(b_{\phi\mathcal{O}})^2$ are positive coefficients  due to the OPE of $\phi$ with the operators localized in the defect  in the defect channel   \cite{Billo:2016cpy,Lemos:2017vnx}.
            \begin{align}\label{hyper}
          \hat{f}_{\hat{\tau},s}(z,\bar{z})&= \binom{\frac{q}{2}+s-2}{\frac{q}{2}-2}^{-1} C_s^{\left(\frac{q}{2}-1\right)}\left(\frac{\bar{z}+z}{2 \sqrt{z \bar{z}}}\right) \, _2F_1\left(\frac{\Delta }{2}+\frac{1}{2},\frac{\Delta }{2};-\frac{p}{2}+\Delta +1;\frac{4 z \bar{z}}{\left(z \bar{z}+1\right)^2}\right),\nonumber\\
          &=\left(z \right)^{\frac{\Delta-s}{2}}\left(\bar{z} \right)^{\frac{\Delta+s}{2}} \, _2F_1\left(-s,\frac{q}{2}-1,2-\frac{q}{2}-s ;\frac{z}{\bar{z}}\right)\,_2F_1\left(\Delta,\frac{p}{2},\Delta+1-\frac{p}{2};z\bar{z} \right).
        \end{align}
        where $C_s^{\left(\frac{q}{2}-1\right)}$ refers to the  Gegenbauer polynomial.
        For the co-dimension $2$ defect in $d=4$, these functions simplify drastically and we obtain 
          \begin{align} \label{hyper2}
            \hat{f}_{\hat{\tau},s}(z,\bar{z})|_{p =2, d=4} &=\left(z \right)^{\frac{\Delta-s}{2}}\left(\bar{z} \right)^{\frac{\Delta+s}{2}}\frac{1}{1-z\bar{z}}.
        \end{align}
        Let us express the correlation function in (\ref{2ptn2z}) in the form  (\ref{bcftform}) and obtain the
        positive  coefficients $(b_{\phi\mathcal{O}})^2$.
        \begin{align}\label{our}
               G(z, \bar{z})_{\Sigma_2 }&= \frac{1}{16 \pi^2 r_1 r_2}
               \frac{2\sqrt{z\bar{z}}}{(1+\sqrt{z\bar{z}})(1-\sqrt{z})(1-\sqrt{\bar{z}})},\nonumber\\
               &=  \frac{1}{16 \pi^2 r_1 r_2} \frac{2\sqrt{z\bar{z}}(1-\sqrt{z\bar{z}})}{(1-z\bar{z})(1-\sqrt{z})(1-\sqrt{\bar{z}})},\nonumber\\
               &=  \frac{1}{16 \pi^2 r_1 r_2} \sum_{m=0}^{\infty}\sum_{n=0}^{\infty}\frac{2\sqrt{z\bar{z}}(1-\sqrt{z\bar{z}})}{(1-z\bar{z})}z^{\frac{n}{2}}\bar{z}^{\frac{m}{2}},\nonumber\\
               &=  \frac{1}{16 \pi^2 r_1 r_2} ( 2   \sqrt{z \bar z} ) 
               \frac{(1+\sum_{n=1}^{\infty}z^{\frac{n}{2}})(1+\sum_{m=1}^{\infty}\bar{z}^{\frac{m}{2}})-\sum_{m=1}^{\infty}\sum_{n=1}^{\infty}z^{\frac{n}{2}}\bar{z}^{\frac{m}{2}}}{1-z\bar{z}},\nonumber\\
               &=  \frac{1}{16 \pi^2 r_1 r_2}
               \frac{2 }{1-z\bar{z}}  \Big[ ( z\bar z)^{\frac{1}{2}}   + \sum_{n=1}^\infty z^{\frac{n+1}{2}}  \bar{z}^{\frac{1}{2}}  + 
               \sum_{n=1}^\infty  z^{\frac{1}{2}} \bar{z}^{ \frac{n+1}{2}}   \Big].
        \end{align}
        Now comparing (\ref{bcftform}), (\ref{hyper2}) and (\ref{our}) we see that
        \begin{equation}
        b_{\phi{\cal O}}^2 = \frac{1}{8\pi^2}.
        \end{equation}
        for boundary operators ${\cal O}$  with 
        \begin{equation}
        \Delta =1, s=0, \qquad\qquad  \Delta  = \frac{n}{2} + 1, \;s = \pm \frac{n}{2}, \quad n = 1, 2 \cdots
        \end{equation}
        Therefore  the scalar correlator on the cone can be expressed as a $2$ point function of primaries of dimension 
        $1$ in the presence of a co-dimension $2$ defect. 
        
\begin{figure}[htb]
\centering
\begin{subfigure}{.5\textwidth}
  \centering
  \includegraphics[width=1\linewidth]{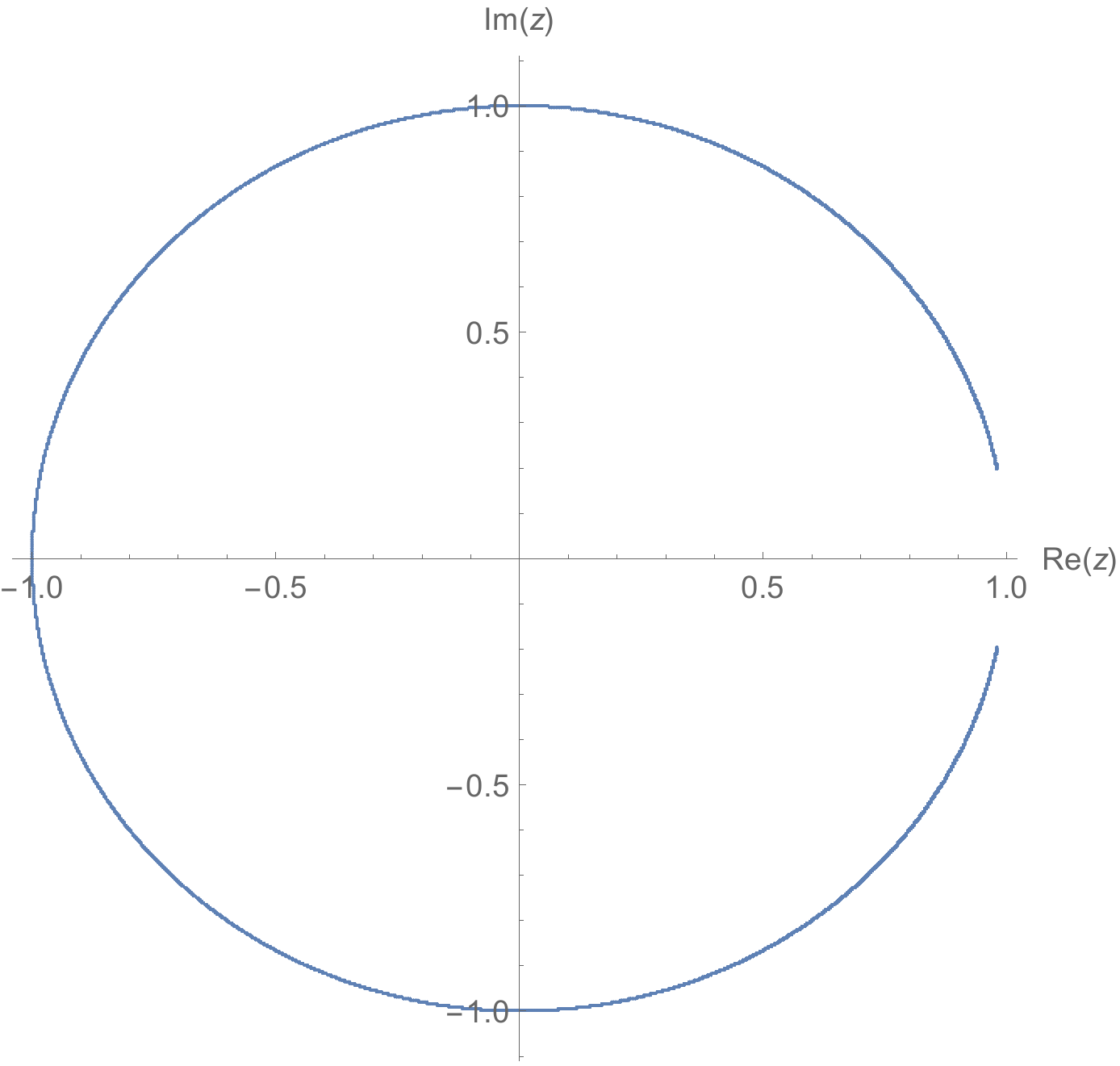}
  \label{1}
\end{subfigure}%
\begin{subfigure}{.5\textwidth}
  \centering
  \includegraphics[width=1\linewidth]{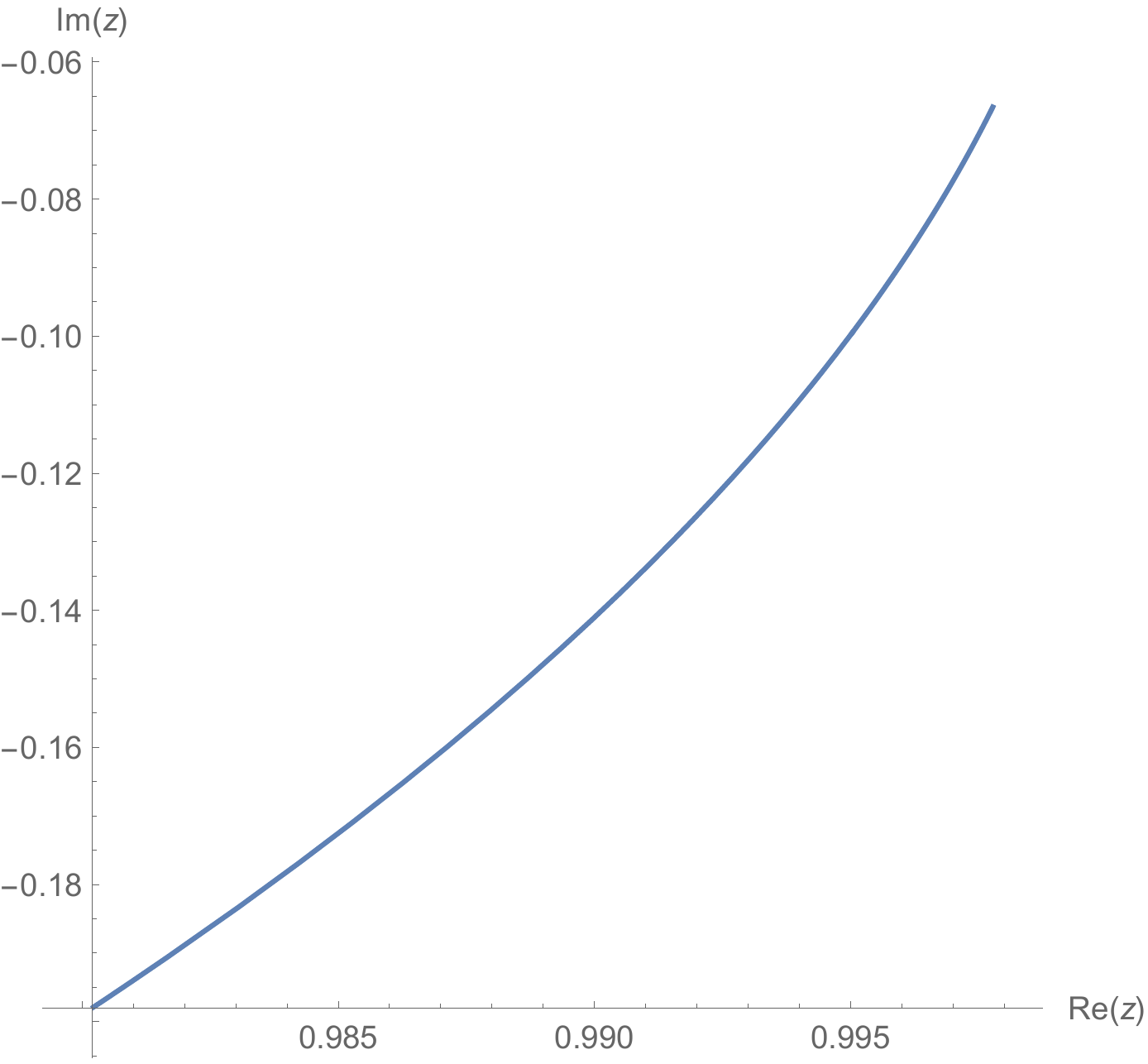}
  \label{2}
\end{subfigure}
\label{zplot}
\caption{We plot $z$ and $\bar{z}$  as we increase $t$ from $0$ to $10$ with  $y_0=5, \epsilon =0.5$
$z$ moves  on the unit circle 
from above the real axis to below the $z$ axis counter clockwise, while $\bar z $ always stays 
below the real axis close to unity. 
} \label{fig7}
\end{figure}

        It is also interesting to track the behaviour of the cross ratios $z, \bar z$ as we vary time given a particular $y_0$. 
        Inverting the relation in (\ref{reltn}) we find
         \begin{eqnarray}
           z&=& \left(\cosh (\eta )+\sqrt{\cosh ^2(\eta )-1}\right) \left(\cos (\theta )+\sqrt{\cos ^2(\theta )-1}\right),  
           \\ \nonumber 
           \bar{z}&=&\frac{\cosh (\eta )+\sqrt{\cosh ^2(\eta )-1}}{\cos (\theta )+\sqrt{\cos ^2(\theta )-1}}.
       \end{eqnarray}
       While performing the inversion of (\ref{reltn}), there  are 4 branches we can choose from. A pair of these  occur because of the symmetry $z\rightarrow \bar z$ of the equations, while a second pair can be thought of as a time 
       reversed version of the quench.  As we will see, the above branch represents the physical situation in which the quench moves from $-y_0$ and the R\'{e}nyi entropy increases as time increases. 
       We can use the relation between $\cosh\eta, \cos\theta$ given in (\ref{defcross}), to relate them to $y, t$ by 
        $r _1^2 =  y_0^2 + (- \epsilon - i t)^2, \tau_1 =  -\epsilon - i t$ and  
        $r _2^2 =  y_0^2 + (\epsilon - i t)^2, \tau_2 =  \epsilon -i t$ with $x_1=x_2, z_1 = z_2$. 
        In figure (\ref{fig7}) we have plotted the behaviour of $z, \bar z$ as we begin with $y_0 =  -5$ with 
       $ \epsilon  = 0.5 $ and $t$ranges from $0 $ to  $10$. 
       We see that $z$ moves on the unit circle from $z\sim 1$ to $e^{2\pi i }z$, while $\bar z $ remains close to one 
       below  the  real axis. 
       This phenomenon is similar to that seen in 2d quenches, see for example in \cite{David:2016pzn} \footnote{For the 2d quenches studied in \cite{David:2016pzn} $\bar z$ remained closed to unity  above the real axis.}.
       Therefore when $t<y_0$, the singularity of the correlator (\ref{2ptn2z}) arises from the factors
       $( 1- \sqrt{z}) ( 1-\sqrt{\bar z})$ in the denominator. 
         While for $t>y_0$,  the singularity in the correlator arises from the factor $( 1+ \sqrt{z\bar z}) ( 1-\sqrt{\bar z})$.

        \section{Details for curvature correlators}\label{dercorgrav}

        We give the details of the derivations of two-point functions on the replica surface for curvature components
        belonging to class 1, 2, 3, 4, 5, 6, 7. 
        We also show that the 2 point functions of curvature components  belonging to the same class are identical. 
         Therefore quenches due to these components will result  in same time dependence 
        of the R'{e}nyi/entanglement entropies.

       \subsection*{Class 1:  $R_{tyty}, R_{xzxz}, R_{tyxz}$}
       
       We have  already demonstrated in the main text, that the two-point functions of the
       curvature components  in class 1  are related. 
       Let us consider $R_{tyty}$ as the representative of this class.
    Here we present the complete expression for the two-point function of the curvature component
    $R_{tyty}$ in terms of the cross ratios $\theta$ and $\eta$. 
     { \small\begin{align} \label{exprtyty}
     &\langle R_{tyty} R_{tyty}\rangle=\frac{1}{32 \pi ^2 n^3 r_1^3 r_2^3 \left(\cos \left(\frac{\theta }{n}\right)-\cosh \left(\frac{\eta }{n}\right)\right)^3\left(\sinh (\eta)\right)^5}\nonumber\\
        &  \times \Bigg[2 n^2 \sinh \left(\eta  \left(2-\frac{3}{n}\right)\right)+6 (n+3) n \sinh \left(\eta  \left(2-\frac{1}{n}\right)\right)-12 \sinh \left(\frac{\eta }{n}\right)\nonumber\\
        &+6 \sinh \left(\eta  \left(\frac{1}{n}-2\right)\right)+6 \sinh \left(\eta  \left(\frac{1}{n}+2\right)\right)+2n \Big(3\big(-6\cos\left(\frac{\theta}{n}\right)+\cosh \left(\frac{\eta +2 i \theta }{n}\right)\nonumber\\
        &+\cosh \left(\frac{\eta -2 i \theta }{n}\right)\big)\sinh (2 \eta )+4 n \sinh \left(2 \eta  \left(\frac{1}{n}-1\right)\right) \cos \left(\frac{\theta }{n}\right)-3 (n-3) \sinh \left(\eta  \left(\frac{1}{n}+2\right)\right)\nonumber\\
        &-4 n (\cosh (2 \eta )+2) \sinh \left(\frac{\eta }{n}\right) \cos \left(\frac{2 \theta }{n}\right)-n \left(12 \sinh \left(\frac{\eta }{n}\right)+\sinh \left(\eta  \left(\frac{3}{n}+2\right)\right)\right)\nonumber\\
        &+16 n \sinh \left(\frac{2 \eta }{n}\right) \cos \left(\frac{\theta }{n}\right)
        -3 \sinh \left(\frac{2 \eta  (n-1)}{n}\right) \cos \left(\frac{\theta }{n}\right)+(4 n-3) \sinh \left(\frac{2 \eta  (n+1)}{n}\right) \cos \left(\frac{\theta }{n}\right)\nonumber\\
        &-4 n \sinh \left(\frac{3 \eta }{n}\right)\Big)+2 \sinh \left(\frac{2 \eta +i \theta }{n}\right)+\sinh \left(\frac{2 \eta  (n-1)-i \theta }{n}\right)-\sinh \left(\frac{2 \eta  (n+1)-i \theta }{n}\right)\nonumber\\
       & +\sinh \left(\frac{2 \eta  (n-1)+i \theta }{n}\right)-\sinh \left(\frac{2 \eta  (n+1)+i \theta }{n}\right)+2 \left(\sinh \left(\frac{2 \eta -i \theta }{n}\right)+\sinh \left(\frac{\eta -2 i \theta }{n}\right)\right)\nonumber\\
       &2 \sinh \left(\frac{\eta +2 i \theta }{n}\right)-\sinh \left(\frac{\eta +2 i \theta +2 \eta  n}{n}\right)+\sinh \left(\frac{\eta  (2 n-1)+2 i \theta }{n}\right)-\sinh \left(\frac{\eta +2 i \theta -2 \eta  n}{n}\right)\nonumber\\
       &-\sinh \left(\frac{\eta -2 i \theta +2 \eta  n}{n}\right)\Bigg].
        \end{align}}

       \subsection*{Class 2 :$\{R_{tytx}, R_{tytz},  R_{yzxz}, R_{yxxz} \}$}
         Let us begin with the curvature tensor $R_{tytx}$, from (\ref{reim}) we have 
         \begin{align}
            R_{tytx}&=\frac{1}{2}\Big[\partial_y\partial_t h_{tx}-\partial_t^2h_{yx}+\partial_t\partial_x h_{yt}-\partial_y\partial_x h_{tt}\Big].
        \end{align}
        Using the graviton propagator given in (\ref{defcor1}), 
          the  two-point function can be evaluated, 
        \begin{align}\label{Rtytx}
           & \qquad\qquad \langle R_{tytx}(x ) R_{tytx}(x')\rangle|_{x_1= x_2, z_1=z_2} = \nonumber \\
           & \qquad\qquad  \frac{1}{4}\Big[\partial_{y_1}\partial_{y_2}\partial_{t_1}\partial_{t_2}\langle h_{tx}h_{tx}\rangle-\partial_{y_1}\partial_{t_1}\partial_{t_2}^2\langle h_{tx}h_{yx}\rangle+\partial_{t_1}^2\partial_{t_2}^2\langle h_{yx}h_{yx}\rangle\nonumber\\
            & \qquad\qquad -\partial_{t_1}^2\partial_{y_2}\partial_{t_2}\langle h_{yx}h_{tx}\rangle+\partial_{x_1}\partial_{x_2}\partial_{t_1}\partial_{t_2}\langle h_{yt}h_{yt}\rangle- \partial_
            {t_1}\partial_{x_1}\partial_{y_2}\partial_{x_2}\langle h_{yt}h_{tt}\rangle\nonumber\\
           &\qquad\qquad  + \partial_{y_1} \partial_{y_2}\partial_{x_1}\partial_{x_2}\langle h_{tt}h_{tt}\rangle-\partial_{y_1}\partial_{t_2}\partial_{x_1}\partial_{x_2}\langle h_{tt}h_{yt}\rangle\Big]|_{x_1= x_2, z_1=z_2}  \nonumber\\
           &\qquad\qquad =\frac{\partial_{x_1}^2(-2\partial_{y_1}\partial_{y_2}+ \partial_{t_1}\partial_{t_2})(\partial_{t_1}^2+\partial_{y_1}^2)(\partial_{t_2}^2+\partial_{y_2}^2)}{\nabla^4}G(x,x')|_{x_1= x_2, z_1=z_2}, \nonumber\\
           &\qquad\qquad =\partial_{x_1}^2(\partial_{t_1}\partial_{t_2}-2\partial_{y_1}\partial_{y_2} )G(x,x')|_{x_1= x_2, z_1=z_2} .
        \end{align}
        To derive the last but one line we use isometry in $x,z$-plane derived in (\ref{isoa1}),  (\ref{isoa2})
        \begin{align}
            \frac{\partial_{x_1}^2}{\nabla^2}G(x, x') |_{x_1= x_2, z_1=z_2} &=  
            \frac{\partial_{z_1}^2}{\nabla^2}G(x,x')|_{x_1= x_2, z_1=z_2}  = \frac{1}{2} G(x, x') |_{x_1= x_2, z_1=z_2} .
        \end{align}
          Let us  compute $\langle R_{tytz}(x ) R_{tytz}(x')\rangle$, the next element in this class. 
          The expression for $R_{tytz}$  from (\ref{reim}) is given by
        \begin{align}
           R_{tytz}&=\frac{1}{2}\Big[\partial_y\partial_t h_{tz}-\partial_t^2h_{yz}+\partial_t\partial_z h_{yt}-\partial_y\partial_z h_{tt}\Big].
        \end{align}
           Using the graviton propagator in (\ref{defcor1}) we compute 
         \begin{align} \label{Rtytz}
           &   \langle R_{tytz}(x ) R_{tytz}(x')\rangle |_{x_1= x_2, z_1=z_2}= \nonumber \\
           & \qquad\qquad \frac{1}{4}\Big[\partial_{y_1}\partial_{y_2}\partial_{t_1}\partial_{t_2}\langle h_{tz}h_{tz}\rangle-\partial_{y_1}\partial_{t_1}\partial_{t_2}^2\langle h_{tz}h_{yz}\rangle+\partial_{t_1}^2\partial_{t_2}^2\langle h_{yz}h_{yz}\rangle\nonumber\\
            &\qquad\qquad -\partial_{t_1}^2\partial_{y_2}\partial_{t_2}\langle h_{yz}h_{tz}\rangle+\partial_{x_1}\partial_{z_2}\partial_{t_1}\partial_{t_2}\langle h_{yt}h_{yt}\rangle- \partial_
            {t_1}\partial_{z_1}\partial_{y_2}\partial_{z_2}\langle h_{yt}h_{tt}\rangle\nonumber\\
           &\qquad\qquad + \partial_{y_1} \partial_{y_2}\partial_{z_1}\partial_{z_2}\langle h_{tt}h_{tt}\rangle-\partial_{y_1}\partial_{t_2}\partial_{z_1}\partial_{z_2}\langle h_{tt}h_{yt}\rangle\Big]|_{x_1= x_2, z_1=z_2}\nonumber\\
           &\qquad\qquad =\frac{\partial_{z_1}^2(-2\partial_{y_1}\partial_{y_2}+ \partial_{t_1}\partial_{t_2})(\partial_{t_1}^2+\partial_{y_1}^2)(\partial_{t_2}^2+\partial_{y_2}^2)}{\nabla^4}G(x,x' ) |_{x_1= x_2, z_1=z_2},\nonumber\\
           &\qquad\qquad =\partial_{z_1}^2(\partial_{t_1}\partial_{t_2}-2\partial_{y_1}\partial_{y_2})G(x ,x' )  |_{x_1= x_2, z_1=z_2}.
         \end{align}
         Again we have used the isotropy property of the $x, z$ plane. 
          Let us compute the two-point function of $R_{yzxz}$.   The expression of $R_{yzxz}$ is given by
        \begin{align}
            R_{yzxz}&=\frac{1}{2}\Big[\partial_z\partial_x h_{yz}-\partial_y\partial_x h_{zz}+\partial_y\partial_z h_{zx}-\partial_z^2 h_{yx}\Big]
        \end{align}
        Using the graviton propagator we obtain
        \begin{align}\label{Ryzxz}
            & \langle R_{yzxz}(x)R_{yzxz}(x')\rangle |_{x_1= x_2, z_1=z_2}\nonumber \\
            &\qquad\qquad =\frac{1}{4}\Big[\partial_{z_1}\partial_{z_2}\partial_{x_1}\partial_{x_2}\langle h_{yz}h_{yz}\rangle-\partial_{z_1}\partial_{x_1}\partial_{z_2}^2\langle h_{yz}h_{yx}\rangle+\partial_{y_1}\partial_{x_1}\partial_{y_2}\partial_{x_2}\langle h_{zz}h_{zz}\rangle\nonumber\\
            &\qquad\qquad  -\partial_{y_1}\partial_{x_1}\partial_{y_2}\partial_{z_2}\langle h_{zz}h_{zx}\rangle+\partial_{y_1}\partial_{z_1}\partial_{y_2}\partial_{z_2}\langle h_{zx}h_{zx}\rangle- \partial_
            {y_1}\partial_{z_1}\partial_{y_2}\partial_{x_2}\langle h_{zx}h_{zz}\rangle\nonumber\\
           &\qquad\qquad  + \partial_{z_1}^2 \partial_{z_2}^2\langle h_{yx}h_{yx}\rangle-\partial_{z_1}^2\partial_{z_2}\partial_{x_2}\langle h_{yx}h_{yz}\rangle\Big]|_{x_1= x_2, z_1=z_2},\nonumber\\
           &\qquad\qquad  =\frac{\partial_{z_1}^2(\partial_{t_1}\partial_{t_2}-2 \partial_{y_1}\partial_{y_2})(\partial_{z_1}^2+\partial_{x_1}^2)^2}{\nabla^4}G(x,x')|_{x_1= x_2, z_1=z_2},\nonumber\\
           &\qquad\qquad  =\partial_{x_1}^2(-2\partial_{y_1}\partial_{y_2}+ \partial_{t_1}\partial_{t_2})G(x ,x')|_{x_1= x_2, z_1=z_2}.
        \end{align}
         Let us consider the last correlator in this class
         $ \langle R_{yxxz}(x)R_{yxxz}(x')\rangle$. The expression of the curvature component $R_{yxxz}$ is given by
      \begin{align}
            R_{yxxz}&=\frac{1}{2}\Big[\partial_x^2 h_{yz}-\partial_y\partial_x h_{xz}+\partial_y\partial_z h_{xx}-\partial_x\partial_z h_{yx}\Big].
        \end{align}
         The two-point function is given by
        \begin{align}\label{Ryxxz}
           & \langle R_{yxxz}(x )R_{yxxz}(x')\rangle|_{x_1= x_2, z_1=z_2} \nonumber \\
            &\qquad\qquad =\frac{1}{4}\Big[\partial_{x_1}^2\partial_{x_2}^2\langle h_{yz}h_{yz}\rangle-\partial_{z_2}\partial_{x_2}\partial_{x_1}^2\langle h_{yz}h_{yx}\rangle+\partial_{y_1}\partial_{x_1}\partial_{y_2}\partial_{x_2}\langle h_{xz}h_{xz}\rangle\nonumber\\
            &\qquad\qquad -\partial_{y_1}\partial_{x_1}\partial_{y_2}\partial_{z_2}\langle h_{xz}h_{xx}\rangle+\partial_{y_1}\partial_{z_1}\partial_{y_2}\partial_{z_2}\langle h_{xx}h_{xx}\rangle- \partial_
            {y_1}\partial_{z_1}\partial_{y_2}\partial_{x_2}\langle h_{xx}h_{xz}\rangle\nonumber\\
           &\qquad\qquad + \partial_{z_1}\partial_{x_1}\partial_{x_2} \partial_{z_2}\langle h_{yx}h_{yx}\rangle-\partial_{z_2}^2\partial_{z_1}\partial_{x_1}\langle h_{yx}h_{yz}\rangle\Big]|_{x_1= x_2, z_1=z_2},\nonumber\\
           &\qquad\qquad =\frac{\partial_{z_1}^2(\partial_{t_1}\partial_{t_2}-2 \partial_{y_1}\partial_{y_2})(\partial_{z_1}^2+\partial_{x_1}^2)^2}{\nabla^4}G(x,x')|_{x_1= x_2, z_1=z_2},\nonumber\\
           &\qquad\qquad =\partial_{x_1}^2(-2\partial_{y_1}\partial_{y_2}+ \partial_{t_1}\partial_{t_2})G(x,x')|_{x_1= x_2, z_1=z_2}.
        \end{align}
       We have used  the isotropy in the  $x,z$ directions.
       
        Therefore, by explicit computation we have seen that 
        \begin{eqnarray}
          \langle  R_{tytx}(x)R_{tytx}(x')\rangle |_{x_1= x_2, z_1=z_2}&=&\langle R_{tytz}(x)R_{tytz}(x')\rangle|_{x_1= x_2, z_1=z_2}, \\ \nonumber
          &=& \langle R_{yzxz}(x)R_{yzxz}(x')\rangle|_{x_1= x_2, z_1=z_2}, \\ \nonumber
          &= &   \langle R_{yxxz}(x)R_{yxxz}(x')\rangle|_{x_1= x_2, z_1=z_2}.
        \end{eqnarray}
        The two-point function of all curvature components in the class are identical. 
        
        \subsection*{Class 3 :$\{ R_{txxz}, R_{tzxz}, R_{tyyx}, R_{tyyz} \}$}

         Let us now compute the correlator of $R_{txxz}$.
        \begin{align}
            R_{txxz}&=\frac{1}{2}\Big[\partial_x^2 h_{tz}-\partial_t\partial_x h_{xz}+\partial_t\partial_z h_{xx}-\partial_x\partial_z h_{tx}\Big].
        \end{align}
         Therefore, we evaluate the two-point function
            \begin{align} \label{Rtxxz}
                 \langle R_{txxz}(x)R_{txxz}(x')\rangle|_{x_1= x_2, z_1=z_2} &  
                =(\partial_{y_1}\partial_{y_2}-2\partial_{t_1}\partial_{t_2})\frac{\partial_{z_1}^2(\partial_{x_1}^2+\partial_{z_1}^2)^2}{\nabla^4}G(x,x')|_{x_1= x_2, z_1=z_2},\nonumber\\
             & =\partial_{x_1}^2(\partial_{y_1}\partial_{y_2}-2 \partial_{t_1}\partial_{t_2})G(x,x')|_{x_1= x_2, z_1=z_2}.
            \end{align}
          To derive the last line we use isometry in $x,z$-plane
        \begin{align}
            \frac{\partial_{x_1}^2}{\nabla^2}G(x,x')|_{x_1= x_2, z_1=z_2}&=  \frac{\partial_{z_1}^2}{\nabla^2}G(x,x')|_{x_1= x_2, z_1=z_2}.
        \end{align}
            Similarly we compute the two-point function of $R_{tzxz}$  and observe that they are identical.
            \begin{align} \label{Rtzxz}
                \langle R_{tzxz}(x)R_{tzxz}(x')\rangle|_{x_1= x_2, z_1=z_2}&=\partial_{x_1}^2(\partial_{y_1}\partial_{y_2}-2 \partial_{t_1}\partial_{t_2})G(x,x')|_{x_1= x_2, z_1=z_2}\nonumber\\
                &=\langle R_{tyyx}(x) R_{tyyx}(x')\rangle|_{x_1= x_2, z_1=z_2}.\nonumber\\
            \end{align}
            To derive the last line we use the isometry of the scalar Green's function in the $x-z$ plane. Let us now compute the two-point functions of $R_{tyyx}$
            \begin{align} \label{Rtyyx}
                  & \langle R_{tyyx}(x) R_{tyyx}(x')\rangle|_{x_1= x_2, z_1=z_2} \nonumber \\
                  &\qquad\qquad\qquad =\frac{1}{4}\Big[\partial_{y_1}^2\partial_{y_2}^2\langle h_{tx}h_{tx}\rangle-\partial_{t_2}\partial_{y_2}\partial_{y_1}^2\langle h_{tx}h_{yx}\rangle+\partial_{y_1}\partial_{t_1}\partial_{y_2}\partial_{t_2}\langle h_{yx}h_{yx}\rangle\nonumber\\
            &\qquad\qquad\qquad+\partial_{t_1}\partial_{x_1}\partial_{t_2}\partial_{x_2}\langle h_{yy}h_{yy}\rangle-\partial_{t_1}\partial_{y_1}\partial_{y_2}^2\langle h_{yx}h_{tx}\rangle- \partial_
            {t_1}\partial_{x_1}\partial_{y_2}\partial_{x_2}\langle h_{yy}h_{ty}\rangle\nonumber\\
           &\qquad\qquad\qquad+ \partial_{y_1}\partial_{x_1}\partial_{y_2} \partial_{x_2}\langle h_{ty}h_{ty}\rangle-\partial_{y_1}\partial_{x_1}\partial_{t_2}\partial_{x_2}\langle h_{ty}h_{yy}\rangle\Big]G(x,x'),\nonumber\\
           &\qquad\qquad=\partial_{x_1}^2(\partial_{y_1}\partial_{y_2}-2 \partial_{t_1}\partial_{t_2})G(x,x')|_{x_1= x_2, z_1=z_2}.
            \end{align}
           Similarly we compute the two-point function of $R_{tyyz}$ and obtain 
           \begin{align} \label{Rtyyz}
                  \langle R_{tyyz}(x) R_{tyyz}(x')\rangle|_{x_1= x_2, z_1=z_2}&=\partial_{z_1}^2(\partial_{y_1}\partial_{y_2}-2 \partial_{t_1}\partial_{t_2})G(x,x')|_{x_1= x_2, z_1=z_2}.
           \end{align}
          From the isotropy of the scalar Green's function in the $x, z$ plane, it is clear that
        \begin{align}
            \langle R_{txxz}(x) R_{txxz}(x')\rangle|_{x_1= x_2, z_1=z_2}&=\langle R_{tzxz}(x) R_{tzxz}(x')\rangle|_{x_1= x_2, z_1=z_2},\nonumber\\
            &=\langle R_{tyyx}(x) R_{tyyx}(x')\rangle|_{x_1= x_2, z_1=z_2},\nonumber\\
            &=\langle R_{tyyz}(x) R_{tyyz}(x')\rangle|_{x_1= x_2, z_1=z_2}.
        \end{align}

        \subsection*{Class 4:$\{ R_{txtx}, R_{yzyz}, R_{tztz}, R_{yxyx} \}$}
        
        We now compute the two-point function of $R_{txtx}$ and show that this is identical to the two-point functions of three other curvature tensors which are $R_{yzyz}$, $R_{tzyz}$ and $R_{yxyz}$.
        \begin{align}
            R_{txtx}&=\frac{1}{2}\Big[2\partial_x\partial_t h_{tx}-\partial_t^2 h_{xx}-\partial_x^2 h_{tt}\Big]
        \end{align}
         The two-point function is given by
        \begin{align}\label{Rtxtx}
           & \langle R_{txtx}(x)R_{txtx}(x')\rangle|_{x_1= x_2, z_1=z_2} \nonumber \\
           & \qquad\qquad =\frac{1}{4}\Big[4\partial_{x_1}\partial_{x_2}\partial_{t_1}\partial_{t_2}\langle h_{tx}h_{tx}\rangle+\partial_{t_1}^2\partial_{t_2}^2 \langle h_{xx} h_{xx}\rangle+\partial_{x_1}^2\partial_{x_2}^2 \langle h_{tt} h_{tt}\rangle,\nonumber\\
            &\qquad\qquad\qquad\qquad+\partial_{x_2}^2\partial_{t_1}^2\langle h_{xx}h_{tt}\rangle+\partial_{x_1}^2\partial_{t_2}^2\langle h_{tt}h_{xx}\rangle\Big]|_{x_1= x_2, z_1=z_2}\nonumber\\
            &\qquad\qquad=\frac{1}{4\nabla^4}\Big(-4\partial_{x_1}^2\partial_{z_1}^2\partial_{t_1}\partial_{t_2}\partial_{y_1}\partial_{y_2}+2\partial_{z_1}^4\partial_{t_1}^2\partial_{t_2}^2+2\partial_{x_1}^4\partial_{y_1}^2\partial_{y_2}^2\nonumber\\
            &\qquad\qquad\qquad\qquad +2\partial_{x_1}^2\partial_{z_1}^2(\partial_{t_1}^2\partial_{y_2}^2+\partial_{y_1}^2\partial_{t_2}^2\Big)G(x,x')|_{x_1= x_2, z_1=z_2},\nonumber\\
           &\qquad\qquad =\frac{1}{16}\Big(3\partial_{t_1}^2\partial_{t_2}^2+3 \partial_{y_1}^2\partial_{y_2}^2+(\partial_{t_1}\partial_{y_2}-\partial_{y_1}\partial_{t_2})^2\Big)G(x,x')|_{x_1= x_2, z_1=z_2}.
        \end{align}
        Here we use the following identities to derive the last line. The proof of these identities are shown in \eqref{isoa3} and \eqref{isoa5}.
        \begin{align}\label{identity}
        \begin{split}
     \frac{\partial_{x_1}^4}{\nabla^4}G(x,x')|_{x_1= x_2, z_1=z_2}&=  \frac{\partial_{z_1}^4}{\nabla^4}G(x,x')|_{x_1= x_2, z_1=z_2},\\
            &=\frac{3}{8}G(x,x')|_{x_1= x_2, z_1=z_2}.\\
              \frac{\partial_{x_1}^2\partial_{z_1}^2}{\nabla^4}G(x,x')|_{x_1= x_2, z_1=z_2}&=\frac{1}{8}G(x,x')|_{x_1= x_2, z_1=z_2}.
        \end{split}
        \end{align}
        These identities arise because of the isotropy of the scalar Greens function.  Therefore we expect that 
        the two point function $\langle R_{tztz}(x)R_{tztz}(x')\rangle$ to be 
        identical to $\langle R_{txtx}(x)R_{txtx}(x')\rangle$ in the coincident limit.
        The expression of $R_{tztz}$ is given by
        \begin{align}
            R_{tztz}&=\frac{1}{2}\Big[2\partial_z\partial_t h_{tz}-\partial_t^2 h_{zz}-\partial_z^2 h_{tt}\Big].
        \end{align}
        The two-point function is evaluated to
        \begin{align}\label{Rtztz}
            & \langle R_{tztz}(x)R_{tztz}(x')\rangle|_{x_1= x_2, z_1=z_2}  \nonumber \\
             &\qquad\qquad =\frac{1}{4\nabla^4}\Big(-4\partial_{x_1}^2\partial_{z_1}^2\partial_{t_1}\partial_{t_2}\partial_{y_1}\partial_{y_2}+2\partial_{z_1}^4\partial_{t_1}^2\partial_{t_2}^2+2\partial_{x_1}^4\partial_{y_1}^2\partial_{y_2}^2\nonumber\\
            &\qquad\qquad\qquad +2\partial_{x_1}^2\partial_{z_1}^2(\partial_{t_1}^2\partial_{y_2}^2+\partial_{y_1}^2\partial_{t_2}^2\Big)G(x,x')|_{x_1= x_2, z_1=z_2},\nonumber\\
           &\qquad\qquad=\frac{1}{16}\Big(3\partial_{t_1}^2\partial_{t_2}^2+3 \partial_{y_1}^2\partial_{y_2}^2+(\partial_{t_1}\partial_{y_2}-\partial_{y_1}\partial_{t_2})^2\Big)G(x,x')|_{x_1= x_2, z_1=z_2}.
        \end{align}
        Similarly we explicitly show that two-point function of $R_{yzyz}$ and $R_{yxyx}$ are identical to the two-point function of $R_{txtx}$ or $R_{tztz}$.
         \begin{align}
            R_{yzyz}&=\frac{1}{2}\Big[2\partial_z\partial_y h_{yz}-\partial_y^2 h_{zz}-\partial_z^2 h_{yy}\Big].
        \end{align}
        The two-point function is given by
        \begin{align}\label{Ryzyz}
          &  \langle R_{yzyz}(x)R_{yzyz}(x')\rangle|_{x_1= x_2, z_1=z_2} \\ \nonumber
            &\qquad\qquad =\frac{1}{4}\Big[4\partial_{x_1}\partial_{x_2}\partial_{y_1}\partial_{y_2}\langle h_{yx}h_{yx}\rangle+\partial_{y_1}^2\partial_{y_2}^2 \langle h_{xx} h_{xx}\rangle+\partial_{x_1}^2\partial_{x_2}^2 \langle h_{yy} h_{yy}\rangle\nonumber\\
            &\qquad\qquad\qquad +\partial_{x_2}^2\partial_{y_1}^2\langle h_{xx}h_{yy}\rangle+\partial_{x_1}^2\partial_{y_2}^2\langle h_{yy}h_{xx}\rangle\Big]|_{x_1= x_2, z_1=z_2},\nonumber\\
            &\qquad\qquad=\frac{1}{4\nabla^4}\Big(-4\partial_{x_1}^2\partial_{z_1}^2\partial_{y_1}\partial_{y_2}\partial_{t_1}\partial_{t_2}+2\partial_{z_1}^4\partial_{y_1}^2\partial_{y_2}^2+2\partial_{x_1}^4\partial_{t_1}^2\partial_{t_2}^2\nonumber\\
            &\qquad\qquad\qquad+2\partial_{x_1}^2\partial_{z_1}^2(\partial_{y_1}^2\partial_{t_2}^2+\partial_{t_1}^2\partial_{y_2}^2\Big)G(x,x')|_{x_1= x_2, z_1=z_2},\nonumber\\
           &\qquad\qquad=\frac{1}{16}\Big(3\partial_{t_1}^2\partial_{t_2}^2+3 \partial_{y_1}^2\partial_{y_2}^2+(\partial_{t_1}\partial_{y_2}-\partial_{y_1}\partial_{t_2})^2\Big)G(x,x')|_{x_1= x_2, z_1=z_2},\nonumber\\
           &\qquad\qquad =\langle R_{yxyx}(x)R_{yxyx}(x')\rangle|_{x_1= x_2, z_1=z_2}.
        \end{align}
      In the last line, we use the identities $\eqref{identity}$. Therefore we find 
      \begin{align}\label{4corsame}
          & \langle R_{txtx}(x)R_{txtx}(x')\rangle|_{x_1= x_2, z_1=z_2}= \langle R_{tztz}(x)R_{tztz}(x')\rangle|_{x_1= x_2, z_1=z_2}\nonumber\\
          &= \langle R_{yxyx}(x)R_{yxyx}(x')\rangle|_{x_1= x_2, z_1=z_2} = \langle R_{yzyz}(x)R_{yzyz}(x')\rangle|_{x_1= x_2, z_1=z_2}.
      \end{align}

      \subsection*{Class 5: $\{R_{txtz}, R_{yxyz} \}$}
      
      We now compute the two-point function of $R_{txtz}$ and $R_{yxyz}$ and show the equality between them. The curvature tensor $R_{txtz}$ in terms of the field variable is given by
         \begin{align}
            R_{txtz}&=\frac{1}{2}\Big[\partial_x\partial_t h_{tz}-\partial_t^2 h_{xz}+\partial_t\partial_z h_{xt}-\partial_z\partial_x h_{tt}\Big].
        \end{align}
        Therefore the two-point function is  given by 
        \begin{align} \label{Rtxtz}
          &  \langle R_{txtz}(x) R_{txtz}(x')\rangle|_{x_1= x_2, z_1=z_2} \\
           &\qquad\qquad= \frac{1}{4}\Big[\partial_{x_1}\partial_{x_2}\partial_{t_1}\partial_{t_2}\langle h_{tz}h_{tz}\rangle+\partial_{x_1}\partial_{z_2}\partial_{t_1}\partial_{t_2}\langle h_{tz}h_{xt}\rangle+\partial_{t_1}^2\partial_{t_2}^2 \langle h_{xz} h_{xz}\rangle\nonumber\\
            &\qquad\qquad\qquad+\partial_{z_1}\partial_{z_2}\partial_{t_1}\partial_{t_2}\langle h_{xt}h_{xt}\rangle+\partial_{t_1}\partial_{z_1}\partial_{t_2}\partial_{x_2}\langle h_{xt}h_{tz}\rangle+\partial_{x_1}^2\partial_{z_1}^2\langle h_{tz}h_{tz}\rangle\nonumber\\
            &\qquad\qquad\qquad +\partial_{t_1}^2\partial_{x_2}\partial_{z_2}\langle h_{xz}h_{tt}\rangle+\partial_{x_1}\partial_{z_1}\partial_{t_2}^2\langle h_{tt}h_{xz}\rangle\Big]|_{x_1= x_2, z_1=z_2},\nonumber\\
            &\quad\quad=\frac{1}{16}\Big(2\partial_{y_1}\partial_{y_2}\partial_{t_1}\partial_{t_2}+\left(\partial_{t_1}\partial_{t_2}-\partial_{y_1}\partial_{y_2}\right)^2
            -\left(\partial_{t_1}\partial_{y_2}-\partial_{y_1}\partial_{t_2}\right)^2\Big)G(x,x')|_{x_1= x_2, z_1=z_2}. \nonumber
        \end{align}
        To derive the second line, we use the identity given in $\eqref{identity}$.
        Similarly we evaluate the correlator of $R_{yxyz}$. The curvature component $R_{yxyz}$ is given by
        \begin{align}
            R_{yxyz}&=\frac{1}{2}\Big[\partial_x\partial_y h_{yz}-\partial_y^2 h_{xz}+\partial_z\partial_y h_{xy}-\partial_{x}\partial_{z}h_{yy}\Big].
        \end{align}
           The correlator is  given by 
        \begin{align}  \label{Ryxyz}
          &  \langle R_{yxyz}(x) R_{yxyz}(x')\rangle|_{x_1= x_2, z_1=z_2} \\
           &\qquad\qquad= \frac{1}{4}\Big[\partial_{x_1}\partial_{x_2}\partial_{y_1}\partial_{y_2}\langle h_{yz}h_{yz}\rangle+\partial_{x_1}\partial_{z_2}\partial_{y_1}\partial_{y_2}\langle h_{yz}h_{xy}\rangle+\partial_{y_1}^2\partial_{y_2}^2 \langle h_{xz} h_{xz}\rangle\nonumber\\
            &\qquad\qquad\qquad+\partial_{z_1}\partial_{z_2}\partial_{y_1}\partial_{y_2}\langle h_{xy}h_{xy}\rangle+\partial_{y_1}\partial_{z_1}\partial_{y_2}\partial_{x_2}\langle h_{xy}h_{yz}\rangle+\partial_{x_1}^2\partial_{z_1}^2\langle h_{yz}h_{yz}\rangle\nonumber\\
            &\qquad\qquad\qquad+\partial_{y_1}^2\partial_{x_2}\partial_{z_2}\langle h_{xz}h_{yy}\rangle+\partial_{x_1}\partial_{z_1}\partial_{y_2}^2\langle h_{yy}h_{xz}\rangle\Big]|_{x_1= x_2, z_1=z_2},\nonumber\\
            &\quad\quad=\frac{1}{16}\Big(2\partial_{y_1}\partial_{y_2}\partial_{t_1}\partial_{t_2}+\left(\partial_{t_1}\partial_{t_2}-\partial_{y_1}\partial_{y_2}\right)^2
            -\left(\partial_{t_1}\partial_{y_2}-\partial_{y_1}\partial_{t_2}\right)^2\Big)G(x,x')|_{x_1= x_2, z_1=z_2}. \nonumber
        \end{align}
        In the last line we use the isotropy of the Green's function along the $x,z$ direction. Therefore  by explicit computation
        we have seen that 
         \begin{align}
           \langle R_{txtz}(x)R_{txtz}(x')\rangle|_{x_1= x_2, z_1=z_2}=\langle R_{yxyz}(x)R_{yxyz}(x')\rangle|_{x_1= x_2, z_1=z_2}.
        \end{align}
        
        \subsection*{Class 6: $\{ R_{txyx}, R_{tzyz} \}$}
        
        Similarly we find two more curvature tensors $R_{txyx}$ and $R_{tzyz}$ having same two-point functions. We compute two-point function of $R_{tzyz}$. The expression of $R_{txyx}$ in terms of field variables is given by
         \begin{align}
            R_{txyx}&=\frac{1}{2}\Big[\partial_x\partial_y h_{tx}-\partial_t\partial_y h_{xx}+\partial_x\partial_t h_{xy}-\partial_{x}^2h_{ty}\Big].
        \end{align}
        The two-point function is evaluated as
        \begin{align}\label{Rtxyx}
           & \langle R_{txyx}(x) R_{txyx}(x')\rangle|_{x_1= x_2, z_1=z_2} \\
            &\qquad\qquad=\frac{1}{4}\Big[\partial_{x_1}\partial_{x_2}\partial_{y_1}\partial_{y_2}\langle h_{tx}h_{tx}\rangle+\partial_{x_1}\partial_{x_2}\partial_{y_1}\partial_{t_2}\langle h_{tx}h_{xy}\rangle+\partial_{t_1}\partial_{t_2}\partial_{y_1}\partial_{y_2} \langle h_{xx} h_{xx}\rangle\nonumber\\
            &\qquad\qquad\qquad+\partial_{x_1}\partial_{x_2}\partial_{t_1}\partial_{t_2}\langle h_{xy}h_{xy}\rangle+\partial_{t_1}\partial_{x_1}\partial_{y_2}\partial_{x_2}\langle h_{xy}h_{tx}\rangle+\partial_{x_1}^2\partial_{x_2}^2\langle h_{ty}h_{ty}\rangle\nonumber\\
        &\qquad\qquad\qquad+\partial_{t_1}\partial_{y_1}\partial_{x_2}^2\langle h_{xx}h_{ty}\rangle+\partial_{x_1}^2\partial_{t_2}\partial_{y_2}\langle h_{ty}h_{xx}\rangle\Big]|_{x_1= x_2, z_1=z_2},\nonumber\\
            &\qquad\qquad =\frac{1}{4}\Big[2\frac{\partial_{x_1}^4}{\nabla^4}\partial_{t_1}\partial_{t_2}\partial_{y_1}\partial_{y_2}+2\frac{\partial_{z_1}^4}{\nabla^4}\partial_{t_1}\partial_{t_2}\partial_{y_1}\partial_{y_2}
            +2\frac{\partial_{x_1}^2\partial_{z_1}^2}{\nabla^4}(\partial_{t_1}^2\partial_{y_2}^2\partial_{y_1}^2\partial_{t_2}^2\nonumber\\
            &\qquad\qquad\qquad -\partial_{t_1}^2\partial_{t_2}^2-\partial_{y_1}^2\partial_{y_2}^2-4\partial_{t_1}\partial_{t_2}\partial_{y_1}\partial_{y_2})\Big]G(x,x')|_{x_1= x_2, z_1=z_2},\nonumber\\
            &\qquad\qquad=\Big[\frac{1}{32}\left(\partial_{t_1}\partial_{y_2}+\partial_{y_1}\partial_{t_2}\right)^2-\frac{1}{32}\left(\partial_{t_1}\partial_{t_2}-\partial_{y_1}\partial_{y_2}\right)^2\Big]G(x,x')|_{x_1= x_2, z_1=z_2}.\nonumber
        \end{align}
        Note that, the second line of equation $\eqref{Rtxyx}$ is symmetric under the exchange of variable $x_1\rightarrow z_1$. Therefore the two-point function of $R_{tzyz}$ is also the same as $R_{txyx}$.
        \begin{align} \label{Rtzyz}
             \langle R_{txyx}(x) R_{txyx}(x')\rangle|_{x_1= x_2, z_1=z_2}=  \langle R_{tzyz}(x) R_{tzyz}(x')\rangle|_{x_1= x_2, z_1=z_2} .
        \end{align}

        \subsection*{Class 7: $\{ R_{txyz} \} $}
        
       Now we are left with only one independent curvature component which is $R_{txyz}$. The curvature tensor $R_{txyz}$ in terms of field variables is given by
       \begin{align} 
            R_{txyz}&=\frac{1}{2}\Big[\partial_x\partial_y h_{tz}-\partial_t\partial_y h_{xz}+\partial_z\partial_t h_{xy}-\partial_{x}\partial_{z}h_{ty}\Big].
        \end{align}
        From the two-point function of the field variables, we compute
        \begin{align}\label{Rtxyz}
           & \langle R_{txyz}(x) R_{txyz}(x')\rangle|_{x_1= x_2, z_1=z_2} \\
           &\qquad\qquad = \frac{1}{4}\Big[\partial_{x_1}\partial_{x_2}\partial_{y_1}\partial_{y_2}\langle h_{tz}h_{tz}\rangle+\partial_{x_1}\partial_{z_2}\partial_{y_1}\partial_{t_2}\langle h_{tz}h_{xy}\rangle+\partial_{t_1}\partial_{t_2}\partial_{y_1}\partial_{y_2} \langle h_{xz} h_{xz}\rangle\nonumber\\
            &\qquad\qquad\qquad+\partial_{z_1}\partial_{z_2}\partial_{t_1}\partial_{t_2}\langle h_{xy}h_{xy}\rangle+\partial_{t_1}\partial_{z_1}\partial_{y_2}\partial_{x_2}\langle h_{xy}h_{tz}\rangle+\partial_{x_1}\partial_{x_2}\partial_{z_1}\partial_{z_2}\langle h_{ty}h_{ty}\rangle\nonumber\\
            &\qquad\qquad\qquad+\partial_{t_1}\partial_{y_1}\partial_{x_2}\partial_{z_2}\langle h_{xz}h_{ty}\rangle+\partial_{x_1}\partial_{z_1}\partial_{t_2}\partial_{y_2}\langle h_{ty}h_{xz}\rangle\Big]|_{x_1= x_2, z_1=z_2},\nonumber\\
            &\qquad\qquad=-\left(\frac{3}{32}\left(\partial_{t_1}\partial_{t_2}-\partial_{y_1}\partial_{y_2}\right)^2+\frac{1}{32}\left(\partial_{y_1}\partial_{t_2}-\partial_{t_1}\partial_{y_2}\right)^2\right)G(x,x')|_{x_1= x_2, z_1=z_2}.
        \end{align}

        \section{Expectation value of the Kretschmann scalar} \label{appendixc}
        
         Since there is no local stress tensor in gravity,  we use Kretschmann scalar to probe the quench created 
         by curvature components. 
        Let us consider the curvature component
         $R_{tyty}$.  We wish to evaluate the $3$ point function
          \begin{eqnarray} \label{k1}
          {\rm Tr} [ K (0, 0) \rho(t, -y_0) ]&=&
          {\rm Tr} [K (0, y_0)  \rho (t, 0)] , \\ \nonumber
          &=& \frac{\langle R_{tyty} ( - \epsilon - i t, 0 ) K (0, y_0)
           R_{tyty} (  \epsilon - i t, 0 ) \rangle}{ \langle R_{tyty} ( - \epsilon - i t, 0 ) R_{tyty} (  \epsilon - i t, 0 ) \rangle},
          \end{eqnarray}
          where
          \begin{equation}
          K = R_{\mu\nu\rho\sigma} R^{\mu\nu\rho\sigma}.
          \end{equation}
          We have not explicitly mentioned the coordinates in the $x, z$ plane since 
         the Kretschman scalar is located  at the same point in the $x, z$ plane at which the quench is created. This makes the computation easier.  Opening up the sum in the Kretschman scalar, we see that we see that by Wick contractions the $3$-point function in the numerator breaks up into product of $2$ point functions. 
         The $2$ point functions involve the component of $R_{tyty}$ with every other component of the Riemann tensors.
         Since we have already grouped the components of the Riemann tensors into classes, we will evaluate 
     these $2$ point functions  by considering one class at a time.

        Consider  components in class 1. 
        \begin{align}
           \langle R_{tyty}  R_{tyty}\rangle|_{x_1= x_2, z_1=z_2}&= \frac{1}{2}\left(\partial_{t_1}^2+\partial_{y_1}^2\right)^2G(x,x')|_{x_1= x_2, z_1=z_2}.\nonumber\\
        \end{align}
        Similarly, one obtains
        \begin{align}
           \langle  R_{tyty}R_{xzxz}\rangle|_{x_1= x_2, z_1=z_2}&= \frac{1}{2}\left(\partial_{t_1}^2+\partial_{y_1}^2\right)^2G(x,x')|_{x_1= x_2, z_1=z_2}.
        \end{align}
    From  (\ref{defcor1}) we see  $\langle h_{ai}h_{c'd'}\rangle=0$,  
    this implies 
    \begin{align}
        \langle R_{tyty} R_{tyxz}\rangle|_{x_1= x_2, z_1=z_2}&=0.
    \end{align}

        We now evaluate the two-point function of $R_{tyty}$ with the field operators belonging to class $2$ and class $3$. Components in this class can be organised as vectors of $SO(2)_T\times SO(2)_L$.  
        \begin{align}
         & \langle R_{tyty}R_{tytx}\rangle|_{x_1= x_2, z_1=z_2}  \nonumber \\
         &\qquad=\frac{1}{4}\Big[2\partial_{y_1}\partial_{t_1}\partial_{t_2}\partial_{x_2}\langle h_{ty}h_{yt}\rangle-2\partial_{y_1}\partial_{t_1}\partial_{y_2}\partial_{x_2}\langle h_{ty}h_{tt}\rangle-\partial_{t_1}^2\partial_{t_2}\partial_{x_2} \langle h_{yy} h_{yt}\rangle\nonumber\\
            &\qquad+\partial_{t_2}^2\partial_{x_2}\partial_{y_2}\langle h_{yy}h_{tt}\rangle-\partial_{y_1}^2\partial_{t_2}\partial_{x_2}\langle h_{tt}h_{yt}\rangle+\partial_{y_1}^2\partial_{x_2}\partial_{y_2}\langle h_{tt}h_{tt}\rangle\Big]|_{x_1= x_2, z_1=z_2},\nonumber\\
            &\qquad=0.
        \end{align}
        The last line follows from the fact that
        $$\partial_{x_2}G(x,x')|_{x_1= x_2, z_1=z_2}=0.$$
          Similarly it is easy to show in the coincident limit,  any odd number of derivative with respect to the coordinate along the direction of the defect acting on the scalar two-point function on the flat space will also vanish.
        We demonstrate one more correlator
        \begin{align}
               \langle R_{tyty}R_{txxz}\rangle|_{x_1= x_2, z_1=z_2}&=-    \frac{1}{2}\partial_{t_2}\partial_{z_2}(\partial_{x_2}^2+\partial_{z_2}^2)G(x,x')|_{x_1= x_2, z_1=z_2},\nonumber\\
            &=0.
        \end{align}
          Therefore, all the two-point functions of $R_{tyty}$ with components in  class $2$ and class $3$ 
          vanishes when we take the positions along $x, z$ directions to coincide.

       We evaluate the correlators with  components in   class $4$ 
        \begin{align}
               \langle R_{tyty}R_{txtx}\rangle|_{x_1= x_2, z_1=z_2}&=   \langle R_{tyty}R_{tztz}\rangle|_{x_1= x_2, z_1=z_2}
             ,  \nonumber \\
           & =   \langle R_{tyty}R_{yxyx}\rangle|_{x_1= x_2, z_1=z_2}\nonumber \\
            &=   \langle R_{tyty} R_{yzyz}\rangle|_{x_1= x_2, z_1=z_2},\nonumber\\
            &=    \frac{1}{2}\partial_{x_1}^2(\partial_{t_2}^2+\partial_{y_2}^2)G(x,x')|_{x_1= x_2, z_1=z_2},\nonumber\\
            &=-\frac{1}{4}(\partial_{t_1}^2+\partial_{y_1}^2)^2G(x,x')|_{x_1= x_2, z_1=z_2}.
        \end{align}
        To obtain the last line we have used (\ref{isoa1}) and the on-shell condition of the scalar Greens function. 
       
      We evaluate the two-point functions of $R_{tyty}$ with curvature components in  class $5$. But note that, these components $R_{txtz}$ and $R_{yxyz}$, 
       have two indices in the $x,z$   directions, which are parallel direction to the defect. 
       Therefore all   terms in the two-point functions will involve   two derivatives with respect to $x$ and $z$. 
      But in the coincident limit
      \begin{equation}
       \partial_{x_2}\partial_{z_2}G(x,x')|_{x_1= x_2, z_1=z_2}=0.
       \end{equation}
       Therefore, all the two-point functions  $R_{tytty}$ with curvature 
        components in class $5$ vanish in the coincident limit.
      
      Consider two-point functions  of $R_{tyty}$ with components in class $6$. 
      \begin{align}
            \langle R_{tyty}R_{txyx}\rangle|_{x_1= x_2, z_1=z_2}&=  \langle R_{tyty}R_{tzyz}\rangle|_{x_1= x_2, z_1=z_2},\nonumber\\
           &=  \frac{1}{2}\partial_{t_2}\partial_{y_2}(\partial_{x_1}^2-\partial_{z_1}^2)G_{f}(x,x')|_{x_1= x_2, z_1=z_2}|_{x_1= x_2, z_1=z_2},\nonumber\\
           &=0.
           \end{align}
           The last line follows from the isotropy relations (\ref{isoa1}) and (\ref{isoa2}).
          Therefore components in class  $6$ do not contribute to the expectation value of the Kretschman scalar.

           We are left with one more correlator from class 7,  which is $\langle R_{tyty} R_{tyxz}\rangle$.
            But, note that  all terms in 
            this two-point function  have  a single  derivative each  in both $x$ and $z$ directions. 
           Therefore in the coincident limit this too will vanish.
           \begin{align}
                 \langle R_{tyty}R_{txyz}\rangle|_{x_1= x_2, z_1=z_2}&=0.
           \end{align}

       Using all these two-point functions, we evaluate \eqref{k1} in the coincident limit in the $x,z$ direction
       \begin{align} \label{kexpect}
        \frac{\langle R_{tyty} ( - \epsilon - i t, 0 ) K (0, y_0)
           R_{tyty} (  \epsilon - i t, 0 ) \rangle}{ \langle R_{tyty} ( - \epsilon - i t, 0 ) R_{tyty} (  \epsilon - i t, 0 ) \rangle}
            &=\frac{6144 \epsilon ^6}{\left(t^4+2 t^2 \left(\epsilon ^2-y_0^2\right)+\left(y_0^2+\epsilon ^2\right)^2\right)^3}.
       \end{align}

        \section{Relations due to isotropy}\label{idproof}
        
        In this appendix we prove a few relations which relate derivatives on the Green's function along the $x$ and $z$ directions at the coincident limit along the directions parallel to the defect. 
        The first relation we prove is the isotropy relation in (\ref{isotropy}). 
        For this it is useful to consider the Green's function (\ref{sclcor}) 
         in Fourier space along the $x, z$ direction which is 
        defined by 
        \begin{equation}
        \tilde G(r_1, r_2, \theta_1, \theta_2 , k_x, k_z )  = \int_{-\infty}^\infty 
        dx dz  G( r_1, r_2, \theta_1, \theta_2, x, z) e^{ - i ( k_x  x + k_z z ) }.
        \end{equation}
        There is translation invariance in $x, z$ directions, so we  have replaced $x_1 -x_2, z_1 -z_2$ by  $x, z$ respectively. 
        It is easy to show that due to rotational invariance in the directions $x, z$, the  Fourier transform 
        $  \tilde G(r_1, r_2, \theta_1, \theta_2 , k_x, k_z ) $ is only a function of the form 
        $  \tilde G(r_1, r_2, \theta_1, \theta_2 , k_x^2 + k_z^2  ) $  or   
        $  \tilde G(r_1, r_2, \theta_1, \theta_2 , k    ) $, where $k = \sqrt{  k_x^2 + k_z^2}$. 
        Now let us consider 
        \begin{eqnarray} \label{gnft1}
        \partial_x^2  G( r_1, r_2, \theta_1, \theta_2, x, z)|_{kx=z=0} &=& 
     \frac{1}{( 2\pi)^2}  \int_0^\infty k dk   \int_0^{2\pi}   d\phi  k_x^2    \tilde G(r_1, r_2, \theta_1, \theta_2 , k   ), 
     \\ \nonumber
     &=& \frac{2\pi}{2}   \frac{1}{( 2\pi)^2} \int_0^\infty k dk   k^2 \tilde G(r_1, r_2, \theta_1, \theta_2 , k   )
        \end{eqnarray}
       To obtain the last line we have substituted $k_x = k \cos\phi$ and performed the angular integral. 
      Compare this to action of the Laplacian along the longitudinal directions
        \begin{eqnarray} \label{gnft2}
         \nabla^2  G( r_1, r_2, \theta_1, \theta_2, x, z)|_{x=z=0} &=& 
     \frac{1}{( 2\pi)^2}  \int_0^\infty k dk   \int_0^{2\pi}   d\phi  ( k_x^2   + k_z^2)   \tilde G(r_1, r_2, \theta_1, \theta_2 , k   ), 
      \nonumber \\
     &=&  (2\pi)    \frac{1}{( 2\pi)^2} \int_0^\infty k dk   k^2 \tilde G(r_1, r_2, \theta_1, \theta_2 , k   )
        \end{eqnarray}
     Comparing  (\ref{gnft1}) and (\ref{gnft2}), we obtain 
        \begin{eqnarray} \label{isoa1}
         \partial_x^2  G( r_1, r_2, \theta_1, \theta_2, x, z)|_{x=z=0} &=&
          \frac{1}{2} \nabla^2  G( r_1, r_2, \theta_1, \theta_2, x, z)|_{x=z=0}
          \end{eqnarray}
          A very similar analysis shows
              \begin{eqnarray} \label{isoa2}
         \partial_z^2  G( r_1, r_2, \theta_1, \theta_2, x, z)|_{x=z=0} &=&
          \frac{1}{2} \nabla^2  G( r_1, r_2, \theta_1, \theta_2, x, z)|_{x=z=0}
          \end{eqnarray}
          
          Now consider 
            \begin{eqnarray} \label{gnft3}
        \partial_x^4  G( r_1, r_2, \theta_1, \theta_2, x, z)|_{x=z=0} &=& 
     \frac{1}{( 2\pi)^2}  \int_0^\infty k dk   \int_0^{2\pi}   d\phi  k_x^4    \tilde G(r_1, r_2, \theta_1, \theta_2 , k   ), 
      \nonumber \\
     &=& \frac{3}{8} \times  2\pi  \frac{1}{( 2\pi)^2} \int_0^\infty k dk   k^4 \tilde G(r_1, r_2, \theta_1, \theta_2 , k   )
        \end{eqnarray}
        To obtain the last line we have replaced $k_x^4 = k^4 \cos^4\phi$ and performed the integral over $\phi$. 
        We can compare this to the action of the square of the Laplacian
             \begin{eqnarray} \label{gnft4}
         \nabla ^4 G( r_1, r_2, \theta_1, \theta_2, x, z)|_{x=z=0} &=& 
     \frac{1}{( 2\pi)^2}  \int_0^\infty k dk   \int_0^{2\pi}   d\phi  ( k_x ^2 + k_z^2 )^2   
      \tilde G(r_1, r_2, \theta_1, \theta_2 , k   ), 
      \nonumber \\
     &=&   2\pi  \frac{1}{( 2\pi)^2} \int_0^\infty k dk   k^4 \tilde G(r_1, r_2, \theta_1, \theta_2 , k   )
        \end{eqnarray}
        Comparing (\ref{gnft3}) and (\ref{gnft4}), we see that 
        \begin{equation} \label{isoa3}
              \partial_x^4  G( r_1, r_2, \theta_1, \theta_2, x, z)|_{x=z=0} =  \frac{3}{8}     \nabla ^4 G( r_1, r_2, \theta_1, \theta_2, x, z)|_{x=z=0} .
              \end{equation}
              Again similarly we obtain 
               \begin{equation} \label{isoa4}
              \partial_z^4  G( r_1, r_2, \theta_1, \theta_2, x, z)|_{x=z=0} =  \frac{3}{8}     \nabla ^4 G( r_1, r_2, \theta_1, \theta_2, x, z)|_{x=z=0} .
              \end{equation}
              
              Finally consider 
            \begin{eqnarray} \label{gnft5}
        \partial_x^2 \partial_z^2   G( r_1, r_2, \theta_1, \theta_2, x, z)|_{x=z=0} &=& 
     \frac{1}{( 2\pi)^2}  \int_0^\infty k dk   \int_0^{2\pi}   d\phi  k_x^2  k_z^2   \tilde G(r_1, r_2, \theta_1, \theta_2 , k   ), 
      \nonumber, \\
     &=& \frac{1}{8} \times  2\pi  \frac{1}{( 2\pi)^2} \int_0^\infty k dk   k^4 \tilde G(r_1, r_2, \theta_1, \theta_2 , k   ).
        \end{eqnarray}
        where we obtain the last line by replacing $k_x^2 k_z^2 = k^4 \cos^2\phi \sin^2\phi$ and performing the integral
        over $\phi$. 
        Thus comparing (\ref{gnft5}) and (\ref{gnft4}) we obtain 
        \begin{equation} \label{isoa5}
             \partial_x^2 \partial_z^2   G( r_1, r_2, \theta_1, \theta_2, x, z)|_{x=z=0} = \frac{1}{8}     \nabla ^4 G( r_1, r_2, \theta_1, \theta_2, x, z)|_{x=z=0} .
             \end{equation}
    
\bibliographystyle{JHEP}
\bibliography{references} 
\end{document}